\documentclass[prx,twocolumn,eqsecnum,amsmath,amssymb,superscriptaddress,floatfix,english,nofootinbib]{revtex4-1}
\usepackage{epsfig, graphicx,graphics,amsmath,amssymb,float}
\usepackage[T1]{fontenc}
\usepackage[latin9]{inputenc}
\usepackage{amsmath}
\usepackage{amssymb}

\usepackage{amscd}
\usepackage{bm}
\usepackage{psfrag}
\usepackage{bbm} 
\usepackage{babel}
\usepackage{wasysym }
\usepackage{mathrsfs}
\usepackage{color}

\definecolor{green}{RGB}{22,140,22}
\definecolor{reddish}{RGB}{140,22,22}

\newcommand{\be}{\begin{equation}}
\newcommand{\ee}{\end{equation}}
\newcommand{\bea}{\begin{eqnarray}}
\newcommand{\eea}{\end{eqnarray}}

\newcommand{\kb}{k_{\rm W}}
\newcommand{\kkB}{\bar k_{\rm W}}
\newcommand{\kkkB}{\tilde k_{\rm W}}
\newcommand{\vklabel}{{\bf k}}
\newcommand{\vkplabel}{{\bf k}'}
\newcommand{\vkvector}{{\bf k}}
\newcommand{\vkpvector}{{\bf k}'}
\newcommand{\vqvector}{{\bf q}}
\newcommand{\vvvector}{{\bf v}}
\newcommand{\vkparallel}{{\bf k}_{\parallel}}
\newcommand{\vkpparallel}{{\bf k}'_{\parallel}}
\newcommand{\vqparallel}{{\bf q}_{\parallel}}
\newcommand{\vrvector}{{\bf r}}
\newcommand{\vrparallel}{{\bf r}_{\parallel}}
\newcommand{\kkvector}{k}
\newcommand{\qqvector}{q}
\newcommand{\kkparallel}{k_{\parallel}}
\newcommand{\qqparallel}{q_{\parallel}}

\begin{document}
	
	\title{Phonon-limited Transport and Fermi Arc Lifetime in Weyl Semimetals}
	
	\author{Francesco Buccheri}
	\affiliation{Institut f\"ur Theoretische Physik,
		Heinrich-Heine-Universit\"at, D-40225  D\"usseldorf, Germany}
	\author{Alessandro De Martino}
	\affiliation{Department of Mathematics, City, University of London, EC1V 0HB London, UK}
	\author{Rodrigo G. Pereira}
	\affiliation{International Institute of Physics,  Universidade Federal do Rio Grande do Norte, 
		Natal, RN, 59078-970, Brazil}
	\affiliation{Departamento de F\'isica Te\'orica
		e Experimental, Universidade Federal do Rio Grande do Norte, 
		Natal, RN, 59078-970, Brazil}	
	\author{Piet W. Brouwer}
	\affiliation{Dahlem Center for Complex Quantum Systems and Institut f\"ur Physik,
		Freie Universit\"at Berlin, Arnimallee 14, D-14195 Berlin, Germany}
	\author{Reinhold Egger}
	\affiliation{Institut f\"ur Theoretische Physik,
		Heinrich-Heine-Universit\"at, D-40225  D\"usseldorf, Germany}
	
	\begin{abstract}
		Weyl semimetals harbor topological Fermi-arc surface states which determine the 
		nontrivial charge current response to external fields. 
		We here study the quasiparticle decay rate of Fermi arc states arising 
		from their coupling to acoustic phonons, as well as the phonon-limited 
		conductivity tensor for a clean Weyl semimetal slab.
		Using the phonon modes for an isotropic elastic continuum with a deformation potential coupling to electrons,
		we determine the temperature dependence of the quasiparticle decay rate, both near and far away from the arc termination points.
		By solving the coupled Boltzmann equations for the bulk and arc state distribution functions in the slab geometry, we show how the linear response
		conductivity depends on key parameters such as the temperature, the chemical potential,
		the geometric shape of the Fermi arcs, or the slab width.
		The chiral nature of Fermi arc states causes an enhancement of the longitudinal conductivity along the chiral direction at low 
		temperatures, together with a $1/T^2$ scaling regime at intermediate temperatures without counterpart for the conductivity 
		along the perpendicular direction. 
	\end{abstract}
	\date{\today}
	\maketitle
	
	\section{Introduction}\label{sec1}
	
	Reaching a firm understanding of three-dimensional Weyl semimetal (WSM) materials 
	represents an important goal of modern condensed matter physics
	\cite{Hosur2013,Burkov2015,Burkov2016,Yan2017,Hasan2017,Armitage2018,Burkov2018}. 
	WSMs are characterized by a gapped quasiparticle spectrum throughout the Brillouin zone, with the exception of
	an even number of nondegenerate band touchings called Weyl nodes.  
	Due to the breaking of inversion and/or time-reversal symmetry, 
	Kramers degeneracy is absent and relativistic Weyl fermions represent the relevant low-energy degrees of freedom.  
	The Weyl nodes act as sources (or sinks) of Berry curvature 
	and thus can be associated with a topological charge \cite{Nielsen1981b}.
	As a consequence, many interesting physical effects of topological origin 
	have been predicted and observed in WSMs.
	For instance, WSMs allow for striking manifestations of the chiral Adler-Bell-Jackiw anomaly \cite{Adler1969,Bell1969}, 
	such as a negative magnetoresistivity in parallel electric and magnetic fields
	\cite{Zyuzin2012,Baireuther2016,Son2013b,Burkov2014,Gorbar2014,Spivak2016,Andreev2018,Ishizuka2019}.
	The relativistic low-energy Weyl cone spectrum and the associated nontrivial response 
	to external electromagnetic fields \cite{Goswami2013,Vazifeh2013} represent clear hallmarks of Weyl materials. 
	
	The gapless bulk Weyl nodes must coexist on general grounds with gapless and topologically protected
	\emph{Fermi-arc  surface states}, which connect the projections on the 
	surface Brillouin zone of different Weyl nodes. 
	These surface states are chiral, i.e., have a unidirectional sense of propagation, and define open curves as Fermi surface segments  
	\cite{Wan2011,Hosur2012,Haldane2014,Gorbar2016,Wilson2018,Resta2018,Dwivedi2018,Breitkreiz2019,Gorbar2019,Chatto2019,Mukherjee2019,Behrends2019,Chen2020,Murthy2020}. The Fermi arc parts of the Fermi surface seamlessly
	merge with the bulk quasiparticle parts at the arc termination points \cite{Haldane2014,Armitage2018,Li2020b,Bovenzi2018}.   
	Upon approaching the latter points, the penetration depth of the Fermi-arc surface states into the bulk diverges. 
	
	Angle-resolved photoemission spectroscopy (ARPES) and scanning tunneling microscopy (STM)
	experiments have confirmed the existence of Fermi arc states in various transition metal compounds such as
	TaAs, TaP, NbAs, or NbP \cite{Yang2011,Yang2015,Weng2015,Xu2015,Huang2015,Lv2015a,Lv2015,Inoue2016,Yan2017,Hasan2017,Yang2019,Min2019}.
	Moreover, for the magnetic WSM material Co$_3$Sn$_2$S$_2$, Fermi arc states have been observed by ARPES and STS \cite{Liu2019b,Morali2019}.
	These surface-sensitive probe techniques have shown that the geometric shape and the corresponding spin 
	orientations of a constant-energy arc in the surface Brillouin zone depends on the specific WSM material.  
	At the same time, however, Fermi-arc surface states are directly responsible for a plethora
	of universal (material-independent) phenomena, e.g., density of states oscillations \cite{Hosur2012b}, 
	supercurrent oscillations \cite{Li2020}, 
	unusually quantized semiclassical orbits in a magnetic field \cite{Potter2014,Parameswaran2014}, or
	anomalous charge \cite{Hosur2012,Shekhar2015,Breitkreiz2019,Zhang2019} and heat transport  \cite{Xiang2019,Burrello2019}. Fermi arc states are also connected to the
	anomalous Hall effect in magnetic WSMs \cite{Burkov2011a,Gorbar2014,Li2020b}. The latter has
	recently been observed experimentally \cite{Liu2018,Wang2018,Chen2019,Shen2020}.
	
	Based on topological arguments, one may expect that arc states give rise to non-dissipative transport phenomena.  
	However, the gapless nature of WSMs implies that this is a rather subtle issue.
	Indeed, if arc and bulk states are connected by some arc-bulk scattering mechanism, 
	arc transport will generally be dissipative.  Such a mechanism has been identified in 
	terms of elastic disorder scattering in Ref.~\cite{Gorbar2016}, see also Ref.~\cite{Perez2021}. We note that Weyl points survive the presence of weak disorder
	\cite{Buchhold2018a,Buchhold2018b}, which also implies that arc states remain well-defined \cite{Slager2017,Wilson2018}.
	Disorder can arise due to randomly distributed impurities or due to sample inhomogeneities  \cite{Shapourian2016,Chen2015}. However, ultrahigh mobilities have been reported for WSM materials, e.g., NbP \cite{Shekhar2015}, and disorder could even be eliminated altogether in fully controlled artificial (metamaterial) WSM realizations \cite{Weststroem2017}.
	
	In this work, we study the quasiparticle decay rate of Fermi arc states 
	and the temperature-dependent conductivity for a clean (disorder-free) WSM slab, assuming that acoustic phonons
	provide the most important electron scattering and equilibration mechanism.
	We note that this quasiparticle decay rate also governs the energy transfer between electrons and phonons \cite{Lundgren2015}.
	We will not take into account optical phonons, which have recently been studied both theoretically~\cite{Song2016,Rinkel2017,Liu2017,Gordon2018,Rinkel2019,Coulter2019} and experimentally \cite{Zhang2020,Hein2020,Osterhoudt2021}, but instead focus on acoustic phonons which dominate at low temperatures. 
	While we investigate phonon-induced effects on the electronic properties of WSMs, 
	it is also of significant interest to study electron-induced effects on phonon observables. 
	For instance, recent works have addressed the Kohn anomaly \cite{Xue2019,Nguyen2020,Yue2020}, 
	quantum oscillations of the sound velocity \cite{Laliberte2020,Zhang2020b} in WSMs, 
	and the phonon magnetochiral effect where one finds a direction-dependent sound velocity in a magnetic field  \cite{Sengupta2020,Antebi2021,Sukhachov2021}.
	Future theoretical work could study such phenomena using the framework presented below.
	
\begin{figure}
		\includegraphics[width=\columnwidth]{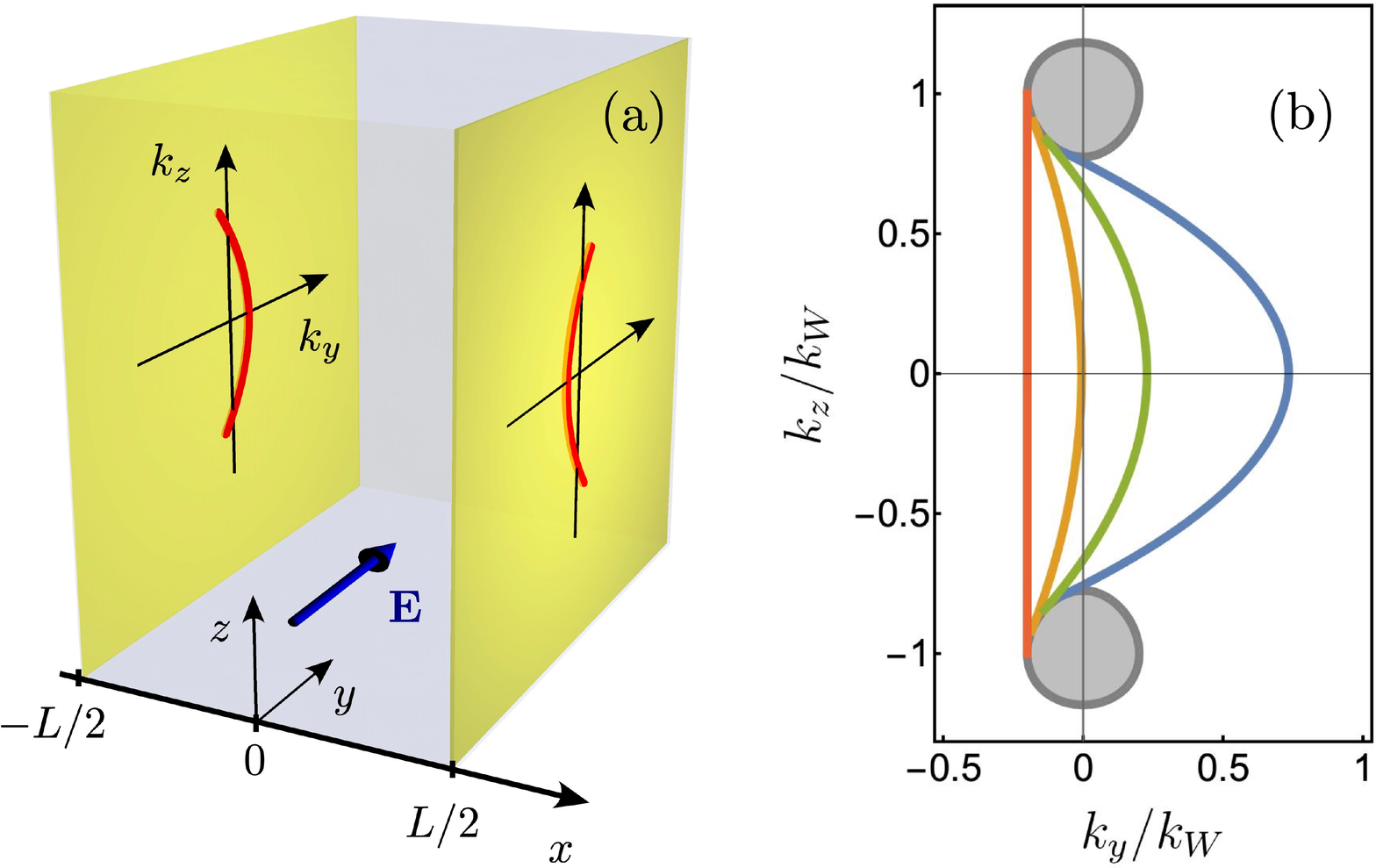}
		\caption{ WSM slab geometry. (a) The sample is infinitely extended along the $(y,z)$-directions, with conserved momentum $\vkparallel=(k_y,k_z)$, and has the transverse width $L$ along the $x$-direction.  Fermi-arc surface states are sketched in the surface Brillouin zone for $\alpha>0$ and constant energy $\varepsilon>0$.  (b)  Arc shapes in the $(k_y,k_z)$-plane, see Eq.~\eqref{eq:k2karch},
			for the surface state at $x=-L/2$ with constant energy $\varepsilon=0.2 k_{\rm W}v$ and different angles $\alpha=0, 0.4, 0.8, 1.2$ (from left to right).  The shaded discs indicate phase space regions where bulk states are present. They appear slightly elongated in the $z$-direction
			due to the anisotropy of the bulk dispersion relation.
		}  
		\label{fig1}
	\end{figure}

	Let us next describe the structure of this article, along with a summary of our main results.
	In Sec.~\ref{sec2}, we describe our model.  The electronic properties of a WSM  
	are modeled in terms of a well-known inversion-symmetric two-band model with broken time reversal symmetry, featuring just two Weyl nodes \cite{Okugawa2014}. We consider a slab geometry with finite width $L$ along the $\hat x$ direction, see Fig.~\ref{fig1}, where
	the Weyl points are separated by the vector $2k_{\rm W}\hat z$ in the bulk Brillouin zone and the chiral direction is
	denoted by the unit vector $\hat y$.  In Sec.~\ref{sec2a} we diagonalize the electronic problem 
	with boundary conditions parametrized by a phenomenological angle $\alpha$ \cite{Burrello2019}.  
	At fixed energy, we obtain chiral Fermi-arc surface states whose dispersion generally has a curved geometrical shape in the surface Brillouin zone: For $\alpha=0$, one finds straight arcs, while $\alpha\to \pi/2$ corresponds to widely open arcs. 
	Next, in Sec.~\ref{sec2b}, we specify our model for the phonon Hamiltonian based on isotropic elastic continuum theory, see also Refs.~\cite{Landau1986,Bannov1994,Giraud2012}.  We assume that acoustic phonons couple to electrons via the deformation potential, see Sec.~\ref{sec2c}. (Our theory can also be adapted to other phonon models, e.g., 
	as obtained from \emph{ab initio} calculations \cite{Garcia2020}.) 
	We introduce the relevant Bloch-Gr\"uneisen temperature scales  in Sec.~\ref{sec2d}.  
	
	In Sec.~\ref{sec3}, we apply Boltzmann theory to the case of phonon-induced transport in the WSM slab geometry
	of Fig.~\ref{fig1}.   In Sec.~\ref{sec3a}, we present the Boltzmann equations for the bulk and arc state
	distribution functions.  We here focus on the linear response regime, where a linearized 
	version of the Boltzmann equations is sufficient, see Sec.~\ref{sec3b}.  Since sound velocities are typically two
	orders of magnitude below the Fermi velocity, we also implement a quasi-elastic approximation. 
	Finally, in Sec.~\ref{sec3c}, we discuss the decay rate for bulk quasiparticles and the applicability conditions 
	for our theory.
	
	In Sec.~\ref{sec4}, we address the temperature-dependent decay rate $\Gamma$ of Fermi arc states.  
	This rate receives contributions from arc-arc scattering, see Sec.~\ref{sec4a}, 
	and from arc-bulk scattering, see Sec.~\ref{sec4b}, and it may 
	be observed through the linewidth of ARPES peaks \cite{Behrends2016}. 
	We find  different temperature scaling regimes which depend on the position along the arc.
	At low temperatures and away from the arc edges, the arc-arc contribution dominates and yields 
	$\Gamma\propto T^3$ because arc-bulk scattering is activated in general.
	However, the activation energy for arc-bulk scattering vanishes upon approaching the arc edges,
	where we find a low-temperature regime with $\Gamma\propto T^{5/2}$.
	
	In Sec.~\ref{sec5}, we discuss the temperature dependence of the conductivity tensor. 
	We first provide qualitative arguments for the longitudinal conductivity along the
	chiral direction ($\sigma_{yy}$), see Sec.~\ref{sec5a}, and along the perpendicular direction  ($\sigma_{zz}$), 
	see Sec.~\ref{sec5b}.  We find the same power law scaling for both conductivities at very low ($\sigma_{jj}\propto 1/T^5$) and at 
	high $(\sigma_{jj}\propto 1/T$) temperatures. However, the chirality of arc states admits
	an intermediate regime with $\sigma_{yy}\propto 1/T^2$ which has no counterpart in $\sigma_{zz}$.
	We then describe a numerical solution of the coupled Boltzmann 
	integral equations in Sec.~\ref{sec5d}. The corresponding results confirm our qualitative analysis in Secs.~\ref{sec5a} and \ref{sec5b}.  
	Apart from the temperature dependence of $\sigma_{jj}$, we study the effects of changing the surface parameter $\alpha$,
	the chemical potential $\mu$, or the slab width $L$. 
	
	The paper concludes with an outlook in Sec.~\ref{sec6}.  Technical details have been relegated 
	to several Appendices. Throughout, the electron charge is denoted by $e<0$ and we often set $k_{\rm B}=\hbar=1$. 
	
	\section{Model}\label{sec2}
	
	In this section, we describe the model employed in our study. 
	In Sec.~\ref{sec2a}, we introduce a two-band model for electrons in a WSM slab 
	with only two Weyl nodes \cite{Okugawa2014}. 
	We impose boundary conditions which depend on a phenomenological angle $\alpha$ quantifying
	the curvature of topological Fermi arc states in the surface Brillouin zone \cite{Burrello2019}. 
	A description of acoustic phonons using elastic continuum theory \cite{Landau1986,Bannov1994,Giraud2012} is given in
	Sec.~\ref{sec2b}. To allow for a theoretical description of phonon-mediated scattering of arc as well as bulk electron states, 
	in Secs.\ \ref{sec2a} and \ref{sec2b} we give expressions for electron wavefunctions and phonon displacement fields in the bulk of the slab as well as near its surfaces.
	In Sec.\ \ref{sec2c} we discuss electron-phonon coupling in the framework of the
	deformation potential, using the wavefunctions and displacement fields calculated in Secs.\ \ref{sec2a} and \ref{sec2b} to construct the electron-phonon matrix elements.
	We discuss characteristic temperatures of our model in Sec.~\ref{sec2d}.

	\subsection{Electronic model and Fermi arc states}   \label{sec2a}
	
	We start from a well-known two-band model for a WSM with only two Weyl nodes \cite{Vazifeh2013,Okugawa2014,Gorbar2016,Burrello2019} separated in the $z$ direction in reciprocal space. We consider a slab geometry for which the system is taken as infinitely extended in the $y$ and $z$ directions, see Fig.~\ref{fig1}. We use the notation $\vrvector = (x,\vrparallel)$, where $\vrparallel = (y,z)$ contains the in-plane coordinates, and use $\hat x$, $\hat y$, and $\hat z$ to denote the unit vectors in the directions of the coordinate axes. In the same way, the momentum is written as $\vkvector = (k_x,\vkparallel)$, where the in-plane momentum $\vkparallel = (k_y,k_z)$ is conserved because of translation symmetry in the $y$ and $z$ directions. We write $\kkvector = |\vkvector|$ and $\kkparallel = |\vkparallel|$. In the $x$-direction, the slab has the width $L$. We choose the origin of the coordinate system such that the surfaces of the slab are at $x = \pm L/2$. 
	
	The electrons are described by the two-band Hamiltonian \cite{Vazifeh2013,Okugawa2014,Gorbar2016,Burrello2019}
	\begin{eqnarray}\nonumber
	H_0(\vkvector) & = & v\left(\sigma_{x}k_{x}+\sigma_{y}k_{y}\right)+m\left(k_{z}\right)\sigma_{z},\\
	m\left(k_z\right) & =& \frac{v}{2\kb}\left(k_{z}^{2}-\kb^{2}\right),\label{eq:Hden-1}
	\end{eqnarray}
	where $k_x = -i \partial/\partial x$, $\sigma_{x,y,z}$ are Pauli matrices acting in a combined spin-orbital space, and $v$ is the Fermi velocity. 
	The time-reversal symmetry breaking parameter $\kb>0$ determines the distance between the two Weyl points in momentum space, which are at $\vkvector = (0,0,\pm \kb)$ and energy $\varepsilon = 0$. (The separation between the two Weyl points is assumed to be parallel to the sample surfaces.) 
	A lattice model that has Eq.~(\ref{eq:Hden-1}) as its low-energy limit was considered in Ref.~\cite{Bovenzi2018}.
	One easily checks that $H_0$ is invariant under inversion $I$, 
	\be
	\label{eq:inversion}
	H_0(k_x,\vkparallel)=\sigma_{z}H_0\left(-k_x,-\vkparallel\right)\sigma_{z},
	\ee
	and under the magnetic twofold rotation symmetry $C_{2x}T$,
	\be
	\label{eq:rotation}
	H_0(k_x,\vkparallel) = \sigma_z H_0^*(-k_x,\vkparallel) \sigma_z.
	\ee
	Although these symmetries are not essential for the temperature dependence of the low-temperature conductivity, 
	their presence helps to simplify our expressions.
	
	
	At $x = \pm L/2$ we impose boundary conditions for the two-component spinor $|\Phi(\vrvector)\rangle$, parameterized by the angles $\alpha_{\pm}$ \cite{Witten2016}
	\begin{eqnarray}\label{eq:bc}
	|\Phi(\pm L/2,\vrparallel)\rangle &=& \pm M(\alpha_{\pm}) |\Phi(\pm L/2,\vrparallel)\rangle,
	\\
	M(\alpha)  &=&  \sigma_{y}\cos\alpha+\sigma_{z}\sin\alpha . \nonumber
	\end{eqnarray}
	Equation \eqref{eq:bc} automatically ensures that the transverse component of the
	charge current vanishes at the slab surfaces.  Inversion symmetry imposes the condition $\alpha_- = - \alpha_+ \equiv \alpha$. 
	Similar boundary conditions have been used before for related WSM models \cite{Gorbar2016} and for different geometries \cite{Erementchouk2018,Burrello2019,DeMartino2021}. The boundary condition \eqref{eq:bc} is compatible with the 
	magnetic twofold rotation symmetry (\ref{eq:rotation})
	and forces the pseudospin for $x = \pm L/2$ to be in the $yz$-plane, at an angle $\alpha_{\pm}$ with the positive ($+$) or negative ($-$) $y$ axis. Specifically, setting $\alpha_- = -\alpha_+ = \alpha$, Eq.\ \eqref{eq:bc} implies that 
	\be
	\label{eq:Phiprop}
	\langle \xi_{\pm}(\alpha) | \Phi(\pm L/2)\rangle = 0,
	\ee
	where
	\be
	| \xi_{\pm}(\alpha) \rangle =
	\left( \begin{array}{cc} \pm \sin(\pi/4 + \alpha/2) \\
		-i \cos(\pi/4 + \alpha/2) \end{array} \right).
	\ee
	
	\subsubsection{Bulk states}
	
	The bulk spectrum of the Hamiltonian \eqref{eq:Hden-1} is
	\begin{equation}\label{eq:eenrgy}
	\varepsilon^{(b)}_{\vkvector,\eta=\pm} = \eta \sqrt{v^2(k_x^2 + k_y^2) + m^2(k_z)}.
	\end{equation} 
	The corresponding bulk eigenstates are 
	\footnote{In this section, for the sake of notational simplicity, 
		we omit various $2\pi$ normalization factors of the plane wave states which are fully restored from Sec.~\ref{sec3} on. 
		Our expressions for the coupling matrix elements ${\cal G}$ describing the various scattering processes  are not affected by this notational simplification.}
	\begin{equation}
	\label{eq:PhiBulk}
	|\Phi^{(b)}_{\vkvector,\eta}(\vrvector)\rangle = e^{i \vkvector \cdot \vrvector}\, |\xi_{\vkvector,\eta}\rangle,
	\end{equation}
	with the normalized two-component spinors $|\xi_{\vkvector,\eta} \rangle$
	\bea
	|\xi_{\vkvector,\eta}\rangle &=& \frac{1}{\sqrt{(\varepsilon_{\vkvector,\eta}^{(b)} +m(k_z))^2 + v^2(k_x^2+k_y^2)}} \nonumber \\ && \mbox{} \times
	\left( \begin{array}{cc} \varepsilon_{\vkvector,\eta}^{(b)} + m(k_z) \\ v(k_x+ik_y) \end{array} \right).
	\eea
	The bulk states have velocity
	\bea
	\label{eq:bulkvelocities} 
	\vvvector^{(b)}_{\eta}(\vkvector) &=& \partial_{\vkvector} \varepsilon^{(b)}_{\vkvector,\eta}
	\nonumber \\ &=&
	\frac{v^2(k_x \hat x + k_y \hat y)}{\varepsilon^{(b)}_{\vkvector,\eta}}
	+ \frac{v k_z m(k_z) \hat z}{\kb \varepsilon^{(b)}_{\vkvector,\eta}}.
	\eea
	In this work, we consider a positive chemical potential 
	$\mu$ much larger than temperature, so that the negative energies $\varepsilon^{(b)}_{\vkvector,-}$ can be disregarded; we will drop the index $\eta$ henceforth. For energies $0 < \varepsilon < v \kb/2$, which is the range corresponding to well-separated Weyl nodes, the bulk density of states (DoS) is
	\be\label{bulkdos}
	n_b\left(\varepsilon \right)= 
	\frac{\kb\varepsilon}{2\pi^2 v^2} \left( \sqrt{1+\frac{2\varepsilon}{v\kb}} - \sqrt{1-\frac{2\varepsilon}{v\kb}} 
	\right).
	\ee  
	
	With the help of the boundary condition \eqref{eq:bc}, we may find expressions for the bulk eigenstates of $H_0$ near the surfaces of the slab at $x = \pm L/2$. Labeling the states at the boundary by the in-plane momentum $\vkparallel$ and a positive transverse momentum $k_x > 0$, we write the bulk states near the surface at $x = \pm L/2$ as the combination of an incident plane wave at transverse momentum $\pm k_x$ and a reflected wave at transverse momentum $\mp k_x$,
	\bea
	\label{eq:PhiInterface}
	|\Phi_{k_x,\vkparallel}^{(b)\pm}(\vrvector)\rangle &=& 
	e^{\pm i k_x (x \mp L/2)+i \vkparallel \cdot \vrparallel}
	|\xi_{(\pm k_x,\vkparallel)}\rangle
	\\ && \nonumber \mbox{}
	- r_{k_x,\vkparallel}^{\pm} e^{\mp i k_x (x \mp L/2)+i \vkparallel \cdot \vrparallel}
	|\xi_{(\mp k_x,\vkparallel)}\rangle, 
	\eea
	where 
	\be\label{reflmat}
	r_{k_x,\vkparallel}^{\pm} =
	\frac{\langle \xi_{\pm}(\alpha) | \xi_{(\pm k_x,\vkparallel)}\rangle}{\langle \xi_{\pm}(\alpha)
		| \xi_{(\mp k_x,\vkparallel)}\rangle}
	\ee
	is the (unitary) reflection amplitude. For $k_x = 0$ (bulk modes propagating parallel to the surfaces), one has $r^{\pm}_{k_x,\vkparallel} = 1$, while for $k_y = m(k_z) = 0$ (bulk modes propagating perpendicular to the interface), one finds $r^{\pm}_{k_x,\vkparallel} = i e^{- i \alpha}$.

	\subsubsection{Fermi-arc surface states}
	
	In addition to the bulk solutions \eqref{eq:PhiInterface}, which have real transverse momentum $k_x$, there are Fermi-arc surface states localized at $x=\pm L/2$ with 
	imaginary $k_x=i\kappa_\pm (\vkparallel)$ \cite{Okugawa2014}.  
	The corresponding two-component spinor eigenstates decay exponentially 
	away from the respective surface with decay length $\kappa_{\pm}^{-1}(\vkparallel)$,  
	\be\label{eq:boundarystte}
	|\Phi^{(s)\pm}_{\vkparallel}(\vrvector)\rangle = \sqrt{2\kappa_\pm(\vkparallel)}\,
	e^{\pm \kappa_{\pm}(\vkparallel ) (x \mp L/2) + i \vkparallel  \cdot \vrparallel }  
	\left|\xi_{\mp}(-\alpha)\right\rangle,
	\ee
	where we find
	\be\label{kappadef}
	\kappa_\pm(\vkparallel )= \mp k_y \sin\alpha -\frac{m(k_z)}{v} \cos\alpha.
	\ee
	The arc state \eqref{eq:boundarystte} exists only for $\vkparallel$ for which $\kappa_{\pm}(\vkparallel) > 0$, and the
	dispersion relation is given by
	\be\label{eq:Es}
	\varepsilon^{(s)\pm}_{\vkparallel} = \pm vk_{y}\cos\alpha - m(k_{z})\sin\alpha.
	\ee
	The velocity of the arc states near the surface at $x= \pm L/2$ is locally orthogonal to the constant-energy arc in the surface Brillouin zone,
	\be
	\label{eq:arcvelocities}
	\vvvector^{(s)\pm}_{\parallel}(\vkparallel)  = \pm v \cos (\alpha) \hat y  -\frac{k_z}{\kb} v \sin( \alpha) \hat z.
	\ee
	
	In later calculations, we will find it convenient to employ the variables $(\varepsilon,k_z)$ instead of $\vkparallel=(k_y,k_z)$ to parameterize the arc states at the surface at $x = \pm L/2$. 
	Using Eq.~\eqref{eq:Es}, we see that constant-energy arc states form an open curve in the 
	surface Brillouin zone. The termination points correspond to an inverse penetration depth $\kappa_{\pm}(\vkparallel ) \to 0$, so that the arc states spread over the entire sample and surface and bulk states merge \cite{Witten2016}. 
	The arc at constant energy $\varepsilon > 0$ extends in the interval 
	\be  \label{eq:kkB}
	-\kkB(\varepsilon)\le k_z \le \kkB(\varepsilon), \quad
	\kkB(\varepsilon) = \kb\sqrt{1 - \frac{2\varepsilon}{v\kb}\sin\alpha}. 
	\ee
	If the variables $(\varepsilon,k_z)$ are used, the $y$-component of the momentum as a function of the energy and $k_z$ is given by
	\be\label{eq:k2karch}
	k_{y}^{\pm}(\varepsilon,k_{z})= \pm \frac{\varepsilon+m(k_z)\sin\alpha}{v\cos\alpha}.
	\ee
	and the inverse decay length becomes
	\be\label{eq:kapp2}
	\kappa_\pm(\varepsilon,k_{z})= \frac{\kkB(\varepsilon)^2-k_z^2}{2\kb\cos\alpha}.
	\ee
	When expressed in terms of the energy, $\kappa_\pm$ is the same for both surfaces, so that we may omit the index $\pm$ if we use the variable combination $(\varepsilon,k_z)$.
	
	For $\alpha=0$, one obtains a straight Fermi arc with $-\kb\le k_z\le \kb$ for all $\varepsilon$ and $k_y^{\pm}=\pm \varepsilon/v$.  
	Moreover, for $\varepsilon \to 0$, we observe that $\kkB(\varepsilon)= \kb$ for arbitrary $\alpha$. 
	Arc shapes in the surface Brillouin zone for a few characteristic values of ($\alpha,\varepsilon)$ are illustrated  in Fig.\ \ref{fig1}(b).
	The $k_z$-resolved DoS associated with Fermi arc states is given by
	\be
	n_{{\rm FA}}(\varepsilon,k_z)  = \int \frac{d k_y}{2\pi}\delta
	\left(\varepsilon-\varepsilon^{(s)\pm}_{\vkparallel}\right) = \frac{1}{2\pi v \cos\alpha},\label{eq:nFAkz}
	\ee
	for $-\kkB(\varepsilon) < k_z < \kkB(\varepsilon)$. The total DoS of the arc states is obtained by integrating Eq.~\eqref{eq:nFAkz} over $k_z$,
	\be
	n_{{\rm FA}}(\varepsilon) = \frac{\kkB(\varepsilon)}{2 \pi^2 v \cos \alpha}.\label{eq:nFA}
	\ee
	Equation~\eqref{eq:nFA} predicts a very large DoS for widely open Fermi arc curves with $\alpha$ approaching $\pi/2$, which arises because the total arc length diverges in this somewhat artificial limit.
	
	While the penetration depth diverges when approaching the arc ends for $k_z\to\pm \kkB(\varepsilon)$,
	the minimal penetration depth occurs at the arc center. 
	For $k_z=0$, we find $\kappa^{-1}\approx (2\cos\alpha)/\kb$ at small energies, see Eq.~(\ref{eq:kapp2}). 
	Throughout, we assume that $\kb L\gg 1$, so that Fermi arc states on opposite surfaces have exponentially small overlap
	away from the arc termination points. 
	Large WSM crystals of dimensions up to $1.5$~mm have been reported in the literature \cite{Lv2015}, corresponding to $\kb L\sim 10^{6}$ for typical values of $\kb$. This justifies the neglect of the overlap of the arc states of opposing surfaces.
	
	\subsection{Phonon model}\label{sec2b}
	
	In this work, we study how low-energy quasiparticles in WSMs, arc states as well as bulk states, are scattered by acoustic phonons in the slab geometry of Fig.~\ref{fig1}(a). 
	Within the isotropic elastic continuum description, the properties of acoustic phonons are determined by only two elastic constants, the longitudinal and transverse sound velocities $c_l$ and $c_t$, where $c_t < c_l$ \cite{Landau1986}. The sound velocities $c_l$ and $c_t$ are typically much smaller than the Fermi velocity $v$ of the electrons. For example, for TaAs, one has $c_{l}\simeq 2\times10^{3}$~m$/$s and a Fermi velocity \mbox{$v\simeq 1.16\times10^{5}$~m$/$s} \cite{Peng2016}, so that $c_{l,t}/v \sim 10^{-2}$. We note that the optical phonon gap in WSMs is typically of order 
	$10$~meV. For instance, density functional calculations for the magnetic WSM material ZrCo$_2$Sn find an 
	optical phonon gap $\sim 15$~meV \cite{Sklyadneva2021}. Our theory neglects optical phonons and  
	holds for energies well below this scale. 
	
	Quite generally, we may distinguish three types of acoustic phonons in the slab geometry: longitudinal bulk phonon modes of wavevector $\vqvector = (q_x,\vqparallel)$, for which the displacement field ${\bf u}_{\vqvector}$ is collinear with $\vqvector$;
	transverse bulk phonon modes, for which ${\bf u}_{\vqvector}$ is perpendicular to $\vqvector$;
	and Rayleigh modes, which are exponentially localized at one of the two surfaces at $x = \pm L/2$. For each three-dimensional wavevector $\vqvector$, there are one longitudinal and two transverse bulk modes with frequencies $\Omega^{(l)}_{\vqvector} = c_l \qqvector$ and $\Omega^{(t)}_{\vqvector} = c_t \qqvector$, respectively; for each in-plane wavevector $\vqparallel$, there is additionally one Rayleigh mode at each interface.
	
	While the full phonon spectrum for an isotropic elastic continuum in the slab geometry is known \cite{Landau1986,Bannov1995,Giraud2012}, for a theory of phonon-mediated scattering between bulk electrons, between arc states, and between arc and bulk states, it is sufficient to know the phonon modes in the bulk and in the vicinity of the surfaces at $x = \pm L/2$, respectively. For this purpose, it suffices to consider an infinite or semi-infinite geometry, which considerably simplifies calculations. Note that the same approach was taken in Sec.\ \ref{sec2a} for the electronic wavefunctions.
	
	The equation of motion for the displacement field ${\bf u}(\vrvector,t) = {\bf u}(\vrvector) e^{- i \Omega t}$ is
	\be
	\label{eq:ueom1}
	- \Omega^2 {\bf u} = c_t^2 \nabla^2 {\bf u} 
	+ (c_l^2-c_t^2) \nabla (\nabla \cdot {\bf u}).
	\ee
	Since the transverse phonon modes have no deformation potential, in the interior of the slab only the longitudinal acoustic mode couples to the electrons. For a longitudinal mode with wavevector $\vqvector$, the displacement is in the direction 
	\be
	\hat u^{(l)}(\vqvector) = (q_x \hat x + \vqparallel)/\qqvector.
	\ee
	The corresponding displacement field is
	\be
	\label{eq:ubulk}
	{\bf u}(\vrvector) = \int d\vqvector \frac{
		e^{i \vqvector \cdot \vrvector}}{\sqrt{2 \rho_M \Omega^{(l)}_{\vqvector}}} 
	\left [
	a_{\vqvector}^{(l)} \hat u^{(l)}({\vqvector})
	+ a_{-\vqvector}^{(l)\dagger} \hat u^{(l)}({-\vqvector})
	\right  ],
	\ee
	where $\rho_{M}$ is the volume mass density and $a_{\vqvector}^{(l)}$ is the annihilation operator for the longitudinal phonon mode.
	
	At the surfaces with $x = \pm L/2$, we apply stress-free boundary conditions \cite{Landau1986},
	\bea
	\label{eq:phononbc}
	(c_l^2 - 2 c_t^2) \nabla \cdot {\bf u} &=& -2 c_t^2 \partial_x u_x, \nonumber \\
	0 &=& \partial_x u_y + \partial_y u_x, \nonumber \\
	0 &=& \partial_x u_z + \partial_z u_x.
	\eea
	The boundary conditions are compatible with the inversion symmetry $I$ and the
	magnetic twofold rotation symmetry $C_{2x} T$.
	To find the displacement fields in the vicinity of the surfaces at $x = \pm L/2$, it is necessary to consider the transverse phonon modes, too. The reason is that the boundary condition at $x = \pm L/2$ couples longitudinal and transverse modes. The transverse modes can be separated into a mode for which the displacement ${\bf u}$ is in the plane spanned by $\hat x$ and $\hat q_{\parallel} = \vqparallel/\qqparallel$, and a mode for which ${\bf u}$ is perpendicular to both $\hat x$ and $\hat q_{\parallel}$. The boundary condition at $x = \pm L/2$ only mixes the first of these two transverse modes with the longitudinal mode. The displacement of this transverse mode is in 
	the direction $(q_{\parallel} \hat x - q_x \hat q_{\parallel})/\qqvector$ and we choose
	\be
	\hat u^{(t)}(\vqvector) = i(q_{\parallel} \hat x - q_x \hat q_{\parallel})/\qqvector
	\ee
	so that we have
	\be
	\hat u^{(\lambda)}(-\vqvector)^* = - \hat u^{(\lambda)}(\vqvector),
	\quad \lambda = l,t.
	\ee
	The second transverse phonon mode is a horizontal shear wave, which is not mixed with the longitudinal mode upon reflection at the surface. This mode has no associated deformation potential and, hence, need not be discussed further.
	
	Upon reflection from the interface, the in-plane wavevector $\vqparallel$ and the frequency $\Omega$ are conserved, but the transverse wavevector component $q_x$ is not. Specifically, a transverse mode with transverse wavevector component $\pm q_x$ (with $q_x > 0$) incident on the surface at $x = \pm L/2$ is reflected as a superposition of a transverse mode with $\mp q_x$ and a longitudinal mode with transverse wavevector component 
	\be
	\label{eq:qxl}
	q_x^{(l,t)} \equiv \frac{1}{c_l} \sqrt{\qqvector^2 c_t^2- q_{\parallel}^2 c_l^2}.
	\ee
	In the same way, a longitudinal mode with transverse wavevector component $\pm q_x$ incident on the surface at $x = \pm L/2$ is reflected as a superposition of a longitudinal mode with $\mp q_x$ and a transverse mode with transverse wavevector component 
	\be
	q_x^{(t,l)} \equiv \frac{1}{c_t} \sqrt{\qqvector^2 c_l^2 -q_{\parallel}^2 c_t^2}.
	\ee
	The normalized displacement field ${\bf w}^{(\lambda)\pm,{\rm in}}_{q_x \vqparallel}$ of a mode incident on the surface at $x = \pm L/2$ with a longitudinal ($\lambda = l$) or transverse ($\lambda = t$) polarization contains contributions from the incident and the reflected waves,
	\begin{widetext}
		\bea
		{\bf w}^{(\lambda)\pm,{\rm in}}_{q_x,\vqparallel}(x) &=& e^{\pm i q_x (x \mp L/2)} \hat u^{(\lambda)}{(\pm q_x,\vqparallel)} 
		+ \sum_{\lambda'} s^{(\lambda',\lambda)\pm}_{q_x,\vqparallel}
		e^{\mp i q_{x}^{(\lambda',\lambda)} (x \mp L/2)} \hat u^{(\lambda')}{(\mp q_x^{(\lambda',\lambda)},\vqparallel)},
		\label{eq:ulambda}
		\eea
		where $q_x^{(\lambda,\lambda)} = q_x$. Explicit expressions for the reflection amplitudes $s^{\pm}$ are given in App.\ \ref{appA}. In the same way, one finds that the normalized displacement field of a phonon reflected from the surface at $x = \pm L/2$ in mode $\lambda$ is
		\bea
		&&{\bf w}^{(\lambda)\pm,{\rm out}}_{q_x,\vqparallel}(x) = e^{\mp i q_x (x \mp L/2)} \hat u^{(\lambda)}{(\mp q_x,\vqparallel)}
		+\sum_{\lambda'} s^{(\lambda',\lambda)\pm*}_{q_x,\vqparallel}
		e^{\pm i q_{x}^{(\lambda',\lambda)} (x \mp L/2)} \hat u^{(\lambda')}{(\pm q_x^{(\lambda',\lambda)},\vqparallel)}.
		\eea
	\end{widetext}
	Since $c_t < c_l$, the transverse wavenumber $q_x^{(l,t)}$ of the longitudinal mode may be imaginary. If that is the case, the longitudinal phonon mode decays exponentially away from the surface at $x = \pm L/2$. Equation \eqref{eq:ulambda} also holds in this case, provided the square root with positive imaginary part is chosen in Eq.\ \eqref{eq:qxl}.
	
	
	Equation \eqref{eq:ueom1} also allows for solutions that are exponentially localized at the surfaces at $x = \pm L/2$. These are called Rayleigh modes and their frequency is \cite{Auld1973v2,Giraud2011,Giraud2012}
	\be
	\Omega^{(R)}_{\vqparallel} = c_R q_{\parallel},
	\ee
	where $c_R < c_t$. The precise value of the Rayleigh-mode velocity $c_R$ depends on the ratio $c_l/c_t$ \cite{Giraud2011}. The Rayleigh mode may be considered as a superposition of longitudinal and transverse phonon modes with imaginary $q_x$, 
	\be
	q_x^{(\lambda,R)} = i \frac{q_{\parallel}}{c_{\lambda}} \sqrt{c_{\lambda}^2 - c_{R}^2},\quad \lambda=l,t.
	\ee
	The normalized displacement field for the Rayleigh mode at the surface at $x = \pm L/2$ then reads
	\bea\nonumber
	{\bf w}_{\vqparallel}^{(R)\pm}(x) &=&
	\sum_{\lambda'}
	s^{(\lambda',R)\pm}_{\vqparallel} e^{\pm |q^{(\lambda',R)}_x| (x \mp L/2)} \\ &\times&
	\hat u^{(\lambda')}{(\mp q_x^{(\lambda',R)},\vqparallel)},
	\eea
	where the coefficients $s^{(\lambda',R)\pm}_{\vqparallel}$ are determined by the boundary conditions. The normalization of the displacement field for the Rayleigh modes is chosen such that $\int dx |{\bf w}^{(R)\pm}_{\vqparallel}(x)|^2 = 1$. We again refer to App.\ \ref{appA} for detailed expressions for the coefficients.
	
	Combining contributions from bulk and surface modes, we may write the displacement field in the vicinity of the interface at $x = \pm L/2$ as
	\begin{widetext}
		\bea
		\label{eq:uinmodes}
		{\bf u}(\vrvector) &=& 
		\sum_{\lambda=l,t} 
		\int_0^{\infty} dq_x \int d\vqparallel
		\frac{e^{i \vqparallel \cdot \vrparallel}}{\sqrt{2 \rho_M \Omega^{(\lambda)}_{\vqvector}}}
		\left[
		{\bf w}^{(\lambda)\pm,{\rm in}}_{q_x,\vqparallel}(x)
		a^{(\lambda)\pm,{\rm in}}_{q_x,\vqparallel}
		+  {\bf w}^{(\lambda)\pm,{\rm out}*}_{q_x,-\vqparallel}(x) 
		a^{(\lambda)\pm,{\rm out}\dagger}_{q_x,-\vqparallel} \right]
		\nonumber \\ && \mbox{}  +
		\int d\vqparallel
		\frac{e^{i \vqparallel \cdot \vrparallel}}{\sqrt{2 \rho_M \Omega^{(R)}_{\vqparallel}}}
		\left[ 
		{\bf w}^{(R)\pm}_{\vqparallel}(x) a^{(R)\pm}_{\vqparallel}
		+ {\bf w}^{(R)\pm*}_{-\vqparallel}(x) a^{(R)\pm\dagger}_{-\vqparallel} \right],
		\eea
	\end{widetext}
	where $a^{(\lambda)\pm,{\rm in}}_{q_x,\vqparallel}$ and $a^{(\lambda)\pm,{\rm out}}_{q_x,\vqparallel}$ are the annihilation operators for a longitudinal ($\lambda = l$) or transverse $(\lambda = t$) bulk phonon of in-plane wavevector $\vqparallel$ and transverse wavector component $\pm q_x$ incident on or reflected from the interface at $x = \pm L/2$, respectively, and $a^{(R)\pm}_{\vqparallel}$ is the annihilation operator for a Rayleigh surface phonon of in-plane wavevector $\vqparallel$ at the interface at $x = \pm L/2$.
	
	The phonon model described here captures the essential physics of phonon-induced 
	scattering in WSMs while allowing for analytical progress.  
	Since WSMs are typically found in anisotropic materials \cite{Lv2015}, where also so-called chiral phonon modes are possible \cite{Chernodub2019}
	(for experiments on WSMs with broken inversion symmetry, see Ref.~\cite{Zhu2018}), this model may be too simple to allow for a detailed quantitative comparison with experimental data.
	However, many low-$T$ transport phenomena are directly linked to scattering properties of topological arc states which are 
	largely independent of the detailed phonon model.  
	
	\subsection{Electron-phonon interaction} \label{sec2c}
	
	Next we address the coupling between electrons and phonons, where  
	we focus on the electron-phonon interaction Hamiltonian $H_{\rm ep}$ resulting from the deformation potential.
	Other coupling mechanisms or more exotic vibrational modes, such as chiral phonons \cite{Chernodub2019,Song2019}, may also emerge in low-energy WSM theories. For example,
	the coupling to unconventional pseudoscalar phonons can generate a deformation potential that is different for Weyl nodes of opposite chiralities \cite{Song2016,Rinkel2017}. Furthermore,
	elastic gauge field interactions (``pseudo-magnetic fields'') have been addressed in Refs.~\cite{Cortijo2015,Cortijo2016,Rinkel2017,Arjona2018,Heidari2020,Shapourian2015},
	and piezoelectric interactions can be important in WSMs with broken inversion symmetry \cite{Pereira2019}. 
	With minor modifications, such types of couplings can be included in our theory, see also Ref.~\cite{Bannov1995}. For definiteness, however, we focus on the deformation potential which often gives the dominant electron-phonon coupling in WSM materials \cite{Vogl1976,Lundgren2015,Rinkel2017}.

	We here assume that the electron-phonon interaction is diagonal in spin-orbital space and given by \cite{Bannov1994,Bannov1995}
	\be\label{eq:Hep}
	H_{\rm ep}  =  g_0  \mathbf{\nabla}\cdot{\bf u}\left({\bf r}\right),
	\ee
	with a deformation potential coupling $g_0$ (of dimension energy).
	The displacement field  ${\bf u}({\bf r})$ is expressed in terms of phonon creation and annihilation operators as in Eq.~\eqref{eq:uinmodes}.
	Using Thomas-Fermi theory for a simple estimate \cite{Mahan,Lundgren2015}, one obtains $g_0\sim n/n_b(\mu)$, where $n$ is the electron density and $n_b(\varepsilon)$ the bulk DoS, see Eq.~\eqref{bulkdos}. We therefore expect large values of $g_0$ for $0<\mu\ll v \kb$.
	However, since $g_0$ is affected by screening processes, it is difficult to reliably estimate its value for realistic materials.
	Strong electron-phonon couplings have recently been reported for the type-II WSM material WP$_2$ \cite{Osterhoudt2021}. 
	
	For phonon-mediated scattering of electrons in the interior of the slab, there is a contribution from longitudinal phonons only. Upon substitution of Eqs.\ \eqref{eq:PhiBulk} and \eqref{eq:ubulk} into Eq.\ \eqref{eq:Hep}, we find that the matrix element for scattering between bulk states $|\Phi^{(b)}_{\vklabel}\rangle$ is of the form \cite{Resta2018}
	\be
	\langle \Phi^{(b)}_{\vkplabel} | H_{\rm ep} | \Phi^{(b)}_{\vklabel} \rangle =
	{\cal G}^{(bbl)}_{\vkplabel,\vklabel} \left(
	a^{(l)}_{\vqvector} - a_{-\vqvector}^{(l)\dagger} \right)
	\ee
	with $\vqvector = \vkpvector - \vkvector$. The bulk-bulk amplitudes ${\cal G}_{\vklabel,\vkplabel}$ are
	\be
	\label{bulkbulkmatrixelem}
	{\cal G}^{(bbl)}_{\vkplabel,\vklabel} =
	i \frac{g_0 \sqrt{\Omega^{(l)}_{\vqvector}}}{c_l \sqrt{2 \rho_{M}}}
	\langle \xi_{\vkplabel}|\xi_{\vklabel} \rangle.
	\ee
	We note in passing that the factor $\langle \xi_{\vkplabel}|\xi_{\vklabel} \rangle$ in 
	Eq.\ \eqref{bulkbulkmatrixelem} is characteristic for Weyl fermions and causes a suppression of intra-node backscattering. 
	Since this factor is absent for conventional fermions, the bulk-bulk decay rate is reduced by a factor $1/2$ in the Weyl case, see
	Eq.\ \eqref{eq:bulkrate3} below.
	
	Taking the phonon modes to be in thermal equilibrium at temperature $T$, according to Fermi's Golden Rule, the corresponding bulk-bulk transition rate is
	\begin{widetext}
		\bea
		\label{eq:Wbb}
		W_{\vkplabel,\vklabel}^{(bb)} &=&
		2 \pi |{\cal G}^{(bbl)}_{\vkplabel,\vklabel}|^2 
		\left\{  
		n_{\rm B}(\Omega^{(l)}_{\vqvector})
		\delta(\varepsilon^{(b)}_{\vkplabel}-\varepsilon^{(b)}_{\vklabel} - \Omega^{(l)}_{\vqvector})
		+
		[n_{\rm B}(\Omega^{(l)}_{\vqvector})+1] 
		\delta(\varepsilon^{(b)}_{\vkplabel}-\varepsilon^{(b)}_{\vklabel} + \Omega^{(l)}_{\vqvector})
		\right\},
		\eea
		where $n_B(\Omega)$ is the Planck function (Bose-Einstein function at zero chemical potential).
		
		To obtain the matrix elements for scattering between arc states, or between arc and bulk states, at the surface at $x = \pm L/2$, 
		we substitute Eqs.\ \eqref{eq:PhiInterface} and \eqref{eq:boundarystte} for the electronic states and Eqs.\ \eqref{eq:uinmodes} for the displacement field. This gives arc-bulk interaction matrix elements of the form
		\be
		\langle \Phi_{k_x',\vkpparallel}^{(b)\pm} | H_{\rm ep} | \Phi_{\vkparallel}^{(s)\pm} \rangle
		=   \sum_{\lambda=l,t} \int_{0}^{\infty} dq_x
		{\cal G}^{(bs\lambda)\pm}_{\vkplabel,\vkparallel,q_x}
		\left(
		a^{(\lambda)\pm,{\rm in}}_{q_x,\vqparallel}
		-
		a^{(\lambda)\pm,{\rm out} \dagger}_{q_x,-\vqparallel}
		\right)  
		+ {\cal G}^{(bsR)\pm}_{\vkplabel,\vkparallel}
		\left(
		a^{(R)\pm}_{\vqparallel} 
		-
		a^{(R)\pm \dagger}_{-\vqparallel}
		\right)   
		\label{ephmatrixelements}
		\ee
		with $\vqparallel = \vkpparallel-\vkparallel$.
		Because $H_{\rm ep}$ is local, there is no direct phonon-induced scattering between arc states at different surfaces. Of course, electrons may transition between different surfaces via intermediate bulk states. Such processes are accounted for in the Boltzmann theory that will be developed in Sec.~\ref{sec3}.
		Detailed expressions for the arc-arc amplitudes ${\cal G}^{(ss)\pm}$ and the arc-bulk amplitudes ${\cal G}^{(sb)\pm}$ and ${\cal G}^{(bs)\pm}$ are given in App.\ \ref{appB}, where we also specify the matrix elements of $H_{\rm ep}$ for bulk-arc and arc-arc scattering.
		
		The arc-arc, arc-bulk, and bulk-arc transition rates have contributions from scattering mediated by longitudinal bulk, transverse bulk, and Rayleigh phonons.  In particular, the arc-bulk transition rate has the form
		\bea
		\label{eq:Wbs}
		W_{\vkplabel,\vkparallel}^{(bs)\pm} &=& 
		2 \pi \sum_{\lambda=l,t} \int_0^{\infty} \frac{dq_x}{2 \pi}
		|{\cal G}^{(bs\lambda)\pm}_{\vkplabel,\vkparallel,q_x}|^2
		\left\{ 
		n_{\rm B}(\Omega^{(\lambda)}_{\vqvector})
		\delta(\varepsilon^{(b)}_{\vkplabel}-\varepsilon^{(s)\pm}_{\vkparallel} - \Omega^{(\lambda)}_{\vqvector}) 
		+
		[n_{\rm B}(\Omega^{(\lambda)}_{\vqvector})+1]
		\delta(\varepsilon^{(b)}_{\vkplabel}-\varepsilon^{(s)\pm}_{\vkparallel} + \Omega^{(\lambda)}_{\vqvector}) \vphantom{|{\cal G}^{\pm}_{{\bf k}'s,\vkparallel,q_x\lambda}|^2} \right\}
		\nonumber  \\ && \mbox{}
		+
		2 \pi |{\cal G}^{(bsR)\pm}_{\vkplabel,\vkparallel}|^2
		\left\{
		n_{\rm B}(\Omega^{(R)}_{\vqparallel})
		\delta(\varepsilon^{(b)}_{\vkplabel}-\varepsilon^{(s)\pm}_{\vkparallel} - \Omega^{(R)}_{\vqparallel})
		+
		[n_{\rm B}(\Omega^{(R)}_{\vqparallel})+1]
		\delta(\varepsilon^{(b)}_{\vkplabel}-\varepsilon^{(s)\pm}_{\vkparallel} + \Omega^{(R)}_{\vqparallel})
		\right\},
		\eea
	\end{widetext}
	where $\vqparallel = \vkpparallel - \vkparallel$ and $\vqvector = (q_x,\vqparallel)$.
	Expressions for the bulk-arc and arc-arc transition rates are given in App.\ \ref{appB}.
	
	\subsection{Characteristic temperatures} \label{sec2d}
	
	For conventional electron-phonon coupled systems, the crossover temperature separating the low- and high-$T$ regimes is the Bloch-Gr\"uneisen temperature $T_{\rm BG}= 2 c_{\rm ph} k_F$, where $c_{\rm ph}$ is the sound velocity and $k_F$ the Fermi momentum \cite{Ashcroft}. Only phonons with momentum $q \sim 2k_F$ can efficiently backscatter electrons. The frequency of such phonons is $\sim 2 c_{\rm ph} k_F$. Clearly, for $T\ll T_{\rm BG}$, such processes are rare events, while they proliferate for $T\gg T_{\rm BG}$.
	
	For the WSM model considered in this paper, it is useful to introduce an ``effective Bloch-Gr\"uneisen'' crossover temperature as 
	\begin{equation}
	k_{\rm B} T_{\rm BG}= c_{l} \kb.\label{eq:TBG}
	\end{equation}
	The rationale behind Eq.~\eqref{eq:TBG} is that momentum exchange for arc-arc scattering and for scattering between the Weyl cones is naturally limited by $q\alt \kb$, since $2 \kb$ is the distance between the Weyl points and the ``length'' of the Fermi arcs in reciprocal space. For $T\ll T_{\rm BG}$, phonons with $q\ll \kb$ dominate the phonon-induced scattering. For arc states, scattering is then local in reciprocal space. For bulk states, the two Weyl points are effectively decoupled if $T \ll T_{\rm BG}$.
	
	In practice, $T_{\rm BG}$ can weakly depend on other parameters, e.g., the angle $\alpha$, the chemical potential $\mu$ or the relevant phonon mode. We here disregard such details and use the longitudinal sound velocity $c_l$ to define the effective Bloch-Gr\"uneisen temperature. 
	(Note that the various phonon mode velocities differ by factors of order one,  which is not relevant for the definition of a crossover scale.)
	To give an estimate, for the WSM material TaAs, the closest pair of Weyl nodes is separated by $\kb \approx 0.1 \pi/a_0$, with the lattice constant $a_0\simeq 3.4\times10^{-10}$~m \cite{Lv2015}. Using $c_l\approx 2\times 10^3$~m$/$s \cite{Peng2016,Buckeridge2016}, we obtain $T_{\rm BG}\approx 13$~K.
	
	For bulk quasiparticles, there is a second characteristic temperature for \emph{intra-node} scattering processes, which is the conventional Bloch-Gr\"uneisen temperature corresponding to the radius $\mu/v$ of the Fermi surface at each Weyl node,
	\begin{equation}\label{eq:tbgb}
	k_{\rm B} T_{\rm BG}^{(b)} = 2 c_l \mu/v.
	\end{equation}
	We observe that for $0<\mu \ll v \kb$, an intermediate temperature regime opens up,
	\begin{equation}\label{intermediateTb}
	T_{\rm BG}^{(b)}\ll T\ll T_{\rm BG},
	\end{equation}
	where the phonon-induced inter-node backscattering of bulk quasiparticles is frozen out but backscattering processes within a given Weyl node can proliferate at the same time.
	
	\section{Boltzmann theory}\label{sec3}
	
	We now describe the Boltzmann approach \cite{Mahan} for the calculation of phonon-induced electronic transport observables in a clean WSM slab, using the model discussed in Sec.~\ref{sec2}. The applicability of the Boltzmann equation requires that the slab width $L \gg v/\mu$ is much larger than the electron wavelength and that $L \gg c_l/T$ is much larger than the thermal phonon wavelength. We consider a chemical potential $T \ll \mu \ll \kb v$, so that the two Fermi surfaces at the two Weyl nodes are well separated in reciprocal space. This condition automatically ensures that the slab width is much larger than the typical transverse width of the arc states. 
	
	We consider the linear response of the system to a homogeneous electric field 
	${\bf E}_{\parallel}  = E_y\hat y+E_z\hat z$ applied parallel to the surface, see Fig.~\ref{fig1}(a).
	Along the transverse direction, the current must vanish ($J_x=0$) such that the induced transverse gradient of the electrochemical potential will be implicitly determined by $E_{y,z}$. The relation between the in-plane electrical field ${\bf E}_{\parallel} $ and the in-plane current density ${\bf J}_{\parallel}$ defines the conductivity tensor
	${\bf J}_{\parallel} =\hat\sigma {\bf E}_{\parallel}.$
	The resistivity tensor is then given by $\hat \rho=\hat\sigma^{-1}$.  
	
	In Sec.~\ref{sec3a}, we discuss the Boltzmann equations for a WSM slab in a uniform in-plane electric field, using the transition rates for phonon-induced scattering from Sec.\ \ref{sec2c}.
	Since we study only the linear transport regime in this work, it is sufficient to linearize the distribution functions and the collision integrals with respect to the applied electric field, see Sec.~\ref{sec3b}. 
	In addition, the smallness of the ratio between sound and Fermi velocities allows us to employ a quasi-elastic approximation such that the linearized Boltzmann equations can be solved for each electron energy $\varepsilon$ separately.
	Conditions for the validity of the Boltzmann approach and the various approximations used are discussed in Sec.\ \ref{sec3c}.
	
	\subsection{Boltzmann equation}\label{sec3a}
	
	The Boltzmann equation describes the dynamics and the spatial variations of the distribution functions $f^{(b)}_{\vklabel}(\vrvector,t)$ and $f^{(s)\pm}_{\vkparallel}(\vrparallel,t)$ of bulk electrons and arc states. For the case of a time-independent homogeneous in-plane electric field ${\bf E}_\parallel$, the distribution functions are independent of $\vrparallel$ and $t$. Moreover, we assume that the typical time for bulk-arc scattering is large in comparison to the  transit time across the width of the slab, so that the bulk distribution function is also independent of the transverse coordinate $x$. (The precise conditions are discussed in Sec.\ \ref{sec3c}.) With these simplifications, the Boltzmann equation takes the form 
	\bea
	\label{eq:dtf}
	e {\bf E}_{\parallel} \cdot \partial_{\vkvector}
	f^{(b)}_{\vklabel} &=&
	{\cal I}_{\vklabel}^{(bb)} + \frac{1}{L} \sum_{\pm} {\cal I}_{\vklabel}^{(sb)\pm}
	\\
	\label{eq:dtf2}
	e {\bf E}_{\parallel}  \cdot \partial_{\vkparallel} f^{(s)\pm}_{\vkparallel} &=&
	{\cal I}_{\vkparallel}^{(ss)\pm} + {\cal I}_{\vkparallel}^{(bs)\pm},
	\eea
	where ${\cal I}^{(bb)}$, ${\cal I}^{(sb)\pm}$, ${\cal I}^{(bs)\pm}$, and ${\cal I}^{(ss)\pm}$ are the collision integrals for bulk-bulk, bulk-arc, arc-bulk, and arc-arc scattering, respectively\footnote{In addition, there 
		is a Berry curvature component along the $x$-direction. However, this term does not generate a net Hall response  and can
		safely be ignored here.}. The factor $1/L$ in Eq.\ \eqref{eq:dtf} arises from a proper consideration of the normalization of bulk and surface electron states. Specular reflection at the surfaces of the slab at $x = \pm L/2$ implies the conditions
	\be \label{eq:fbc}
	f^{(b)}_{(k_x,\vkparallel)} = f^{(b)}_{(-k_x,\vkparallel)}, 
	\ee
	consistent with the form \eqref{eq:PhiInterface} of the bulk states at the sample surfaces. We will use Eq.\ \eqref{eq:fbc} to restrict consideration of the bulk distribution function $f^{(b)}_{\vklabel}$ to positive values of $k_x$.
	
	The collision integrals are then expressed in terms of the transition rates discussed in Sec.\ \ref{sec2c}.  For the bulk-bulk and arc-bulk 
	collision integrals, we have
	\begin{widetext}
		\bea \nonumber
		{\cal I}^{(bb)}_{\vklabel} &=&
		\int_{-\infty}^{\infty} \frac{dk_x'}{2 \pi} \int \frac{d\vkpparallel}{(2 \pi)^2}
		\left\{ W^{(bb)}_{\vklabel,\vkplabel}
		f^{(b)}_{\vkplabel}(1-f^{(b)}_{\vklabel}) -
		W^{(bb)}_{\vkplabel,\vklabel}
		f^{(b)}_{\vklabel}(1-f^{(b)}_{\vkplabel}) \right\}, \\ \label{collint}
		{\cal I}_{\vkparallel}^{(bs)\pm} &=&
		\int_0^{\infty} \frac{dk_x'}{2 \pi}
		\int \frac{d\vkpparallel}{(2 \pi)^2} \left\{
		W_{\vkparallel,\vkplabel}^{(sb)\pm} f^{(b)}_{\vkplabel} (1 - f_{\vkparallel}^{(s)\pm})
		- W_{\vkplabel,\vkparallel}^{(bs)\pm} f_{\vkparallel}^{(s)\pm} (1 - f^{(b)}_{\vkplabel})
		\right\}.
		\eea
	\end{widetext}
	The transition rates $W^{(bb)}$ and $W^{(bs)}$ are given in Eqs.\ (\ref{eq:Wbb}) and \eqref{eq:Wbs}, respectively.
	The transition rate $W^{(sb)}$ is given in Eq.\ (\ref{eq:Wsb}), and
	the collision integrals for bulk-arc and arc-arc scattering can be found in App.\ \ref{appC}.
	Our expressions for the transition rates ensure that these collision integrals vanish identically in thermal equilibrium.
	
	Next we observe that the in-plane current density 
	\be
	{\bf J}_{\parallel} = {\bf J}^{(b)}_{\parallel}  + \sum_{\pm} {\bf J}^{(s)\pm}_{\parallel}
	\label{eq:Jparallel}
	\ee
	has contributions from the bulk and from the current carried by the arc states,
	\bea
	\label{eq:J}
	{\bf J}^{(b)}_{\parallel} &=& e L \int \frac{d\vkvector}{(2 \pi)^3}
	\vvvector_{\parallel}^{(b)}(\vkvector) f_{\vkvector}^{(b)}, \\
	{\bf J}^{(s)\pm}_{\parallel}  &=& e \int \frac{d\vkparallel}{(2 \pi)^2}
	\vvvector^{(s)\pm}_{\parallel}(\vkparallel) f_{\vkparallel}^{(s)\pm}.
	\eea
	Here $\vvvector_{\parallel}^{(b)}(\vkvector)$ is the in-plane velocity of bulk states, see Eq.\ \eqref{eq:bulkvelocities}, and $\vvvector_{\parallel}^{(s)\pm}(\vkparallel)$ is the velocity of the arc states at the surface at $x = \pm L/2$, see Eq.\ \eqref{eq:arcvelocities}.
	The factor $L$ in Eq.\ \eqref{eq:J}  arises because we consider two-dimensional current densities, 
	so the bulk current density is actually integrated over $x$, which produces the factor $L$.
	The transverse response to the applied field is characterized by the difference $eV_{\perp}$ 
	of the chemical potentials for the surface states at the surfaces at $x=L/2$ and $x=-L/2$,
	\be
	\label{eq:Vperp}
	eV_{\perp} = \int \frac{d\vkparallel}{(2 \pi)^2} \frac{f_{\vkparallel}^{(s)+} - f_{-\vkparallel}^{(s)-}}{n_{\rm FA}(\varepsilon^{(s)+}_{\vkparallel})},
	\ee
	where $n_{\rm FA}(\varepsilon)$ is the DoS of arc states in Eq.\ \eqref{eq:nFA}.
	
	\subsection{Linear response and quasi-elastic approximation} \label{sec3b}
	
	To linear order in the applied electric field, we may expand the arc-state distribution function as 
	\be
	f_{\vkparallel}^{(s)\pm} = n_{\rm F}(\varepsilon_{\vkparallel}^{(s)\pm}) +
	\varphi_{\vkparallel}^{(s)\pm} \left( - \frac{d n_{\rm F}(\varepsilon_{\vkparallel}^{(s)\pm})}{d \varepsilon_{\vkparallel}^{(s)\pm}} \right),
	\ee
	where $n_{\rm F}(\varepsilon) = 1/(e^{(\varepsilon - \mu)/ T} + 1)$ is the Fermi-Dirac distribution function.
	Similarly, the linearized bulk distribution function $f^{(b)}_{\vklabel}$ is encoded by $\varphi^{(b)}_{\vklabel}$.
	With this Ansatz, the collision integrals are expanded to linear order in $\varphi^{(b)}_{\vklabel}$ and $\varphi_{\vkparallel}^{(s)\pm}$, whereas the distribution functions on the left-hand side of Eqs.~\eqref{eq:dtf} and \eqref{eq:dtf2} can be replaced by the equilibrium distribution functions.
	
	After linearization, the Boltzmann equation becomes
	\bea
	-e {\bf E}_{\parallel} \cdot \vvvector^{(b)}_{\parallel}({\vklabel}) &=&
	{\cal J}^{(bb)}_{\vklabel} + \frac{1}{L} \sum_{\pm} {\cal J}^{(sb)\pm}_{\vklabel}, \label{eq:BE1}
	\\
	-e {\bf E}_{\parallel} \cdot \vvvector_{\parallel}^{(s)\pm}(\vkparallel) &=&
	{\cal J}^{(bs)\pm}_{\vkparallel} + {\cal J}^{(ss)\pm}_{\vkparallel},
	\label{eq:BE2}
	\eea
	where the linearized collision integrals for bulk-bulk and arc-bulk scattering are given by 
	\bea
	\label{eq:calJ}
	{\cal J}^{(bb)}_{\vklabel} &=&
	\int_{-\infty}^{\infty} \frac{dk_x'}{2 \pi} \int \frac{d\vkpparallel}{(2 \pi)^2}
	{\cal W}_{\vkplabel,\vklabel}^{(bb)}
	(\varphi^{(b)}_{\vkplabel} - \varphi^{(b)}_{\vklabel}), \\
	{\cal J}^{(bs)\pm}_{\vkparallel} &=&
	\int_{0}^{\infty} \frac{dk_x'}{2 \pi} \int \frac{d\vkpparallel}{(2 \pi)^2}
	{\cal W}_{\vkplabel,\vkparallel}^{(bs)\pm} 
	(\varphi^{(b)}_{\vkplabel} - \varphi_{\vkparallel}^{(s)\pm}). \nonumber 
	\eea
	The bulk-bulk kernel ${\cal W}_{\vkplabel,\vklabel}^{(bb)}$ reads
	\begin{widetext}
		\bea
		\label{eq:calWbb}
		{\cal W}_{\vkplabel,\vklabel}^{(bb)}
		&=&
		2 \pi
		|{\cal G}^{(bbl)}_{\vkplabel,\vklabel}|^2
		\left\{
		[n_{\rm B}(\Omega^{(l)}_{\vqvector}) + n_{\rm F}(\varepsilon^{(b)}_{\vklabel} + \Omega^{(l)}_{\vqvector})]
		\delta(\varepsilon^{(b)}_{\vkplabel} - \varepsilon^{(b)}_{\vklabel} - \Omega^{(l)}_{\vqvector})
		\right. \nonumber \\ && \left. \ \ \ \ \mbox{}
		+
		[n_{\rm B}(\Omega^{(l)}_{\vqvector}) + 1 - n_{\rm F}(\varepsilon^{(b)}_{\vklabel} - \Omega^{(l)}_{\vqvector})]  
		\delta(\varepsilon^{(b)}_{\vkplabel} - \varepsilon^{(b)}_{\vklabel} + \Omega^{(l)}_{\vqvector})
		\right \},
		\eea
		where $\vqvector = \vkpvector - \vkvector$, and the arc-bulk kernel is given by
		\bea\label{eq:calWbs}
		{\cal W}^{(bs)\pm}_{\vkplabel,\vkparallel} &=&
		2 \pi
		\sum_{\lambda} \int_0^{\infty} \frac{dq_x}{2 \pi}
		|{\cal G}^{(bs\lambda)\pm}_{\vkplabel,\vkparallel,q_x}|^2
		\left\{
		[n_{\rm B}(\Omega^{(\lambda)}_{\vqvector}) + n_{\rm F}(\varepsilon_{\vkparallel}^{(s)\pm} + \Omega^{(\lambda)}_{\vqvector})] 
		\delta(\varepsilon^{(b)}_{\vkplabel} - \varepsilon_{\vkparallel}^{(s)\pm} - \Omega^{(\lambda)}_{\vqvector})
		\right.  \nonumber  \\ && \left. \ \ \ \ \mbox{}
		+
		[n_{\rm B}(\Omega^{(\lambda)}_{\vqvector}) + 1 - n_{\rm F}(\varepsilon_{\vkparallel}^{(s)\pm} - \Omega^{(\lambda)}_{\vqvector})]
		\delta(\varepsilon^{(b)}_{\vkplabel} - \varepsilon_{\vkparallel}^{(s)\pm} + \Omega^{(\lambda)}_{\vqvector})  
		\right\}
		\nonumber \\ && \mbox{} 
		+ 2 \pi 
		|{\cal G}^{(bsR)\pm}_{\vkplabel,\vkparallel}|^2
		\left\{ 
		[n_{\rm B}(\Omega^{(R)}_{\vqparallel}) + n_{\rm F}(\varepsilon_{\vkparallel}^{(s)\pm} + \Omega^{(R)}_{\vqparallel})] 
		\delta(\varepsilon^{(b)}_{\vkplabel} - \varepsilon_{\vkparallel}^{(s)\pm} - \Omega^{(R)}_{\vqparallel})
		\right. \nonumber \\ && \left. \ \ \ \ \mbox{} 
		+ 2 \pi  [n_{\rm B}(\Omega^{(R)}_{\vqparallel}) + 1 - n_{\rm F}(\varepsilon_{\vkparallel}^{(s)\pm} - \Omega^{(R)}_{\vqparallel})]
		\delta(\varepsilon^{(b)}_{\vkplabel} - \varepsilon_{\vkparallel}^{(s)\pm} + \Omega^{(R)}_{\vqparallel})  
		\right\}
		,
		\eea
	\end{widetext}
	where $\vqparallel = \vkpparallel - \vkparallel$.  Contributions involving phonon emission 
	or absorption can easily  be identified in the above expressions.
	Expressions for the bulk-arc and arc-arc collision terms along with the associated kernels 
	are specified in App.\ \ref{appC}, see Eq.\ \eqref{eq:calWssApp}.
	
	Since the sound velocities $c_l$, $c_t$, and $c_R$ are generally much smaller than the Fermi velocity $v$, the typical change of the electronic energy is small in comparison to the characteristic energy scales $\mu$ and $\kb v$ for chemical potential $T \ll \mu \ll \kb v$. This motivates the ``quasi-elastic'' approximation, in which the phonon energies are dropped from the energy-conserving delta functions in Eqs.\ \eqref{eq:calWbb}, \eqref{eq:calWbs} and \eqref{eq:calWssApp}. 
	The phonon energies are retained in the arguments of the Planck functions $n_{\rm B}$ and Fermi-Dirac functions $n_{\rm F}$.
	
	Because energy is conserved after making the quasi-elastic approximation, it is advantageous to use the energy $\varepsilon$ to label bulk states and arc states. For the bulk states, we eliminate the transverse momentum $k_x$ in favor of $\varepsilon$; for the arc states, we replace $k_y$ in favor of $\varepsilon$. The integration over $k_x'$ (for bulk states) and over $k_y'$ (for arc states) in the expressions for the collision integrals can be performed with the help of the delta functions of energy in the scattering kernels of Eqs.\ \eqref{eq:calWbb}, \eqref{eq:calWbs} and \eqref{eq:calWssApp}, using
	\bea
	\int_0^\infty \frac{dk_x'}{2 \pi} \delta(\varepsilon^{(b)}_{\vkplabel} - \varepsilon) &=& \frac{1}{2 \pi |v^{(b)}_{x}(\varepsilon,\vkpparallel)|}, \nonumber \\
	\int \frac{dk_y'}{2 \pi} \delta(\varepsilon_{\vkpparallel}^{(s)\pm} - \varepsilon) &=& \frac{1}{2 \pi |v^{(s)\pm}_{y}(\varepsilon,k_z')|},
	\eea
	where $v^{(b)}_{x}({\varepsilon,\vkparallel}) = \partial \varepsilon^{(b)}_{\vklabel}/\partial k_x$ and $v_{y}^{(s)\pm}({\varepsilon,k_z)} = \partial \varepsilon_{\vkparallel}^{(s)\pm}/\partial k_y$ are the $x$ and $y$ components of the velocities of bulk and arc states, respectively, see Eqs.\ \eqref{eq:bulkvelocities} and \eqref{eq:arcvelocities}. The range of $k_z$-values at energy $\varepsilon$ is $-\kkB(\varepsilon) \le k_z \le \kkB(\varepsilon)$, see Eq.\ \eqref{eq:kkB}.
	
	In this notation, the equations of motion for the linear-response corrections to the distribution function read
	\bea \label{eq:BE4}
	-e {\bf E}_{\parallel} \cdot \vvvector^{(b)}_{\parallel}(\varepsilon,\vkparallel) &=&
	{\cal J}^{(bb)}_{\varepsilon,\vkparallel} + \frac{1}{L} \sum_{\pm} {\cal J}^{(sb)\pm}_{\varepsilon,\vkparallel}, \\
	-e {\bf E}_{\parallel} \cdot \vvvector_{\parallel}^{(s)\pm}(\varepsilon,k_z) &=&
	{\cal J}^{(bs)\pm}_{\varepsilon,k_z} + {\cal J}^{(ss)\pm}_{\varepsilon,k_z}, \label{eq:BE5}
	\eea
	where the collision integrals are obtained from Eqs.\ \eqref{eq:calJ} using the procedure described above.  We refer to App.\ \ref{appC} for a detailed discussion of these collision integrals.
	The expressions \eqref{eq:Jparallel}--\eqref{eq:Vperp} for the in-plane current density and the transverse voltage then become 
	\bea
	\label{eq:J_lin}
	{\bf J}^{(b)}_{\parallel} &=& 2e L \int \frac{d\vkparallel}{(2 \pi)^2} \int \frac{d\varepsilon}{2 \pi} \frac{\vvvector^{(b)}_{\parallel}(\varepsilon,\vkparallel)}{|v^{(b)}_{x}(\varepsilon,\vkparallel)|}
	\left( - \frac{d n_{\rm F}(\varepsilon)}{d\varepsilon} \right)
	\varphi^{(b)}_{\varepsilon,\vkparallel} , \nonumber \\ 
	{\bf J}^{(s)\pm}_{\parallel} &=& e \int \frac{d k_z}{2 \pi} \int \frac{d\varepsilon}{2 \pi} \frac{\vvvector^{(s)\pm}_{\parallel}(\varepsilon,\vkparallel)}{|v^{(s)\pm}_{y}(\varepsilon,k_z)|}
	\left( - \frac{d n_{\rm F}(\varepsilon)}{d\varepsilon} \right) 
	\varphi^{(s)\pm}_{\varepsilon,k_z} ,  \nonumber \\ &&
	\eea
	and
	\be
	\label{eq:Vperp_lin}
	e V_{\perp}= \int \frac{d k_z}{2 \pi} \int \frac{d\varepsilon}{2 \pi} 
	\frac{\varphi^{(s)+}_{\varepsilon,k_z} - \varphi^{(s)-}_{\varepsilon,-k_z}}{n_{\rm FA}(\varepsilon) |v^{(s)+}_{y}(\varepsilon,k_z)|}
	\left( - \frac{d n_{\rm F}(\varepsilon)}{d\varepsilon} \right).
	\ee
	Since in the quasi-elastic approximation the energy $\varepsilon$ is conserved, the label $\varepsilon$ will be dropped from the expressions if no confusion is possible. 
	Finally, we note that the inversion symmetry of the problem allows to reduce 
	the number of degrees of freedom by one half. 
	The corresponding symmetry relations are summarized in App.~\ref{appC}. 
	
	\subsection{Scattering rate for bulk electrons}\label{sec3c}
	
	The scattering rate for bulk electrons is given by 
	\be\label{eq:bulkbulkrate}
	\Gamma^{(bb)}_{\vklabel} = \int \frac{dk_x'}{2 \pi} \int \frac{d\vkparallel'}{(2 \pi)^2}
	{\cal W}^{(bb)}_{\vkplabel,\vklabel},
	\ee
	see Eq.\ \eqref{eq:calWbb}.
	The probability of bulk-arc scattering for an electron incident on the surface at $x = \pm L/2$ is given by
	\be
	P_{\vklabel}^{(sb)\pm} = \frac{1}{|v^{(b)}_x(\vklabel)|}
	\int \frac{d\vkpparallel}{(2 \pi)^2} 
	{\cal W}_{\vkpparallel,\vklabel}^{(sb)\pm},
	\ee
	where $v^{(b)}_{x}(\vklabel)$ is the transverse velocity, see Eq.\ \eqref{eq:bulkvelocities}.
	A necessary condition for the use of Fermi's Golden Rule to calculate the transition rates is
	\be
	P_{\vklabel}^{(sb)\pm} \ll 1.
	\label{eq:Pineq1}
	\ee
	Since the rates for phonon-induced scattering are strongly temperature dependent, this condition is always satisfied at sufficiently low temperatures.
	We here also assume that
	\be   \label{eq:Pineq2}
	P_{\vklabel}^{(sb)\pm} \ll \frac{|v^{(b)}_x(\vklabel)|}{L \Gamma^{(bb)}_{\vklabel}}.
	\ee
	This inequality ensures that the transit time of electrons between opposite surface is less than the escape time into the arc states, so that the distribution function of electrons in the bulk is uniform across the cross section of the slab. The inequality \eqref{eq:Pineq2} follows from the inequality \eqref{eq:Pineq1} if the bulk mean free path for electron-phonon scattering is larger than the slab width $L$, but it may still be satisfied if that is not the case.
	(In the ultra-low temperature regime $T \ll T_{\rm BG}^{(b)}$, the rate $\Gamma_{\bf k}^{(bb)}$ in Eq.\ \eqref{eq:Pineq2} should be replaced by the transport scattering rate, which leads to an even weaker condition on the slab width $L$ than Eq.\ \eqref{eq:Pineq2}.)
	
	We now consider the regime $\mu\ll \kb v$, where the intermediate temperature window $T_{\rm BG}^{(b)}\ll T\ll T_{\rm BG}$ opens up. Here 
	the Bloch-Gr\"uneisen temperature $T^{(b)}_{\rm BG}$ for intra-node scattering of bulk quasiparticles has been defined in Eq.\ \eqref{eq:tbgb}.
	Within this temperature window, the scattering rate for bulk quasiparticles with energy $\varepsilon = \mu$ can be calculated from Eq.\ \eqref{eq:bulkbulkrate},
	see also App.\ \ref{appC}, as
	\be\label{eq:bulkrate1}
	\Gamma^{(bb)} = \Gamma_0 \frac{T T_{\rm BG}^{(b)2}}{T_{\rm BG}^3} \Xi^{(bb,2)},
	\ee 
	where $\Gamma_0$ is a characteristic energy scale for the electron-phonon scattering rate,
	\begin{equation}\label{eq:xilambda}
	\Gamma_0 = \frac{g_0^{2} \kb^{3}}{\rho_M c_{l} v},
	\end{equation}
	and $\Xi^{(bb,2)}$ is a dimensionless numerical constant. For $\mu \ll \kb v$,  we find $\Xi^{(bb,2)} = 1/8 \pi$. 
	On the other hand, in the high-temperature limit $T \gg T_{\rm BG}$ (but still $T \ll \kb v$), 
	one arrives at a similar  result as in Eq.\ \eqref{eq:bulkrate1},
	\be\label{eq:bulkrate1high}
	\Gamma^{(bb)} = \Gamma_0 \frac{T T_{\rm B}^{(b)2}}{T_{\rm BG}^3} \Xi^{(bb,1)},
	\ee
	but with a different numerical constant $\Xi^{(bb,1)}$.
	For $\mu \ll \kb v$, we obtain  $\Xi^{(bb,1)} = 2 \Xi^{(bb,2)}$ since now bulk quasiparticles can also be efficiently scattered between different Weyl nodes.
	In the low temperature limit $T \ll T_{\rm BG}^{(b)}$, inter-node scattering is suppressed and intra-node scattering implies
	\begin{equation}
	\label{eq:bulkrate3}
	\Gamma^{(bb)} =  \Gamma_0 \frac{T^3}{T_{\rm BG}^3}\Xi^{(bb,3)},
	\end{equation}
	with the numerical constant $ \Xi^{(bb,3)} =7\zeta(3)/4\pi$. 
	Here $\zeta(s)$ is the Riemann zeta function and $\zeta(3)\simeq 1.202$.
	
	Using $g_0\sim 1$~eV and the above-mentioned TaAs material parameters with $\rho_M\approx 12$~g$/$cm$^{3}$ for a rough estimate, we find $\Gamma_0\sim 80$~GHz and hence $\Gamma_0/ T_{\rm BG}\sim 0.3$. The linear temperature dependence in Eq.~\eqref{eq:bulkrate1} reflects the fact that intra-node backscattering processes proliferate in the temperature regime \eqref{intermediateTb}.
	The Boltzmann approach requires a sufficiently high charge carrier density, $\mu \gg \Gamma^{(bb)}_{\vklabel}$ \cite{Son2013b}. 
	This situation is commonly realized in experiments \cite{Lv2015,Xu2015}. 
	We note that very low carrier densities $\sim 10^{19}/$cm$^{3}$ have been reported for TaAs \cite{Luo2016}. For $\mu\to 0$, a conductivity minimum is then expected, where disorder may cause puddles with local variations of the chemical potential  \cite{Ramakrishnan2015}. Such effects are beyond the scope of Eqs.\ \eqref{eq:dtf} and \eqref{eq:dtf2}.
	Using the above parameter estimates, Eq.~\eqref{eq:bulkrate1} predicts a bulk mean free path
	$\ell_{\rm ph}\agt 1$~mm for $T\alt T_{\rm BG}$. We conclude that
	in a disorder-free WSM slab at low temperatures, electrons move ballistically
	along the transverse $(\hat x)$ direction, i.e., $L\ll \ell_{\rm ph}(T)$, for slab width $L\alt 1$~mm, such that the condition \eqref{eq:Pineq2} is satisfied.
	
	\section{Fermi arc quasiparticle decay rate}\label{sec4}
	
	In this section, we discuss the phonon-induced scattering rate for Fermi arc states, which is the sum of the rate $\Gamma_{\vkparallel}^{(ss)\pm}$ for arc-arc scattering and the rate $\Gamma_{\vkparallel}^{(bs)\pm}$ for arc-bulk scattering,
	\bea
	\Gamma_{\vkparallel}^{(ss)\pm} &=& \int \frac{d\vkparallel'}{(2 \pi)^2} {\cal W}^{(ss)\pm}_{\vkparallel',\vkparallel}, \nonumber \\ \label{ratedef}
	\Gamma_{\vkparallel}^{(bs)\pm } &=& \int_0^{\infty} \frac{dk_x'}{2 \pi} \int \frac{d\vkparallel'}{(2 \pi)^2} {\cal W}^{(bs)\pm}_{\vkparallel',\vkparallel}.
	\eea
	Expressions for the kernels ${\cal W}^{(ss)}$ and ${\cal W}^{(bs)}$ are given in Eqs.\ \eqref{eq:calWssApp} and \eqref{eq:calWbs}, respectively. 
	For each rate in Eq.\ \eqref{ratedef}, contributions from different phonon types add up.
	For simplicity, we focus on the Fermi arc states at the surface $x=+L/2$ and omit the superscript $+$ in what follows.  We then separately describe the scattering rates $\Gamma_{\vkparallel}^{(ss)}$ and $\Gamma_{\vkparallel}^{(bs)}$ and their temperature dependence, where
	(unless noted otherwise) we focus on the regime $|\varepsilon - \mu| \ll  T$.

	\begin{figure*}
		\includegraphics[width=0.465\textwidth]{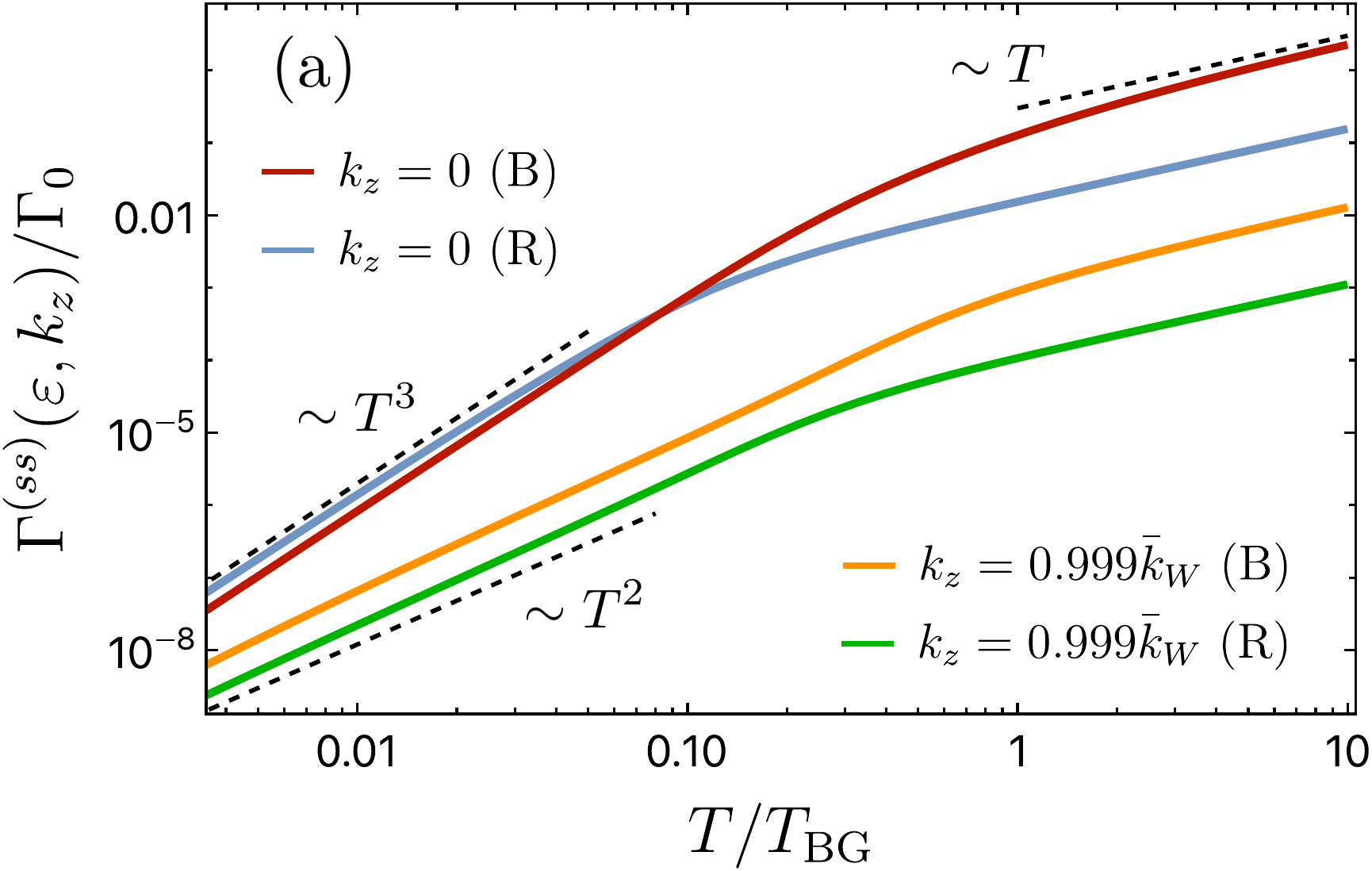} \ \ 
		\includegraphics[width=0.45\textwidth]{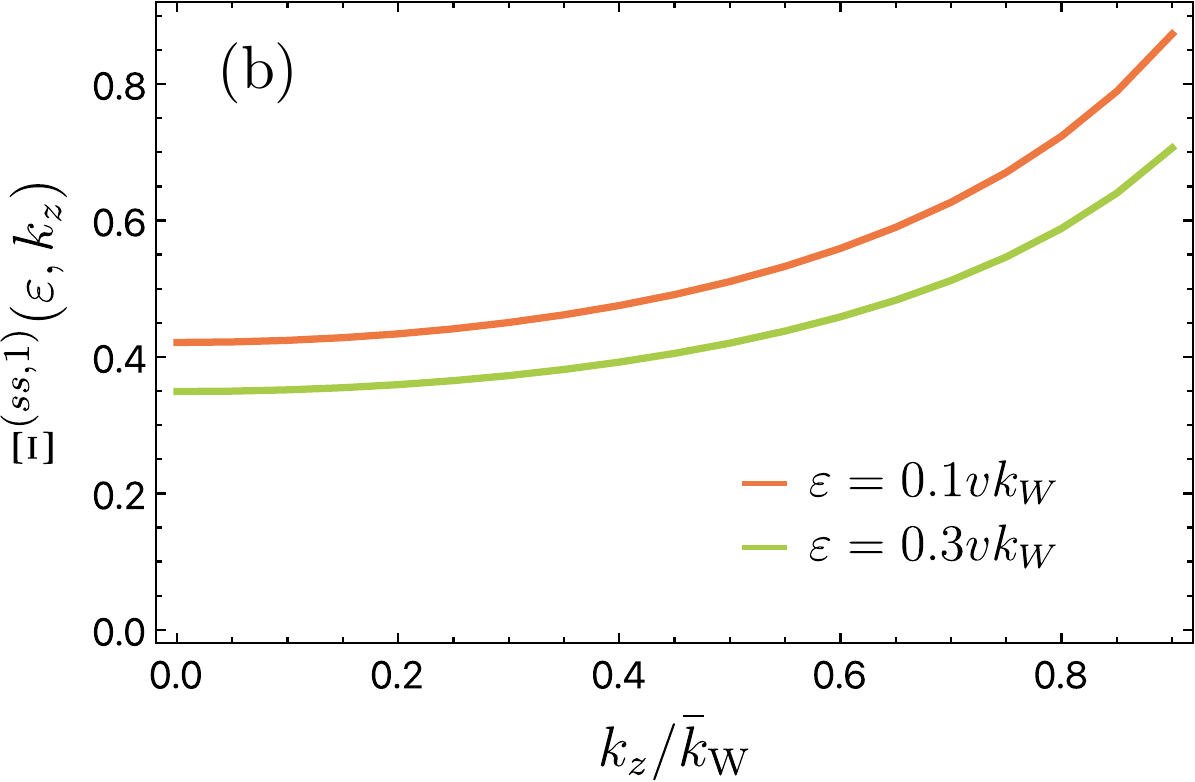}
		\caption{
			Arc-arc decay rate for $\alpha=0.8$.  In all figures, we use $c_l/v=0.0170$ and  $c_t/c_l=0.571$.   
			Panel (a) shows the temperature dependence of $\Gamma^{(ss)}(\varepsilon,k_z)$, see Eq.\ \eqref{eq:Gammass},
			on double-logarithmic scales for  $\varepsilon=0.1 vk_{\rm W}$ and two values  of $k_z$,
			with $\Gamma_0$ in Eq.\ \eqref{eq:xilambda} and $T_{\rm BG}=c_l k_{\rm W}$, see Eq.\ \eqref{eq:TBG}.
			We distinguish contributions from the Rayleigh (R) mode and from bulk (B) phonon modes ($\lambda=l,t)$.  The dashed lines represent the 
			$T^3$ and $T^2$ scaling near the arc center,  see Eq. \eqref{eq:GammassEstimate}, and close to the arc edge, see Eq. \eqref{eq:GammassEstimateEnd}, respectively.
			Panel (b) shows (for two different energies) the $k_z$-dependence of the rate for $T\gg T_{\rm BG}$ as encoded by the 
			coefficient $\Xi^{(ss,1)}(k_z)$ in Eq.\ \eqref{eq:GammassThigh}.   } 
		\label{fig2}
	\end{figure*}

	\subsection{Arc-arc decay rate}\label{sec4a}
	
	We consider first the decay rate of a Fermi-arc surface state with in-plane momentum $\vkparallel$ and energy $\varepsilon = \varepsilon^{(s)+}_{\vkparallel}$ 
	due to phonon-induced arc-arc scattering.
	Using the expressions in Sec.\ \ref{sec2b} and in App.\ \ref{appC}, this rate is given by
	\bea
	\label{eq:Gammass}
	\Gamma_{\vkparallel}^{(ss)} &=&  2 \int \frac{dk_z'}{2 \pi |v_y^{(s)+}(\varepsilon,k_z')|} 
	\Biggl[ \vphantom{\frac{M}{M}} |{\cal G}_{\vkparallel',\vkparallel}^{(ssR)}|^2
	{\cal F}(c_R q_{\parallel}) \nonumber \\ && \  \mbox{} + \sum_{\lambda=l,t}
	\int_0^\infty \frac{dq_x}{2 \pi} |{\cal G}_{\vkparallel',\vkparallel,q_x}^{(ss\lambda)}|^2
	{\cal F}(c_{\lambda} q) \Biggr],
	\eea
	where we defined
	\be
	\label{eq:FdefF}
	{\cal F}(\Omega) = n_{\rm B}(\Omega) + \frac{1}{2} \left[ n_{\rm F}(\varepsilon + \Omega) + 1 - n_{\rm F}(\varepsilon - \Omega) \right ].
	\ee
	For $|\varepsilon-\mu|\ll T$, Eq.\ \eqref{eq:FdefF} can be simplified to ${\cal F}(\Omega)=1/\sinh(\Omega/T)$.  In Eq.\ \eqref{eq:Gammass},
	we use $\vkparallel' = k_y' \hat y + k_z' \hat z$ with $k_y'$ being the solution of $\varepsilon^{(s)+}_{\vkparallel'} = \varepsilon$, $\vqparallel = \vkparallel' - \vkparallel$, and $\vqvector = (q_x,\vqparallel)$. From the Fermi arc dispersion relation \eqref{eq:Es}, we find that
	\be\label{qydef}
	q_y =  \frac{q_z(2 k_z + q_z)}{2 \kb} \tan \alpha 
	\ee
	for the surface at $x = L/2$.  
	
	Representative numerical results for the temperature dependence of the arc-arc decay rate are shown in Fig.\ \ref{fig2}(a).
	In the low-temperature regime $T\ll T_{\rm BG}$, with the effective Bloch-Gr\"uneisen temperature in Eq.\ \eqref{eq:TBG}, we observe that the Rayleigh mode 
	quantitatively gives the largest contribution since this phonon mode has the lowest frequency.  The contributions
	from bulk ($\lambda=l,t)$ phonon modes give the same temperature dependence but come with a smaller prefactor.  
	At high temperatures, $T\gg T_{\rm BG}$, bulk modes give the dominant contribution.
	
	For temperatures $T \gg T_{\rm BG}$, we may approximate ${\cal F}(\Omega) \approx T/\Omega$. After the integration over $k_z'$ in Eq.\ \eqref{eq:Gammass}, one  finds 
	\bea
	\label{eq:GammassThigh}
	\Gamma^{(ss)}_{\vkparallel}&\approx & \Gamma_0 \frac{T T^*(\varepsilon,k_z)}{T_{\rm BG}^2} \Xi^{(ss,1)}(\varepsilon,k_z)\\
	&=& 2\pi \lambda^{(ss)}(\varepsilon,k_z) T, \nonumber
	\eea
	with the rate $\Gamma_0$ in Eq.\ \eqref{eq:xilambda}.  The temperature scale 
	\be\label{tstar}
	T^*(\varepsilon,k_z) = \frac{\kappa(\varepsilon,k_z)}{\kb} T_{\rm BG}
	\ee
	goes to zero at the arc ends, see Eq.\ \eqref{eq:kapp2}, and $\Xi^{(ss,1)}$ is a dimensionless numerical coefficient which depends  
	on the position $k_z$ along the arc.  The second line
	in Eq.\ \eqref{eq:GammassThigh} defines the dimensionless electron-phonon coupling  parameter $\lambda^{(ss)}$ \cite{Mahan,Ashcroft}  due to arc-arc scattering which also depends on $k_z$.
	Using the matrix  elements in App.\ \ref{appB} and typical material parameters, we obtain the 
	numerical results for $\Xi^{(ss,1)}(k_z)$ shown in Fig.\ \ref{fig2}(b).   
	
	In what follows, we discuss the regime $T\ll T_{\rm BG}$ and give explicit expressions for the Rayleigh mode contribution $\Gamma_{\vkparallel}^{(ss,R)}$ only, for which analytical expressions can be obtained.
	Taking the matrix element ${\cal G}^{(ssR)}_{\vkparallel',\vkparallel}$ from Eq.\ \eqref{eq:GssB} and replacing the integration variable in Eq.\ \eqref{eq:Gammass} by $q_z$, we find
	\bea
	\label{eq:GammassR}
	\Gamma_{k_z}^{(ss,R)\pm} &=&
	\frac{4 g_0^2 \xi^{(l)2} c_R}{c_l^2 \rho_M} \int \frac{dq_z}{2 \pi}
	\frac{{\cal F}(c_R q_{\parallel})}{|v_y^{(s)\pm}(k_z+q_z)|}  
	\\ & \times&
	\frac{q_{\parallel}^2 \kappa(k_z) \kappa(k_z+q_z)}{[\kappa(k_z) + \kappa(k_z+q_z) 
		+ \gamma_R q_{\parallel} ]^2}, \nonumber
	\eea  
	where $\xi^{(l)}$ and $\gamma_R=\sqrt{1-(c_R/c_l)^2}$ are dimensionless constants, see Eqs.\ \eqref{eq:xil1}--\eqref{eq:xil3}.
	
	For $T \ll T_{\rm BG}$ and staying away from the arc ends at $k_z = \kkB 
	(\varepsilon)$, we may approximate $k_z+q_z \approx k_z$ 
	and neglect the term proportional to $q_{\parallel}$ in the denominator in Eq.\ \eqref{eq:GammassR}. 
	Performing the integration over $q_z$ and using Eq.\ \eqref{eq:arcvelocities} for $v_y^{(s)\pm}(\varepsilon,k_z')$, we then find a $T^3$ scaling of the rate,
	\be
	\label{eq:GammassEstimate}
	\Gamma_{\vkparallel}^{(ss,R)\pm} \approx
	\Gamma_0   \left( \frac{T}{T_{\rm BG}} \right)^3 \Xi^{(ss,2)}(k_z),
	\ee
	where 
	\be
	\Xi^{(ss,2)}(k_z) = \frac{7 \zeta(3) \xi^{(l)2} c_l^2}{2 \pi c_R^2} \frac{\kb}{
		\sqrt{\kb^2 \cos^2 \alpha + k_z^2 \sin^2 \alpha}}
	\ee
	is a numerical coefficient of order unity that only weakly depends on $k_z$. 
	
	The $T^3$ scaling law in Eq.\ \eqref{eq:GammassEstimate} does not hold if \mbox{$T \gtrsim T^*(\varepsilon,k_z)$}, which occurs in the immediate vicinity of the arc ends. 
	To describe the arc-arc scattering rate for $k_z$ near the arc ends, we note that for 
	\mbox{$T^*(\varepsilon,k_z) \ll T \ll T_{\rm BG}$}, we may approximate \mbox{$\kappa(k_z+q_z) \approx \kkB(\varepsilon) |q_{z}|/(\kb \cos \alpha)$}, see Eq.\ \eqref{eq:kapp2}, and neglect $\kappa (k_z)$ with respect to $\kappa(k_z+q_z)$ and $q_{\parallel}$ in the denominator of Eq.\ \eqref{eq:GammassR}. 
	Performing the integral over $q_{z}$ in Eq.\ \eqref{eq:GammassR}, 
	we now find a $T^2$ scaling law for the rate,
	\bea
	\label{eq:GammassEstimateEnd}
	\Gamma_{\vkparallel}^{(ss,R)} &\approx& 
	\Gamma_0  \left( \frac{T}{T_{\rm BG}} \right)^2  \frac{T^*(\varepsilon,k_z)}{T_{\rm BG}}  \Xi^{(ss,3)}(\varepsilon),
	\eea
	where $\Xi^{(ss,3)}(\varepsilon)$ is a numerical coefficient of order unity,
	\be
	\Xi^{(ss,3)}(\varepsilon) =
	\frac{\pi\xi^{(l)2} c_l }{2 c_R } \frac{\kkB(\varepsilon) \kb}{[\kkB(\varepsilon) + 
		\gamma_R \kkkB(\varepsilon)]^2},
	\ee
	and we use 
	\be
	\kkkB(\varepsilon) = \sqrt{\kb^2 \cos^2 \alpha + \kkB(\varepsilon)^2 \sin^2 \alpha}.
	\ee
	
	\begin{figure*}
		\includegraphics[width=0.45\textwidth]{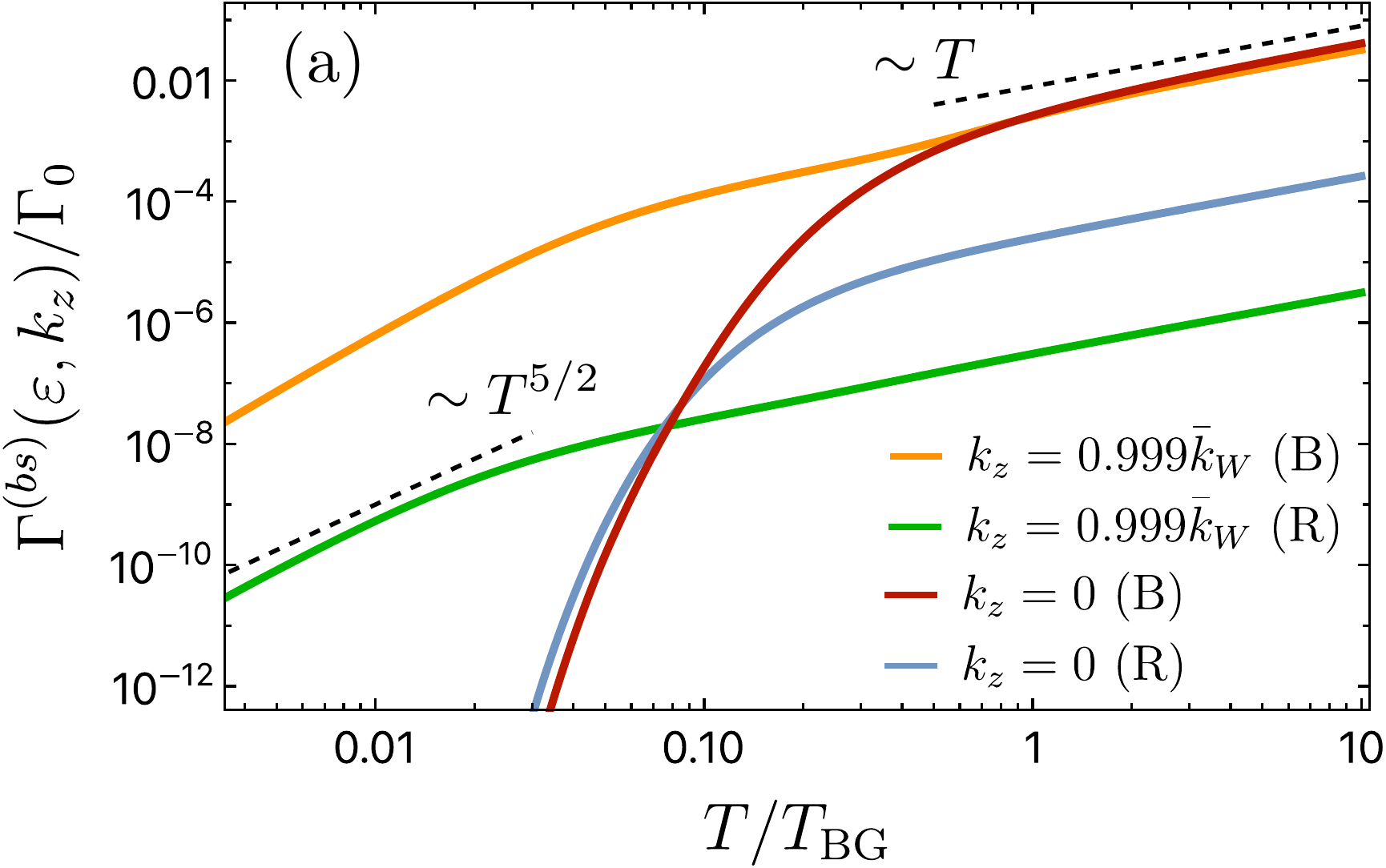}\ \
		\includegraphics[width=0.45\textwidth]{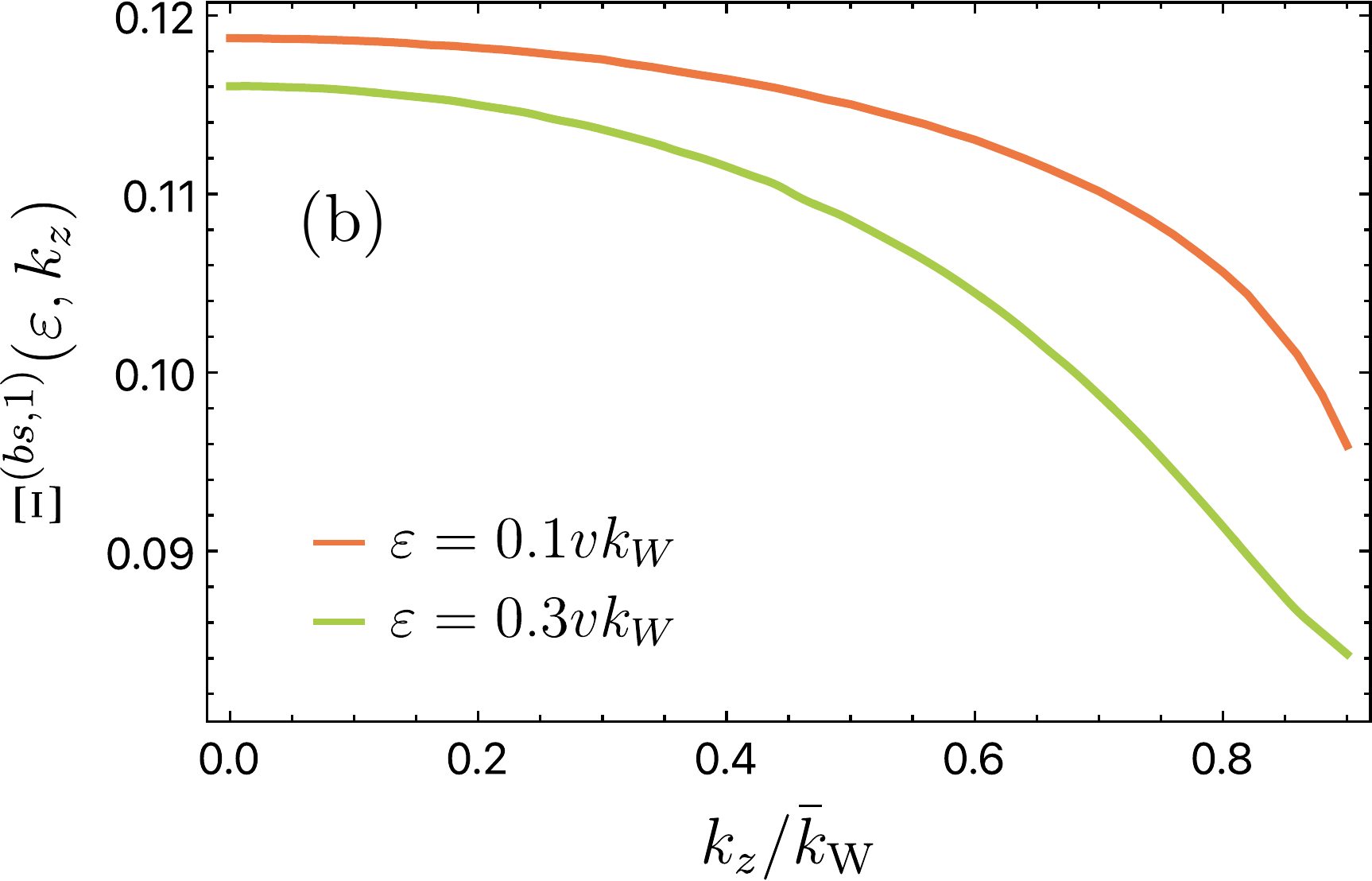}\\
		\caption{
			Arc-bulk decay rate $\Gamma^{(bs)}(\varepsilon,k_z)$ for energy $\varepsilon=0.1v\kb$ and boundary angle $\alpha=0.8$ as obtained by numerical 
			integration of Eq.\ \eqref{eq:Gammabs}.  
			Panel (a) shows the temperature dependence of $\Gamma^{(bs)}$ on double-logarithmic scales for two values of $k_z$. Equation~\eqref{eq:actvani} gives
			$\Omega_{\rm min}/T_{\rm BG}\simeq 4.6\times 10^{-6}$
			for $k_z=0.999 \bar k_{\rm W}$. 
			As in Fig.\ \ref{fig2}(a), we separately show contributions from Rayleigh (R) and bulk (B) phonon modes.  The dashed lines indicate power-law scaling with 
			the respective exponents.  
			Panel (b) shows the $k_z$-dependence of the coefficient $\Xi^{(bs,1)}$ determining the arc-bulk rate for $T\gg T_{\rm BG}$, see Eq.~\eqref{eq:gammabshighT}, for two energies.}
		\label{fig3}
	\end{figure*}
	
	The contributions of bulk ($\lambda=l,t$) phonon modes will 
	renormalize the coefficients $\Xi^{(ss,2)}$ and $\Xi^{(ss,3)}$,
	but they do not affect the overall temperature dependence of the arc-arc scattering rate at low temperatures, cf.\ Fig.~\ref{fig2}(a). 
	Our numerical results show that the Rayleigh contribution to the arc-arc rate is at most slightly larger than the bulk phonon contribution, but both share the same power law temperature dependence of the rate. 
	
	Finally, we remark that similar estimates for the arc-arc rate can be made in the regime $|\varepsilon - \mu| \gg  T$, but with the role of $T$ replaced by $|\varepsilon - \mu|$ and different numerical coefficients $\Xi^{(ss,j)}$ for $j=1,2,3$.
	
	\subsection{Arc-bulk decay rate}\label{sec4b}
	
	We next consider the decay rate of a Fermi arc state with in-plane momentum $\vkparallel$ and energy $\varepsilon = \varepsilon^{(s)+}_{\vkparallel}$ due to phonon-induced arc-bulk scattering. Within the quasi-elastic approximation, the arc-bulk scattering rate is given by the expression
	\bea
	\label{eq:Gammabs}
	\Gamma_{\vkparallel}^{(bs)} &=&
	2 \int \frac{d\vkpparallel}{(2 \pi)^2 |v_x^{(b)}(\varepsilon,\vkpparallel)|} 
	\Biggl [ \vphantom{\frac{M}{M}} |{\cal G}_{\vkpvector,\vkparallel}^{(bsR)}|^2
	{\cal F}(c_R q_{\parallel})  \nonumber \\ &+& \sum_{\lambda=l,t}
	\int_0^\infty \frac{dq_x}{2 \pi} |{\cal G}_{\vkpvector,\vkparallel,q_x}^{(bs\lambda)}|^2
	{\cal F}(c_{\lambda} q) \Biggr].
	\eea
	The matrix elements ${\cal G}_{\vkpvector,\vkparallel}^{(bsR)}$ and ${\cal G}_{\vkpvector,\vkparallel,q_x}^{(bs\lambda)}$ follow from Eqs.\ \eqref{eq:Gsblambda}--\eqref{eq:GbsR}, 
	where $k_x'$ is the positive solution of
	\be
	\label{eq:qconstraint}
	\varepsilon^{(b)}_{(k_x',\vkpparallel)} = \varepsilon,
	\ee
	and $\vqvector = (q_x,\vqparallel)$ with $\vqparallel =\vkpparallel - \vkparallel$. 
	
	The integration in Eq.\ \eqref{eq:Gammabs} is restricted to those values of $\vkpparallel$ for which solutions of Eq.\ \eqref{eq:qconstraint} exist. Using the bulk dispersion \eqref{eq:eenrgy}, Eq.\ \eqref{eq:qconstraint} gives the condition
	\be\label{eq:qmin1}
	v k_x' = \sqrt{\varepsilon^2 - v^2 k_y'^2 - m^2(k_z')},
	\ee
	which, for $\varepsilon \ll \kb v$, effectively restricts the integration over $\vkpparallel$ to an approximately circular
	region of radius $\varepsilon/v$ around the nodal points, see Fig.\ \ref{fig1}(b).
	For energies approaching the neutrality point, $\varepsilon\to 0$, the vanishing bulk DoS 
	$n_b(\varepsilon)\propto \varepsilon^2$ in Eq.~\eqref{bulkdos} thereby implies that the arc-bulk rate approaches zero for all $k_z$.
	The temperature dependence of the arc-bulk rate at finite $\varepsilon$ is illustrated for  $\alpha=0.8$ in Fig.~\ref{fig3}(a), where we separately show the contributions from bulk and from Rayleigh phonons as obtained by numerical integration of Eq.\ \eqref{eq:Gammabs}.  
	
	For $T \gg T_{\rm BG}$ and $|\varepsilon - \mu| \ll T$, we may again use ${\cal F}(\Omega) \approx  T/\Omega$. 
	One then arrives at a linear temperature dependence of the arc-bulk scattering rate, see also Fig.\ \ref{fig3}(a),
	\bea   \label{eq:gammabshighT}
	\Gamma^{(bs)}_{\vkparallel} &\approx&
	\Gamma_0 \frac{T T_{\rm BG}^{(b)2}}{T_{\rm BG}^3} 
	\Xi^{(bs,1)}(\varepsilon,k_z) 
	\nonumber \\ &=&
	2\pi \lambda^{(bs)}(\varepsilon,k_z) T,
	\eea  
	where $\Xi^{(bs,1)}$ is a $k_z$-dependent numerical coefficient which is of order unity away from the arc edges.  
	The effective Bloch-Gr\"uneisen temperature $T^{(b)}_{\rm BG}$ has been defined in Eq.\ \eqref{eq:tbgb}, and the factor $T_{\rm BG}^{(b)2}/T_{\rm BG}^2 \propto (\varepsilon/\kb v)^2$ 
	reflects the suppression of the bulk DoS in the vicinity of the nodal points. With the dimensionless parameter $\lambda^{(bs)}$, 
	the full electron-phonon coupling parameter for Fermi arc states is given by $\lambda=\lambda^{(ss)}+\lambda^{(bs)}$.
	Numerical results for $\Xi^{(bs,1)}$ are shown in Fig.\ \ref{fig3}(b). By comparing these results to those for arc-arc scattering in 
	Fig.\ \ref{fig2}(b), we observe that arc-arc scattering generally dominates over arc-bulk scattering except in the vicinity of the 
	arc termination points at $k_z=\pm \kkB (\varepsilon)$. 
	
	We now turn to the low-temperature regime $T \ll T_{\rm BG}$, where
	it is essential that the arc and bulk states are non-overlapping in the in-plane momentum space, as illustrated in Fig.\ \ref{fig1}(b). 
	As a result, for an arc state at in-plane momentum $\vkparallel$, there is a minimal (in-plane) phonon momentum $q_{\parallel} = |\vkpparallel - \vkparallel|$ required for arc-bulk scattering. (We recall that in the quasi-elastic approximation, the initial arc state at in-plane momentum $\vkparallel$ and the final bulk state at momentum $\vkpvector$ have the same energy $\varepsilon$.) 
	For $\varepsilon \ll \kb v$, the support of bulk states is a disc of radius $\varepsilon/v$ around the nodal points $\pm \kb$, see Fig.\ \ref{fig1}(b).
	Using Eq.\ \eqref{eq:k2karch} for $k_y(\varepsilon,k_z)$, an elementary geometric consideration then 
	shows that the minimum in-plane phonon momentum is given by 
	\be\label{qmin}
	q_{\rm min}(\varepsilon,k_z) = \sqrt{k_y(\varepsilon,k_z)^2 + (|k_z| - \kb)^2} - \frac{\varepsilon}{v}.
	\ee
	If the condition $\varepsilon \ll \kb v$ is not met, $q_{\rm min}$ is determined from the condition \eqref{eq:qconstraint} using a constrained minimization procedure. 
	
	\begin{figure}
		\centering
		\includegraphics[width=0.95\columnwidth]{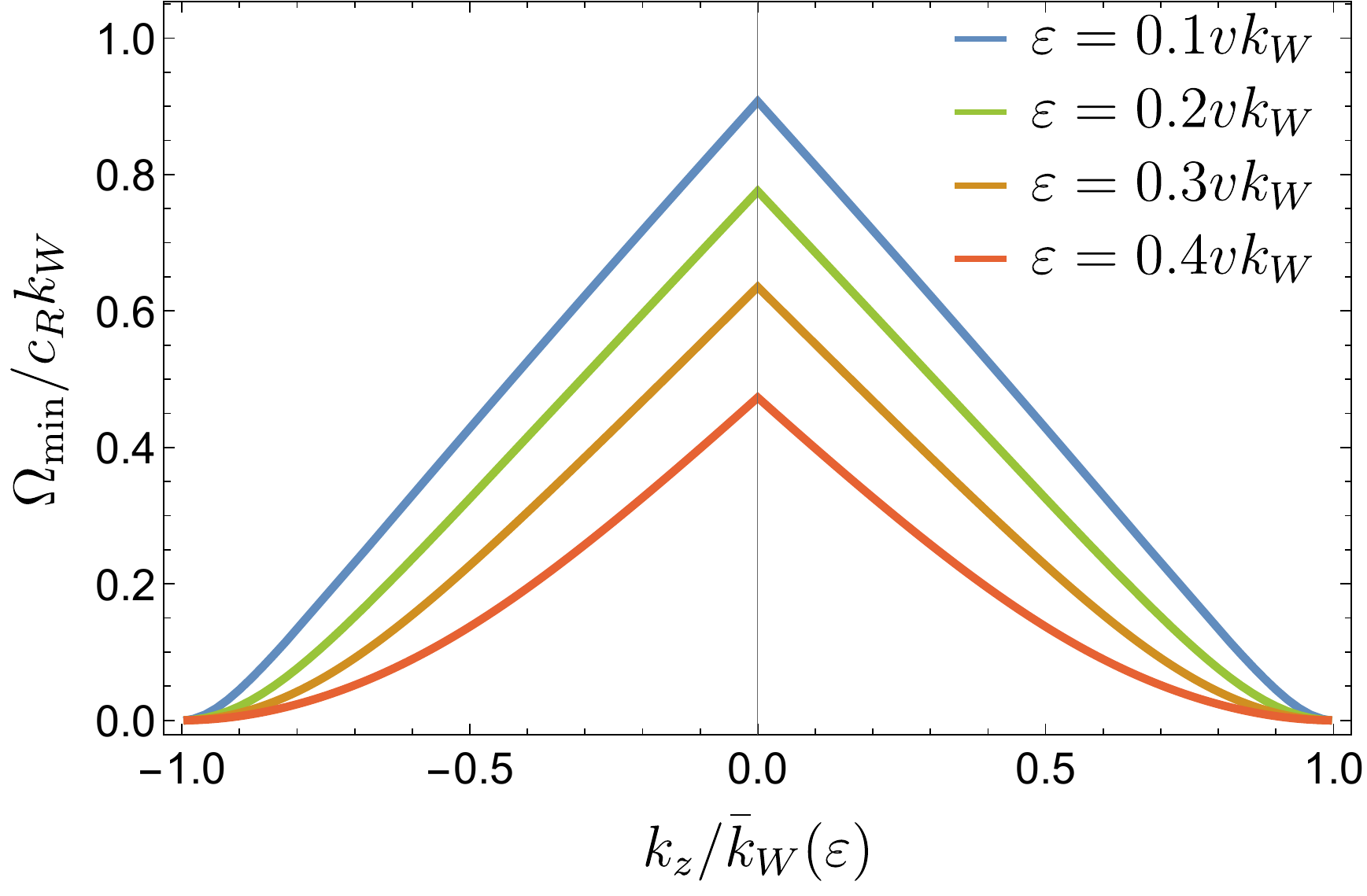}
		\caption{Phonon threshold energy for arc-bulk scattering, $\Omega_{\rm min}(\varepsilon,k_z)$, vs position along the arc, $k_z/\kkB(\varepsilon)$, for $\alpha = 0.5$ and several energies.  
		}
		\label{fig4}
	\end{figure}
	
	\begin{table*}[t!]
		\centering
		\begin{tabular}{ | c || c | c | c | c |}
			\hline
			&$T \ll T^{(b)}_{\rm BG}$  & &  $ T^{(b)}_{\rm BG} \ll T $ \\
			\hline
			$\Gamma^{(bb)}$    &  $T^3$  & &  $T$  \\
			\hline
			\hline
			& $T\ll T^*$ & $ T^* \ll T \ll T_{\rm BG}$ & $ T_{\rm BG} \ll T$   \\
			\hline
			$\Gamma^{(ss)}$   &  $T^3$  &  $T^2$ &   $T$    \\
			\hline
			\hline
			& $ T \ll T^{(b)}_{\rm BG} \ll \Omega_{\rm min}$  
			& $ T^{(b)}_{\rm BG} \ll T \ll \Omega_{\rm min} $ 
			& $ T^{(b)}_{\rm BG} \ll \Omega_{\rm min} \ll T $  \\
			\hline
			$\Gamma^{(bs)}$    away from arc end
			& $T^2 e^{-\Omega_{\rm min}/T}$
			& $e^{-\Omega_{\rm min}/T}$ 
			& $T$    \\
			\hline
			\hline 
			& $T \ll \Omega_{\rm min} \ll T^{(b)}_{\rm BG} $
			& $  \Omega_{\rm min} \ll T \ll T^{(b)}_{\rm BG} $ 
			& $  \Omega_{\rm min} \ll T^{(b)}_{\rm BG} \ll T $  \\
			\hline 
			$\Gamma^{(bs)}$  close to arc end   
			&  $ T^2 e^{-\Omega_{\rm min}/T}$
			& $T^{5/2}$ 
			& $T$   \\
			\hline
		\end{tabular}
		\caption{Temperature dependence of the scattering rates in different parameter regimes.
			Putting $k_{\rm B}=1$,
			the Bloch-Gr\"uneisen temperatures are given by $T_{\rm BG}=c_l k_{\rm W}$ and $T_{\rm BG}^{(b)}=2c_l \mu/v$, see
			Eqs.~\eqref{eq:TBG} and \eqref{eq:tbgb}.  The scale $T^\ast(\varepsilon,k_z)$ has been defined in Eq.~\eqref{tstar} and 
			goes to zero near the arc ends, and the arc-bulk activation energy $\Omega_{\rm min}(\varepsilon,k_z)$ follows from Eqs.~\eqref{qmin} and \eqref{activationgap}.
		}
		\label{table:1}
	\end{table*}

	Since the Rayleigh mode is the energetically lowest phonon branch, the existence of a threshold phonon momentum for arc-bulk scattering $q_{\rm min}$ implies the threshold activation energy
	\be\label{activationgap}
	\Omega_{\rm min}(\varepsilon,k_z)= c_R q_{\rm min}(\varepsilon,k_z).
	\ee
	Representative results for $\Omega_{\rm min}$ are shown in Fig.~\ref{fig4} for different values of $\varepsilon < v \kb/2$. The activation energy $\Omega_{\rm min}$ reaches its maximum value at the arc center, with a cusp-like dependence on $k_z$ for $k_z \to 0$, and vanishes upon approaching the arc edges at $k_z = \pm \kkB(\varepsilon)$,
	\be\label{eq:actvani}
	\Omega_{\rm min}(\varepsilon,k_z) \approx \frac{v c_{\rm R} [|k_z| - \kkB(\varepsilon)]^2}{2 \varepsilon \cos^2 \alpha},
	\ee
	where we have again used $\varepsilon \ll v\kb$.  
	
	The existence of a finite activation gap implies an exponential suppression of the arc-bulk decay rate at temperatures $T \ll \Omega_{\rm min}(\varepsilon,k_z)$.
	As a result, arc-bulk scattering will be appreciable near the arc ends only. 
	Up to numerical prefactors of order unity, analytical estimates for the arc-bulk rate at $T\ll T_{\rm BG}$ can be 
	obtained in a similar way as for the arc-arc rate in Sec.\ \ref{sec4a} by using the expressions in App.\ \ref{appC}.  
	In the remainder of this section, we discuss the arc-bulk rate in different regimes defined by the relation between the temperature $T$, 
	the activation gap $\Omega_{\rm min}(\varepsilon,k_z)$, and the intra-node Bloch-Gr\"uneisen scale $T_{\rm BG}^{(b)}$, where 
	we assume $\varepsilon\ll v \kb$ throughout.
	Those results will also be used in Sec.\ \ref{sec5} below.  
	
	First, let us consider the case away from the arc edges at $k_z=\pm \kkB (\varepsilon)$, where the 
	condition $T_{\rm BG}^{(b)}\ll \Omega_{\rm min}(\varepsilon,k_z)$ is fulfilled.
	For $\Omega_{\rm min} \ll  T$, the arc-bulk scattering rate can be estimated as, cf. Eq.\ \eqref{eq:gammabshighT},
	\be \label{eq:gammasbact0}
	\Gamma^{(bs)}\sim \Gamma_0   \frac{T_{\rm BG}^{(b)2} T}{T_{\rm BG}^3} .
	\ee
	On the other hand, for $T_{\rm BG}^{(b)} \ll T \ll \Omega_{\rm min}$, the activation gap causes an exponential suppression 
	of the arc-bulk rate,
	\be \label{eq:gammasbact}
	\Gamma^{(bs)} \sim \Gamma_0 
	\frac{T_{\rm BG}^{(b)2} \Omega_{\rm min}}{T_{\rm BG}^3}
	e^{-\Omega_{\rm min}/T}.
	\ee
	Finally, for $T \ll T_{\rm BG}^{(b)}$, we obtain  
	\be
	\label{eq:gammasbact4}
	\Gamma^{(bs)} \sim \Gamma_0   \frac{T^{2} \Omega_{\rm min}}{T_{\rm BG}^3} 
	e^{-\Omega_{\rm min}/T}.
	\ee 
	The exponential suppression of the arc-bulk rate away from the arc edges is consistent with the numerical results in Fig.\ \ref{fig3}(a).
	
	We finally discuss what happens very close to the arc edges, where $\Omega_{\rm min}\ll T_{\rm BG}^{(b)}$. For details of the derivation, see App.\ \ref{appD}.
	First, for $T\gg T_{\rm BG}^{(b)}$, a high-temperature regime as in
	Eq.\ \eqref{eq:gammabshighT} is realized,  with 
	\begin{equation}
	\Gamma^{(bs)} \sim \Gamma_0 \frac{ T \Omega_{\rm min}^{1/2} T_{\rm BG}^{(b) 3/2}}{T_{\rm BG}^3}.
	\end{equation}
	Second, for $\Omega_{\rm min}\ll T \ll T_{\rm BG}^{(b)}$,
	we obtain a nontrivial $T^{5/2}$ scaling law for the rate,
	\be
	\label{eq:gammasbact2}
	\Gamma^{(bs)}\sim \Gamma_0 
	\frac{T^{5/2} \Omega_{\rm min}^{1/2}}{T_{\rm BG}^3} .
	\ee 
	This $T^{5/2}$ scaling law is consistent with the numerical results in Fig.\ \ref{fig3}(a).
	These numerical results also show that in this parameter range, the arc-bulk rate is always dominated by bulk phonon contributions.
	Nonetheless, the Rayleigh contribution yields the same power law exponent. 
	Finally, for extremely low temperatures, $T \ll \Omega_{\rm min}$, the rate will again be exponentially suppressed,
	\be
	\label{eq:gammasbact3}
	\Gamma^{(bs)}\sim \Gamma_0   \frac{T^{2} \Omega_{\rm min}}{T_{\rm BG}^3}  e^{-\Omega_{\rm min}/T}.
	\ee
	
	A schematic overview summarizing the temperature dependence of the arc-bulk rate is given in Table \ref{table:1}, which also includes the respective results for the bulk-bulk rate and for the arc-arc rate.
	
	\section{Transport properties} \label{sec5}
	
	In this section, we consider the phonon-limited linear conductivity tensor in a WSM slab for an electric field applied along the $\hat y$ or $\hat z$ direction, 
	i.e., perpendicular or parallel to the separation between the Weyl points. A brief discussion of 
	the Hall response is given in App.~\ref{appE}. We here focus on the longitudinal conductivities 
	$\sigma_{yy}$ and $\sigma_{zz}$. We will first give qualitative considerations for $\sigma_{yy}$ along the chiral direction $\hat y$,
	see Sec.~\ref{sec5a}, and similarly for $\sigma_{zz}$ in Sec.~\ref{sec5b}.  Subsequently, we will summarize the temperature dependence of $\sigma_{yy}$ and $\sigma_{zz}$ in different parameter regions as obtained by a numerical solution of 
	the  Boltzmann equation, see Sec.~\ref{sec5d}. Below we assume $\mu \ll v\kb$ to have a clear separation between the 
	temperature scales $T_{\rm BG}^{(b)}$ and  $T_{\rm BG}$.
	
	\subsection{Diagonal conductivity $\boldsymbol{\sigma_{yy}}$}\label{sec5a}
	
	We first consider an electric field along the chiral direction $\hat y$ and discuss the qualitative behavior of the conductivity $\sigma_{yy}$.
	These arguments are backed up by a numerical analysis of the full problem in Sec.~\ref{sec5d}.
	Since the current is carried by bulk electrons and by Fermi-arc surface states, we can
	write $\sigma_{yy}$ as a sum of bulk and surface contributions,
	\be
	\sigma_{yy} = \sigma_{yy}^{(b)} + \sigma_{yy}^{(s)}.
	\ee
	Each contribution may be estimated using the Drude formula,
	\be\label{Drude}
	\sigma_{yy}^{(b)} \sim \frac{e^2 \mu^2 L}{v} \tau_{{\rm tr},y}^{(b)}, \quad
	\sigma_{yy}^{(s)} \sim e^2 v\kb \tau_{{\rm tr},y}^{(s)},
	\ee
	where $\tau_{{\rm tr},y}^{(b)}$ and $\tau_{{\rm tr},y}^{(s)}$ are transport relaxation time scales for bulk and Fermi arc electrons. 
	Here we have used Eqs.\ \eqref{bulkdos} and \eqref{eq:nFA} for the DoS of bulk and surface electrons, respectively.
	Below we will demonstrate that, up to a numerical factor of order unity,
	\be
	\label{eq:tautrb}
	\frac{1}{\tau_{{\rm tr},y}^{(b)}} \sim 
	\Gamma_0 \times
	\left\{ \begin{array}{ll}
		\displaystyle{\frac{T_{\rm BG}^{(b)2} T}{T_{\rm BG}^3} \vphantom{\int_{M_M^M}^{M^M_M}}} &
		\mbox{if $T \gg T_{\rm BG}^{(b)}$}, \\
		\displaystyle{\frac{T^5}{T_{\rm BG}^{(b)2} T_{\rm BG}^3} \vphantom{\int_{M_M^M}^{M^M_M}}} &
		\mbox{if $T \ll T_{\rm BG}^{(b)}$} ,\end{array} \right.
	\ee
	and
	\begin{equation}
	\label{eq:tautrs}
	\frac{1}{\tau_{{\rm tr},y}^{(s)}} \sim 
	\Gamma_0 \times
	\left\{ \begin{array}{ll}
	\displaystyle{\frac{T_{\rm BG}^{(b)2} T}{T_{\rm BG}^3} \vphantom{\int_{M_M^M}^{M^M_M}}} &
	\mbox{if $T \gg T_{\rm BG}$}, \\
	\displaystyle{\frac{T_{\rm BG}^{(b)2} T^2}{T_{\rm BG}^{4}}
		\vphantom{\int_{M_M^M}^{M^M_M}}} &
	\mbox{if $T'_{\rm BG} \ll T \ll T_{\rm BG}$},  \\
	\displaystyle{\frac{T^5}{T_{\rm BG}^{5}} \vphantom{\int_{M_M^M}^{M^M_M}}} &
	\mbox{if $T \ll T'_{\rm BG}$}, \end{array} \right.
	\end{equation}
	with the temperature scale
	\be \label{Tprime}
	T'_{\rm BG} = \left( T_{\rm BG}^{(b)2} T_{\rm BG}\right)^{1/3}.
	\ee
	
	From Eq.\ \eqref{Drude}, we then find $\sigma_{yy}\propto 1/T$ at high temperatures, $T\gg T_{\rm BG}$. On
	the other hand, for ultra-low temperatures $T\ll T^{(b)}_{\rm BG}$, we obtain $\sigma_{yy}\propto 1/T^5$, see also 
	Ref.~\cite{Resta2018}. Interestingly, Eq.~\eqref{eq:tautrs} admits a third power-law scaling regime 
	with $\sigma_{yy}\propto 1/T^2$ at intermediate 
	temperatures $T'_{\rm BG} \ll T \ll T_{\rm BG}$. This $1/T^2$ scaling is caused by the chirality of Fermi arc states
	and does not occur for $\sigma_{zz}$, see Sec.~\ref{sec5b}.
	However, it is only realized for sufficiently thin slabs.
	Indeed, the above equations imply that with the temperature-dependent crossover length scale
	\begin{equation}\label{Ly}
	L_y = \frac{v^2 k_{\rm W}}{\mu^2} \frac{T_{\rm BG}}{T},
	\end{equation}
	the surface (bulk) contribution will dominate the conductivity $\sigma_{yy}$ for a slab width $L\ll L_y$ ($L\gg L_y)$. 
	The $\sigma_{yy}\propto 1/T^2$ scaling is expected if $L\ll L_y$ holds for temperatures 
	$T'_{\rm BG}\ll T \ll T_{\rm BG}$.  For thicker slabs, the $1/T^2$ dependence is 
	instead replaced by a conventional high-temperature law $\sigma_{yy}\propto 1/T$ 
	for all $T\gg T_{\rm BG}^{(b)}$. In this case, $T'_{\rm BG}$ has no physical importance anymore.
	
	{\em Contribution from bulk electrons.---}For temperatures $T_{\rm BG}^{(b)} \ll T \ll T_{\rm BG}$ and for $T \gg T_{\rm BG}$, the bulk-bulk scattering rate $\Gamma^{(bb)}$ is given by Eqs.\ \eqref{eq:bulkrate1} and \eqref{eq:bulkrate1high}, respectively. Since these rates differ only by a numerical factor of order unity, the first estimate in Eq.\ \eqref{eq:tautrb} follows immediately. For ultra-low temperatures $T \ll T_{\rm BG}^{(b)}$, the rate is instead given by Eq.\ \eqref{eq:bulkrate3}.
	However, $\Gamma^{(bb)}$ differs from the transport rate since the momentum change $\sim k_{\rm B} T/c_l$ for a single electron-phonon scattering event is much less than the momentum change $\sim 2 \mu/v = T_{\rm BG}^{(b)}/ c_l$ required for 
	backscattering. As a result, the transport mean free time is a factor $(T_{\rm BG}^{(b)}/T)^2$ larger than $1/\Gamma^{(bb)}$, and one obtains the standard $T^5$ law of the second estimate in Eq.\ \eqref{eq:tautrb}.
	
	{\em Contribution from Fermi arcs.---}Since all Fermi arc states at the same surface have a velocity component $v_y^{(s)}$ with the same sign, arc-arc scattering alone is not sufficient to relax the current carried by arc states.
	Instead, the transport mean free time $\tau_{{\rm tr},y}^{(s)}$ is determined by the interplay of arc-arc, arc-bulk, and bulk-bulk scattering. The absence of backscattering of Fermi-arc surface states 
	without arc-bulk couplings is a direct consequence of their chirality and therefore 
	holds beyond the specific model studied here.
	We now separately describe this interplay for the three temperature regimes $T \gg T_{\rm BG}$, $T_{\rm BG}^{(b)} \ll T \ll T_{\rm BG}$, and $T \ll T_{\rm BG}^{(b)}$.
	
	For $T \gg T_{\rm BG}$, the arc-arc scattering rate is parametrically larger than the arc-bulk rate, see Eqs.\ \eqref{eq:GammassThigh} and \eqref{eq:gammabshighT}. The fast arc-arc scattering causes equilibration of the arc states, but it does not contribute to backscattering. Instead, backscattering is possible only via arc-bulk scattering. Hence, for $T \gg T_{\rm BG}$, we find $1/\tau_{{\rm tr},y}^{(s)} \sim \Gamma^{(bs)}$. Using Eq.\ \eqref{eq:gammabshighT} for $\Gamma^{(bs)}$ then leads to the first estimate in Eq.\ \eqref{eq:tautrs}.
	
	We next turn to temperatures $T \ll T_{\rm BG}$, where the arc-arc rate $\Gamma^{(ss)}$ acquires a $T^3$ proportionality, 
	see Eq.\ \eqref{eq:GammassEstimate}, except in the immediate vicinity of the arc ends, see Eq.\ \eqref{eq:GammassEstimateEnd}. 
	Moreover, in this temperature range, the typical momentum change upon arc-phonon scattering is smaller by a factor $\sim T/T_{\rm BG}$
	than in the high-temperature regime.  As a result, the rate  $\Gamma^{(ss)}_{\rm eq}$ for full equilibration of the arc states is reduced 
	by a factor $(T/T_{\rm BG})^2$ compared to the typical arc-arc relaxation rate obtained from Eq.\ \eqref{eq:GammassEstimate},
	\be  \label{eq:Gammasseq}
	\Gamma^{(ss)}_{\rm eq} \sim \Gamma_0 \frac{T^5}{T_{\rm BG}^5}.
	\ee
	The linear decay of $\Gamma^{(ss)}_{\vkparallel}$ as one approaches the arc ends, see Eq.\ \eqref{eq:GammassEstimateEnd}, does not affect this estimate. In fact, arc-bulk scattering takes place close to the arc ends only, within a distance such that $\Omega_{\rm min}(\varepsilon,k_z) \lesssim  T$, see Sec.\ \ref{sec4b}. For $T^{(b)}_{\rm BG} \ll T \ll T_{\rm BG}$, only a fraction $T/T_{\rm BG}$ of all arc states are this close to the arc ends. The net arc-bulk transport scattering rate $\Gamma^{(bs)}_{\rm tr}$ is then found by multiplying the arc-bulk decay rate $\Gamma^{(bs)}$ in Eq.\ \eqref{eq:gammasbact0} by a factor $T/T_{\rm BG}$,
	\be
	\label{eq:Gammabstot}
	\Gamma^{(bs)}_{\rm tr} \sim   \Gamma_0 \frac{T_{\rm BG}^{(b)2} T^2}{T_{\rm BG}^4}.
	\ee
	We now observe that both $\Gamma^{(ss)}_{\rm eq}$ and $\Gamma^{(bs)}_{\rm tr}$ are smaller than the bulk-bulk transport relaxation rate 
	in the temperature window $T_{\rm BG}^{(b)}\ll T\ll T_{\rm BG}$. As a consequence, bulk scattering does not restrict the relaxation of Fermi arc states between arcs at opposing surfaces. Comparing Eqs.\ \eqref{eq:Gammasseq} and \eqref{eq:Gammabstot}, we see that $\Gamma^{(bs)}_{\rm tr}$ is smaller than $\Gamma^{(ss)}_{\rm eq}$ for temperatures in the window $T'_{\rm BG} \ll T \ll T_{\rm BG}$. In this case, the relaxation of arc states is dominated by arc-bulk scattering, and we obtain the second estimate in Eq.\ \eqref{eq:tautrs}.
	If, on the other hand, $T_{\rm BG}^{(b)} \ll T \ll T'_{\rm BG}$, it is the intra-arc relaxation that limits the backscattering of arc states, and one arrives at the third estimate in Eq.\ \eqref{eq:tautrs}.  
	For ultra-low temperatures $T \ll T_{\rm BG}^{(b)}$, the estimate for $\Gamma^{(bs)}_{\rm tr}$ will change but arc-arc relaxation remains the limiting factor determining the transport relaxation time $\tau_{{\rm tr},y}^{(s)}$. As a result, the third estimate in Eq.\ \eqref{eq:tautrs} continues to hold for $T \ll T_{\rm BG}^{(b)}$, and therefore for all $T\ll T'_{\rm BG}$.
	
	Let us finally verify the condition $\Gamma^{(ss)}_{\rm eq}\ll \Gamma^{(bs)}_{\rm tr}$ for $T\ll T_{\rm BG}^{(b)}$. We first note that to find $\Gamma^{(bs)}_{\rm tr}$ in this temperature regime, 
	we may set $\Omega_{\rm min} \sim  T$ in Eqs.\ \eqref{eq:gammasbact2} or \eqref{eq:gammasbact3}. Making use of Eq.\ \eqref{eq:actvani}, 
	the fraction of arc states that meet the condition $\Omega_{\rm min}(\varepsilon,k_z) \lesssim  T$ is thus given by $\sim \sqrt{T T_{\rm BG}^{(b)}/T^2_{\rm BG}}$, and we arrive at the estimate
	\be
	\Gamma^{(bs)}_{\rm tr} \sim \Gamma_0 \frac{T^5}{T_{\rm BG}^5} \sqrt{\frac{T_{\rm BG}^2}{T_{\rm BG}^{(b)} T}}.
	\ee
	Clearly, this rate is parametrically larger than $\Gamma^{(ss)}_{\rm eq}$ in Eq.\ \eqref{eq:Gammasseq}.
	
	\begin{figure*}
		\includegraphics[width=0.45\textwidth]{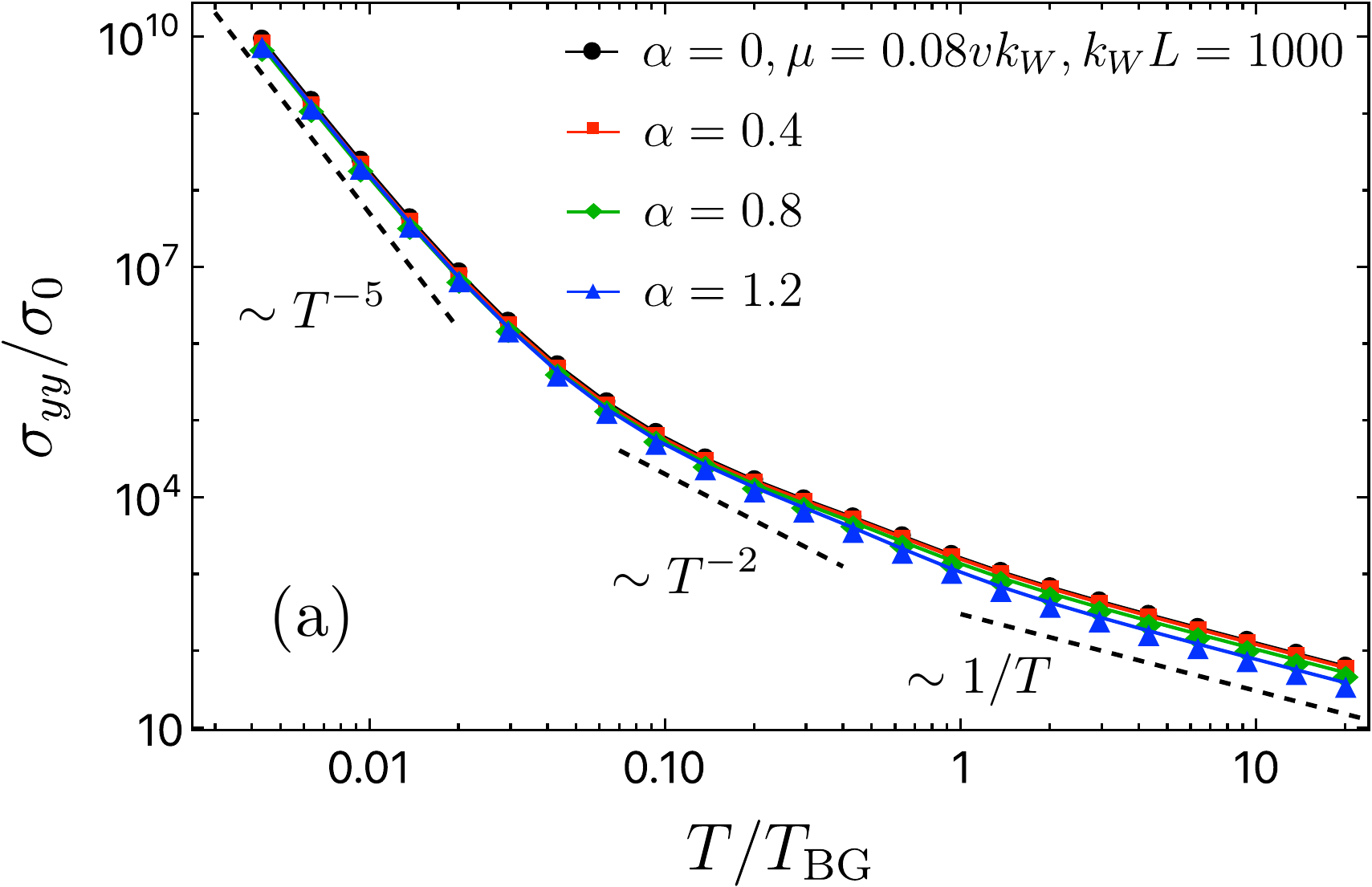}\ \
		\includegraphics[width=0.45\textwidth]{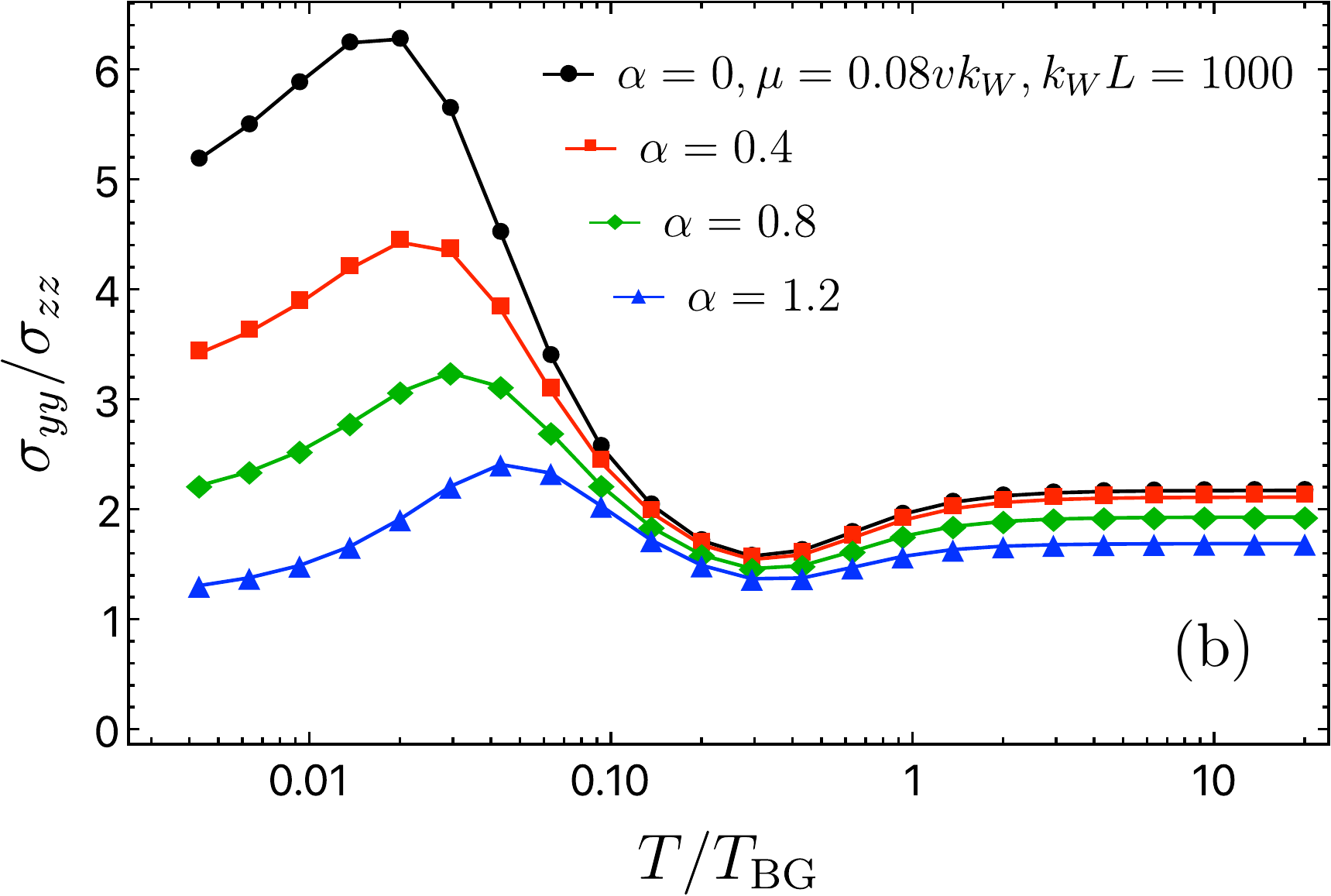}\\
		\caption{
			Temperature dependence of the conductivities $\sigma_{yy}$ and $\sigma_{zz}$ for different values of the 
			boundary angle $\alpha$.  We take $k_{\rm W}L=1000$, and $\mu=0.08vk_{\rm W}$, where $T_{\rm BG}^{(b)}/T_{\rm BG}=0.16$ and $T'_{\rm BG}/T_{\rm BG}\approx 0.29$. The data points have been obtained by numerical solution of 
			the Boltzmann equation; curves are guides to the eye only.
			Panel (a) shows $\sigma_{yy}$ in units of $\sigma_0$, see Eq.~\eqref{sigma0}, vs 
			$T/T_{\rm BG}$ on a double-logarithmic scale.  Straight dashed lines indicate the quoted power laws. 
			Panel (b) shows $\sigma_{yy}/\sigma_{zz}$ vs $T/T_{\rm BG}$ for the same parameters on a semi-logarithmic scale. }
		\label{fig5}
	\end{figure*}
	
	\subsection{Diagonal conductivity $\boldsymbol{\sigma_{zz}}$}\label{sec5b}
	
	The current in the perpendicular direction ($\hat z$) may also be written as the sum of contributions by bulk electrons and by Fermi arc states,
	$\sigma_{zz} = \sigma_{zz}^{(b)} + \sigma_{zz}^{(s)}.$
	We again estimate the respective contributions to $\sigma_{zz}$ by using the Drude formula,
	\be\label{Drudez}
	\sigma_{zz}^{(b)} \sim \frac{e^2 \mu^2 L}{v} \tau_{{\rm tr},z}^{(b)}, \quad
	\sigma_{zz}^{(s)} \sim e^2 v \kb \tau_{{\rm tr},z}^{(s)},
	\ee
	where $\tau_{{\rm tr},z}^{(b)}$ and $\tau_{{\rm tr},z}^{(s)}$ are transport relaxation times for bulk and Fermi arc electrons.
	
	The estimate for the bulk transport mean free time $\tau_{{\rm tr},z}^{(b)}$ is the same as for $\tau_{{\rm tr},y}^{(b)}$ in Eq.\ \eqref{eq:tautrb}, although the numerical coefficient may differ because of the anisotropy in the bulk dispersion if the chemical potential $\mu$ is not very close to zero. For the contribution from Fermi arcs, a key difference with the case of transport in the chiral direction 
	in Sec.~\ref{sec5a} is that for transport along $\hat z$, arc-arc scattering contributes to the transport relaxation rate because 
	Fermi arc states have no net velocity component $v_z^{(s)}$. This argument immediately leads to the estimates
	\be
	\label{eq:tautrsz}
	\frac{1}{\tau_{{\rm tr},z}^{(s)}} \sim   \Gamma_0 \times
	\left\{ \begin{array}{ll}
		\displaystyle{\frac{T}{T_{\rm BG}} \vphantom{\int_{M_M^M}^{M^M_M}}} &
		\mbox{if $T \gg T_{\rm BG}$}, \\
		\displaystyle{\frac{T^5}{T_{\rm BG}^{5}} \vphantom{\int_{M_M^M}^{M^M_M}}} &
		\mbox{if $T \ll T_{\rm BG}$}. \end{array} \right.
	\ee
	As in Sec.~\ref{sec5a}, we thus find $\sigma_{zz}\propto 1/T$ for $T\gg T_{\rm BG}$ and $\sigma_{zz}\propto 1/T^5$ for all $T\ll T_{\rm BG}^{(b)}$. At intermediate temperatures $T_{\rm BG}^{(b)}\ll T\ll T_{\rm BG}$, the surface contribution will dominate 
	for $L\ll L_z$ with
	\begin{equation}\label{Lz}
	L_z = \frac{v^3 k_{\rm W}^2}{\mu^3} \frac{T_{\rm BG}^4}{T^4},
	\end{equation}
	and hence $\sigma_{zz}\propto 1/T^5$.  For $L\gg L_z$, on the other hand, we expect the bulk-dominated high-temperature law $\sigma_{yy}\propto 1/T$.
	
	To conclude, the crossover temperature separating the low-temperature regime $\sigma_{zz}\propto 1/T^5$ from the high-temperature 
	regime 
	$\sigma_{zz}\propto 1/T$ is set by $T_{\rm BG}^{(b)}$ for thick slabs ($L\gg L_z)$, and by $T_{\rm BG}$ for thin slabs.  
	Importantly, the intermediate temperature window with $1/T^2$ scaling found for  
	$\sigma_{yy}$ is \emph{not} expected for $\sigma_{zz}$ anymore.

	\begin{figure*}
		\includegraphics[width=0.45\textwidth]{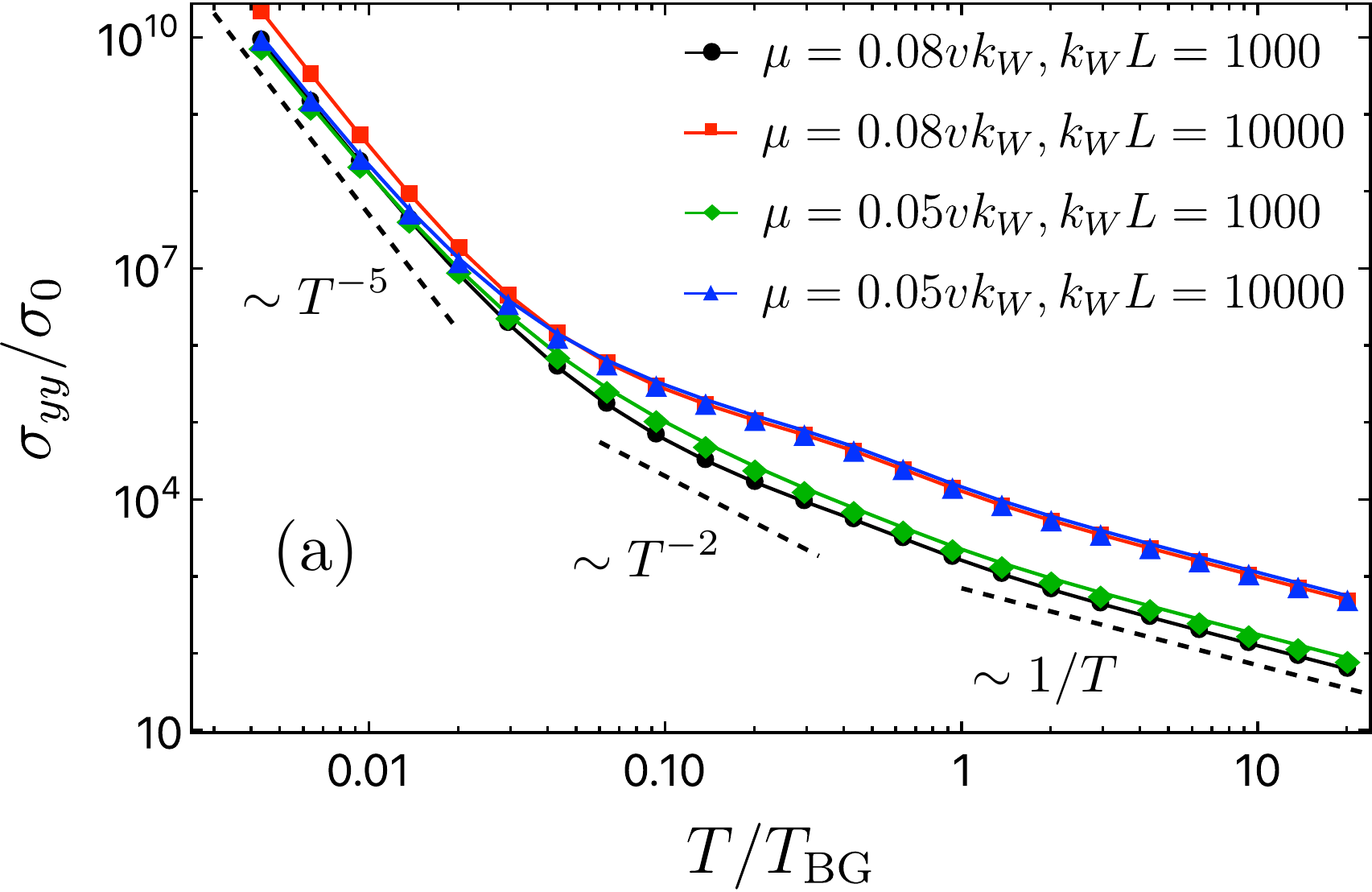}\ \
		\includegraphics[width=0.45\textwidth]{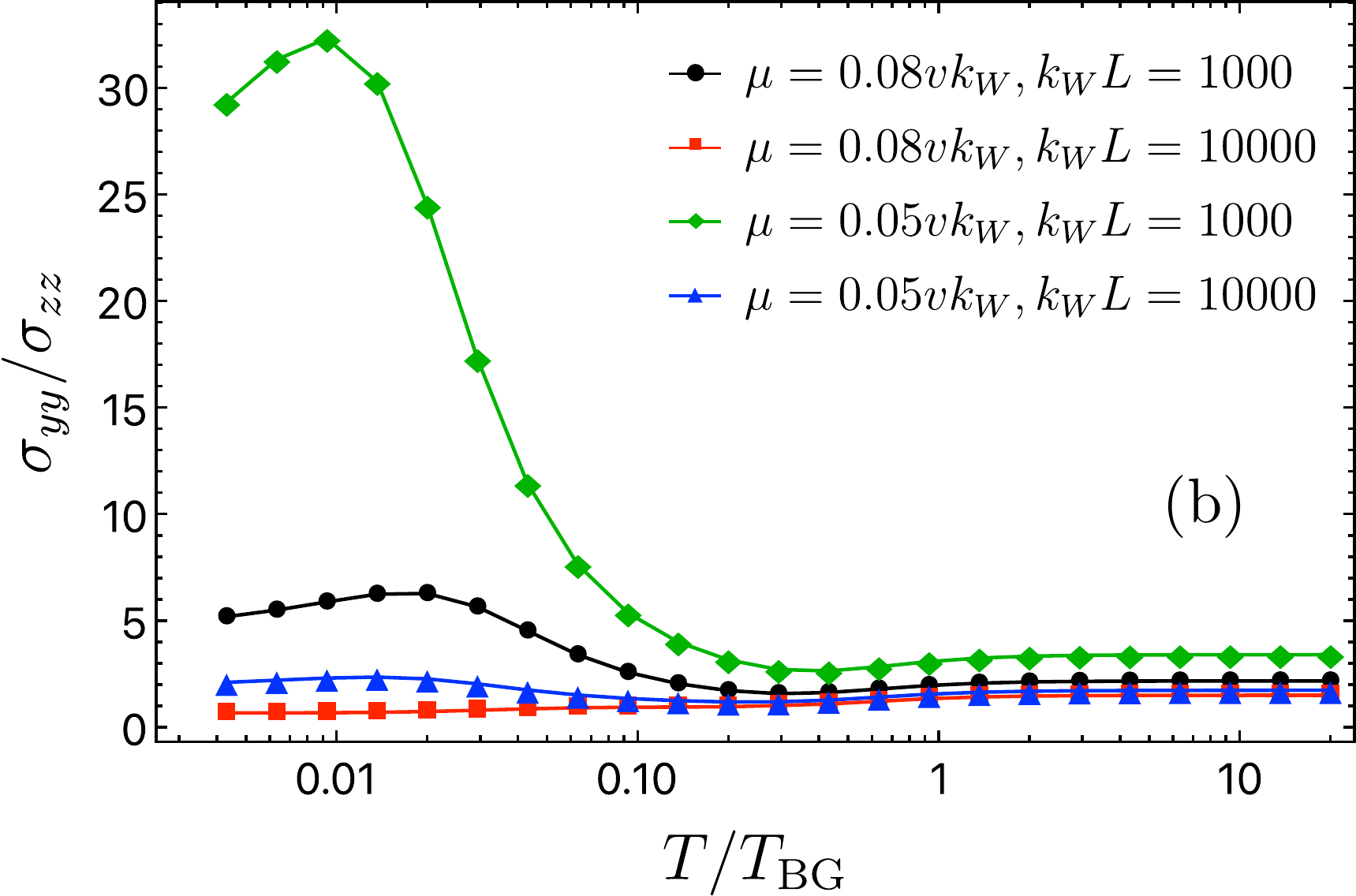}\\
		\caption{
			Temperature dependence of $\sigma_{yy}$ and $\sigma_{zz}$ as in Fig.\ \ref{fig5} but for fixed 
			boundary angle $\alpha=0$ and with $\mu/vk_{\rm W} \in (0.05, 0.08)$ and $k_{\rm W}L \in (1000, 10000)$.
			Panel (a) shows  $\sigma_{yy}/\sigma_0$ vs $T/T_{\rm BG}$ on double-logarithmic scales.   
			Panel (b) shows $\sigma_{yy}/\sigma_{zz}$ vs $T/T_{\rm BG}$ for the same parameters on a semi-logarithmic scale. }
		\label{fig6}
	\end{figure*}

	\subsection{Numerical solution}\label{sec5d}
	
	We now turn to the numerical solution of the coupled Boltzmann equations for the  bulk and arc distribution functions $\varphi^{(b)}_{\vkvector}$ and $\varphi^{(s)}_{\vkparallel}$, respectively, 
	for an electric field along $\hat y$ or $\hat z$. 
	Using the linearized Boltzmann equations in the quasi-elastic approximation, for fixed energy $\varepsilon$,
	we need to solve a set of Fredholm integral equations of the second kind, see Eqs.~\eqref{eq:BE4} and \eqref{eq:BE5}.
	For a numerical solution, we discretize those equations on a momentum-space grid of small linear step-size $\delta_k$ such that 
	integrals over momenta become Riemann sums. At given energy $\varepsilon$, the coupled Boltzmann equations are thereby written as
	non-singular matrix inversion problem which can be numerically solved by standard routines.
	The conductivities $\sigma_{yy}$ and $\sigma_{zz}$ then follow by evaluating the energy integrals in Eq.\ \eqref{eq:J_lin} 
	as Riemann sums as well. Our numerical results include the effects of all phonon modes, where  
	convergence with respect to the step-size $\delta_k$ is typically achieved for $\delta_k\alt 0.002 k_{\rm W}$.   
	Below we use the conductivity reference scale
	\begin{equation}\label{sigma0}
	\sigma_0 = \frac{e^2}{2\pi \hbar} \frac{vk_{\rm W}^2}{\Gamma_0} ,     
	\end{equation}
	where $\Gamma_0$ in Eq.\ \eqref{eq:xilambda} encodes the electron-phonon coupling. 
	While we study small $\mu$ to access the regime $T_{\rm BG}^{(b)}\ll T\ll T_{\rm BG}$, it is worth mentioning that
	the approximations behind our formalism (e.g., $T\ll \mu$) exclude the limit of extremely small $\mu$. 
	
	Figure \ref{fig5} illustrates the temperature dependence of $\sigma_{yy}$ and of the ratio $\sigma_{yy}/\sigma_{zz}$ 
	for different boundary angles $\alpha$, assuming a fixed slab width $k_{\rm W}L = 1000$ and 
	chemical potential $\mu=0.08 vk_{\rm W}$.  The results for $\sigma_{yy}$ in Fig.\ \ref{fig5}(a) 
	clearly show the high-temperature $1/T$ scaling and the low-temperature $1/T^5$ law.  From Eq.\ \eqref{Ly}, 
	we observe that the slab width taken in Fig.\ \ref{fig5} is of the same order as the crossover length $L_y$. 
	Nonetheless, our results for $\sigma_{yy}$ are consistent with $1/T^2$ scaling at intermediate temperatures.
	
	The effects of changing the boundary parameter $\alpha$ are best studied by using 
	the ratio $\sigma_{yy}/\sigma_{zz}$, see Fig.\ \ref{fig5}(b). 
	In a bulk-dominated case with chemical potential near the nodal points, this ratio is expected to
	be close to unity.  We observe from Fig.\ \ref{fig5}(b) that for these parameters, in particular for small $\alpha$, 
	the low-temperature ratio is significantly larger than unity.  For instance,
	the conductivity along the chiral direction is larger by a factor $\approx 5$ for $\alpha=0$ and low temperatures.
	(We recall that $\alpha=0$ corresponds to straight Fermi arc curves connecting the Weyl node projections
	in reciprocal space.)  Chirality effects are then most pronounced for $\alpha=0$, and 
	upon increasing $\alpha$, the conductivity ratio decreases and approaches values near unity, i.e.,
	the chirality-induced anisotropy of the conductivities is washed out.
	We emphasize that a strong dependence on the boundary parameter $\alpha$ represents direct evidence for 
	surface-dominated transport in the slab geometry.
	
	Both at very low, $T\ll T_{\rm BG}^{(b)}$, and at high, $T\gg T_{\rm BG}$, temperatures, the conductivities $\sigma_{yy}$ and
	$\sigma_{zz}$ share the same power-law scaling with temperature. Up to subleading terms, their ratio therefore becomes independent of 
	$T$.  We observe from Fig.\ \ref{fig5}(b) that the intermediate temperature regime is more interesting.  Noting that
	for the  parameters in Fig.\ \ref{fig5}, we have $L\approx L_y$ but $L\ll L_z$, see Eqs.\ \eqref{Ly} and \eqref{Lz},
	the rapid increase of $\sigma_{yy}/\sigma_{zz}$ upon lowering temperature within this regime is in accordance with the qualitative considerations in Secs.\ \ref{sec5a} and \ref{sec5b}. 
	
	Figure \ref{fig6} shows results at fixed boundary angle $\alpha=0$. We here investigate 
	what happens when the slab width $L$ and/or the chemical potential $\mu$ are changed.  
	We first note that the results for $\sigma_{yy}$ in Fig.\ \ref{fig6}(a) show a qualitative
	difference for $k_{\rm W}L=1000$ and $k_{\rm W}L=10000$.  In the latter case, we have $L\gg L_y$ for all temperatures $T_{\rm BG}^{(b)}\ll T\ll T_{\rm BG}$, and there is little room for an intermediate $1/T^2$ scaling law anymore.  Indeed, our numerical results point towards a direct crossover from
	the $1/T^5$ to the $1/T$ scaling around $T\sim T_{\rm BG}^{(b)}$.  The ratio $\sigma_{yy}/\sigma_{zz}$ in Fig.\ \ref{fig6}(b) 
	now reaches values $\approx 30$ for $\mu=0.05 vk_{\rm W}$ and $\alpha=0$ with the same slab width as in Fig.\ \ref{fig5}. 
	By decreasing the chemical potential, the relative importance of the Fermi arc contribution thus has increased.  
	Similarly, for decreasing slab width, one gets a larger ratio $\sigma_{yy}/\sigma_{zz}$. 
	We note that for $k_{\rm W}L=10000$ and $\mu=0.08 vk_{\rm W}$, this ratio is already close to unity 
	for all studied temperatures. Transport is then essentially isotropic as expected in the bulk-dominated case
	with small $\mu$.
	
	Our numerical results show that chirality effects associated with Fermi-arc surface states are 
	most pronounced for small $\alpha$ (corresponding to straight Fermi arcs), for small to intermediate slab width (i.e., $L\ll L_y$ in Eq.\ \eqref{Ly}), and for small values of the chemical potential (where the
	bulk DoS is very small and bulk contributions to the conductivity are suppressed).  
	We conclude that the numerical solution of the Boltzmann equation is consistent with the qualitative considerations in Secs.\ \ref{sec5a} and \ref{sec5b}.   These results also suggest that a convenient way for extracting information about 
	chirality consists of measuring the ratio of conductivities along perpendicular directions (such as $\hat y$ and $\hat z$). 
	This ratio will be maximized if the axis with the larger conductivity is parallel to the chiral direction.
	
	\section{Conclusions and outlook}\label{sec6}
	
	In this paper, we have formulated and studied a low-energy theory of WSMs 
	coupled to acoustic phonons, focusing on the quasiparticle decay rate of Fermi-arc surface 
	states and on transport in a slab geometry.
	While in general the shape of the Fermi arcs in the surface Brillouin  zone is non-universal,
	the phenomenological boundary angle $\alpha$ has allowed us to include such features and their impact
	on transport observables and on the quasiparticle decay rate.  
	We predict transport signatures of chirality from the dependence of the conductivity tensor 
	on key parameters, e.g., on temperature, slab width, or chemical potential.
	Similarly, our predictions for the quasiparticle decay rate may be tested by ARPES or STM experiments.
	
	Experimental reality is inevitably more complex than our theory in several regards.  Let us
	here point towards just a few of these complications, which also provide interesting perspectives for future work.
	(i) In known WSM materials, the band structure is more complicated than assumed in our work, usually featuring several pairs of Weyl nodes and several 
	arcs at a given surface.   Our analysis could be extended to account, e.g., for multiple arcs and the resulting scattering of surface electrons
	between different arc states. (ii) Disorder is typically present in available samples.  We have assumed that phonon scattering 
	dominates over disorder effects. 
	The interplay of disorder and electron-phonon interaction effects, as well as the extension of our theory to diffusive transport along 
	the transverse direction and/or the inclusion of external magnetic fields remain to be addressed.
	(iii) The relevant acoustic phonon modes and their coupling to electrons may differ from our model. We leave studies of more general phonon models
	and other types of electron-phonon coupling to future work.
	(iv) It will be interesting to also study the effects of Coulomb interactions on the temperature-dependent decay rates 
	and resistivities.  We expect that different power laws will govern these quantities in comparison to those reported here for electron-phonon interactions.

	Despite of the above potential complications, on a qualitative level our key predictions are expected to be robust, especially 
	when  they are tied to the topological nature of the Fermi surface. 
	In particular, the various power-law or activated temperature dependences of arc-arc and arc-bulk scattering rates in the respective 
	parameter regimes (as reported in Sec.~\ref{sec4}) only rely on the linearity of the acoustic phonon dispersion at long wavelengths and 
	on the existence of an arc-bulk activation gap.
	The latter, with its strong dependence on the geometric arc shape, is a direct consequence of energy-momentum conservation. 
	Similarly, the chirality-induced anisotropy of the conductivity tensor in thin slabs (at low temperatures and chemical potential near the nodal points)
	and the  intermediate $1/T^2$ scaling regime for transport along the chiral 
	direction should allow for transport signatures of chirality. 
	We are confident that experimental progress will soon lead to the observation of such phenomena in Weyl semimetals.
	
	\begin{acknowledgments}
		We wish to thank M. Breitkreiz and M. Burrello for discussions.
		We acknowledge funding  by the Deutsche Forschungsgemeinschaft (DFG, German Research Foundation) under
		Projektnummer 277101999 - TRR 183 (project A02) and Grant No.~EG 96/12-1, 
		as well as within Germany's Excellence Strategy-Cluster of Excellence ``Matter and Light for
		Quantum Computing'' (ML4Q), EXC 2004/1-390534769.
		We also acknowledge funding by the Brazilian ministries MEC and MCTI and by CNPq. \\
	\end{acknowledgments}
	
	\appendix
	
	\section{Phonon modes} \label{appA}
	
	In this Appendix, we provide details about the phonon modes discussed in Sec.~\ref{sec2b}.
	To find an expression for the phonon reflection amplitudes $s^\pm$, we require that the displacement 
	field \eqref{eq:ulambda} satisfies the boundary condition \eqref{eq:phononbc}. This gives the equations
	\begin{widetext}
		\bea
		(1 + s^{(l,l)\pm}_{q_x,\vqparallel}) \frac{c_l^2 (q_{\parallel}^2 + q_x^{(l,l)2}) - 2 c_t^2 q_{\parallel}^2}{2 c_t^2 \sqrt{q_{\parallel}^2 + q_x^{(l,l)2}}} &=&
		\pm i s^{(t,l)\pm}_{q_x,\vqparallel}  \frac{q_{x}^{(t,l)} q_{\parallel}}{\sqrt{q_{\parallel}^2 + q_x^{(t,l)2}}}, \nonumber \\
		-i  s^{(l,t)\pm}_{q_x,\vqparallel}) \frac{c_l^2 (q_{\parallel}^2 + q_x^{(l,t)2}) - 2 c_t^2 q_{\parallel}^2}{2 c_t^2 \sqrt{q_{\parallel}^2 + q_x^{(l,t)2}}} &=&
		\pm  (s^{(t,t)\pm}_{q_x,\vqparallel} - 1) \frac{q_{x}^{(t,t)} 
			q_{\parallel}}{\sqrt{q_{\parallel}^2 + q_x^{(t,t)2}}}, \nonumber \\
		(1+ s^{(t,t)\pm}_{q_x,\vqparallel}) \frac{q_{\parallel}^2 - q_x^{(t,t)2}}{2 \sqrt{q_{\parallel}^2 + q_x^{(t,t)2}}} &=&
		\mp i s^{(l,t)\pm}_{q_x,\vqparallel}
		\frac{q_x^{(l,t)} q_{\parallel}}{\sqrt{q_{\parallel}^2 + q_x^{(l,t)2}}}, 
		\nonumber\\
		i s^{(t,l)\pm}_{q_x,\vqparallel} \frac{q_{\parallel}^2 - q_x^{(t,l)2}}{2 \sqrt{q_{\parallel}^2 + q_x^{(t,l)2}}} &=&
		\pm (s^{(l,l)\pm}_{q_x,\vqparallel}-1) 
		\frac{q_x^{(l,l)} q_{\parallel}}{\sqrt{q_{\parallel}^2 + q_x^{(l,l)2}}}.
		\label{eq:bcphonon}
		\eea
		For the bulk modes, the coefficients $s^{(\lambda',\lambda)\pm}_{q_x,\vqparallel}$ are dimensionless numbers that depend on the ratio $q_x/q_{\parallel}$ and the mode indices $\lambda'$ and $\lambda$ only.  We find
		\bea
		s^{(ll)\pm}_{q_x,\vqparallel} &=&
		-\frac{\Omega^4 - 4 c_t^2 \Omega^2 q_{\parallel}^2 +
			4 c_t^4 q_{\parallel}^2 (q_{\parallel}^2 - q_x q_x^{(t,l)})}
		{\Omega^4 - 4 c_t^2 \Omega^2 q_{\parallel}^2 +
			4 c_t^4 q_{\parallel}^2 (q_{\parallel}^2 + q_x q_x^{(t,l)})}, \nonumber \\ 
		s^{(tl)\pm}_{q_x,\vqparallel} &=&
		\mp i \frac{4 c_l c_t q_{\parallel} q_x (\Omega^2 - 2 c_t^2 q_{\parallel}^2)}
		{\Omega^4 - 4 c_t^2 \Omega^2 q_{\parallel}^2 +
			4 c_t^4 q_{\parallel}^2 (q_{\parallel}^2 + q_x q_x^{(t,l)})}, \nonumber \\
		s^{(lt)\pm}_{q_x,\vqparallel} &=&
		\mp i \frac{4 c_t^3 q_{\parallel} q_x (\Omega^2 - 2 c_t^2 q_{\parallel}^2)/c_l}
		{\Omega^4 - 4 c_t^2 \Omega^2 q_{\parallel}^2 +
			4 c_t^4 q_{\parallel}^2 (q_{\parallel}^2 + q_x q_x^{(l,t)})}, \nonumber \\
		s^{(tt)\pm}_{q_x,\vqparallel} &=&
		- \frac{\Omega^4 - 4 c_t^2 \Omega^2 q_{\parallel}^2 +
			4 c_t^4 q_{\parallel}^2 (q_{\parallel}^2 - q_x q_x^{(l,t)})}
		{\Omega^4 - 4 c_t^2 \Omega^2 q_{\parallel}^2 +
			4 c_t^4 q_{\parallel}^2 (q_{\parallel}^2 + q_x q_x^{(l,t)})},
		\eea
	\end{widetext}
	where we recall that the phonon frequency is given by
	$\Omega = c_{\lambda} \sqrt{q_x^2 + q_{\parallel}^2}$.  For $\Omega > q_{\parallel} c_{\lambda'}$, we have
	\be
	c_{\lambda'} q_x^{(\lambda',\lambda)} = \sqrt{\Omega^2 - c_{\lambda'}^2 q_{\parallel}^2}.
	\label{eq:qxre}
	\ee
	Similarly, for $\Omega < q_{\parallel} c_{\lambda'}$,
	\be
	c_{\lambda'} q_x^{(\lambda',\lambda)} = i \sqrt{c_{\lambda'}^2 q_{\parallel}^2 - \Omega^2}.
	\label{eq:qxim}
	\ee
	The coefficient $s^{(ll)\pm}_{q_x,\vqparallel}$ is real while 
	$s^{(tl)\pm}_{q_x,\vqparallel}$ is imaginary. They satisfy the 
	flux conservation condition
	\be
	c_l^2 q_{x} |s^{(ll)\pm}_{q_x,\vqparallel}|^2 + c_t^2 q_{x}^{(t,l)} 
	| s^{(tl)\pm }_{q_x,\vqparallel}|^2 = c_l^2 q_{x}.
	\ee
	The coefficients $s^{(tt)\pm}_{q_x,\vqparallel}$ and $s^{(lt)\pm}_{q_x,\vqparallel}$ are respectively real 
	and imaginary for $\Omega > c_l q_{\parallel}$ (because $q_x^{(l,t)}$ is real for these frequencies). 
	In that case, they satisfy the flux conservation condition
	\be
	c_t^2 q_{x} |s^{(tt)\pm }_{q_x,\vqparallel}|^2 + 
	c_l^2 q_{x}^{(l,t)} |s^{(lt)\pm }_{q_x,\vqparallel} |^2= c_t^2 q_{x}. \nonumber
	\ee
	For $c_t q_{\parallel} < \Omega < c_l q_{\parallel}$, the quantity $q_x^{(l,t)}$ is imaginary, so that the coefficients $s^{(tt)\pm}_{q_x,\vqparallel}$ and $s^{(lt)\pm}_{q_x,\vqparallel}$ are complex and satisfy the unitarity condition
	\be
	|s^{(tt)\pm}_{q_x,\vqparallel}|^2 = 1.
	\ee
	
	The same set of equations, but without the inhomogeneous terms
	determine the frequency $\Omega^{(R)}_{\vqparallel}$ and the coefficients $s^{(l,R)\pm}_{\vqparallel}$ and $s^{(t,R)\pm}_{\vqparallel}$ of the Rayleigh surface modes.
	In this case, one has $\Omega<c_t q_\parallel$ and the transverse wavenumbers $q_x^{(l,R)}$ and $q_x^{(t,R)}$ are imaginary,
	\bea
	\label{eq:Req}
	c_{\lambda} q_x^{(\lambda,R)} &=& i \sqrt{c_{\lambda}^2 q_{\parallel}^2 - \Omega^{(R)2}_{\vqparallel}}. \nonumber
	\eea
	The frequency $\Omega^{(R)}_{\vqparallel}$ and the coefficients $s^{(l,R)\pm}_{\vqparallel}$ and $s^{(t,R)\pm}_{\vqparallel}$ satisfy the equation
	\be\label{auxR}
	\Omega^{(R)4}_{\vqparallel} - 4 c_t^2 \Omega^{(R) 2}_{\vqparallel} q_{\parallel}^2 +
	4 c_t^4 q_{\parallel}^2 (q_{\parallel}^2 + q_x^{(l,R)} q_x^{(t,R)}) = 0, 
	\ee
	from which the velocity $c_R$ of the Rayleigh modes can be determined,
	\be
	\Omega_{\vqparallel}^{(R)} = c_R q_{\parallel},
	\ee
	with $c_R < c_t < c_l$. The ratio $c_R/c_l$ depends on the quotient $c_l/c_t$ of longitudinal and transverse sound velocities only. To find the coefficients $s^{(l,R)\pm}_{\vqparallel}$ and $s^{(t,R)\pm}_{\vqparallel}$, we write 
	\bea
	\label{eq:xil1}
	s_{\vqparallel}^{(l,R)\pm} = \xi^{(l)} \sqrt{q_{\parallel}},\ \
	s_{\vqparallel}^{(t,R)\pm} = \pm \xi^{(t)} \sqrt{q_{\parallel}},
	\eea 
	with $\xi^{(l)}$ real and positive and $\xi^{(t)}$ real. (The factors $\sqrt{q_{\parallel}}$ are necessary for normalization.)
	To determine the dimensionless numbers $\xi^{(l)}$ and $\xi^{(t)}$, we note that their ratio follows from Eq.\ \eqref{eq:bcphonon} (without the inhomogeneous terms),
	\bea
	\frac{\xi^{(l)}}{\xi^{(t)}} &=&  \frac{2 c_t^2 \sqrt{c_t^2 - c_R^2}}
	{c_l (2 c_t^2 - c_R^2)} \nonumber \\ &=&
	\frac{2 c_t^2 - c_R^2}{2 c_t \sqrt{c_l^2 - c_R^2}},
	\eea
	where the second equality makes use of  Eq.\ \eqref{auxR}.
	Their magnitude can then be obtained from the normalization condition of the phonon mode, which gives
	\bea
	\label{eq:xil3}
	c_R^2 &=&
	\xi^{(l)2} \frac{c_l(2 c_l^2 - c_R^2)}{2 \sqrt{c_l^2 - c_R^2}} +
	\xi^{(t)2} \frac{c_t(2 c_t^2 - c_R^2)}{2 \sqrt{c_t^2 - c_R^2}} 
	\nonumber \\ && \mbox{}
	- 2 \xi^{(l)} \xi^{(t)} c_l c_t.
	\eea
	Although algebraic expressions for $c_R$, $\xi^{(l)}$ and $\xi^{(t)}$ can be obtained, these expression are not useful for further analytical calculations, and we will work with the dimensionless coefficients $c_R/c_l$, $\xi^{(l)}$, and $\xi^{(t)}$ instead.
	For example, for $c_t/c_l=0.571$, we obtain the numerical values $c_R/c_l\simeq 0.526, \xi^{(l)}\simeq 0.464,$ and $\xi^{(t)}/\xi^{(l)}\simeq 2.583$.
	
	\section{Electron-phonon matrix elements for arc-arc and arc-bulk scattering}\label{appB}
	
	Using that the coefficients $s^{(\lambda',\lambda)\pm}_{q_x,q_{\parallel}}$ depend on $q_x$ and on the magnitude $q_{\parallel} = |\vqparallel|$ only,
	we may write the electron-phonon Hamiltonian near the interface at $x = \pm L/2$ as
	\begin{widetext}
		\bea
		H_{\rm ep} &=& i \frac{g_0}{c_l} \sum_{\lambda}
		\int_0^{\infty} dq_x \int d\vqparallel \sqrt{\frac{\Omega_{\vqvector}^{(\lambda)}}{2 \rho_M}}
		e^{i \vqparallel \cdot \vrparallel} ( e^{\pm i q_x x} \delta_{\lambda,l} + s^{(l,\lambda)\pm}_{q_x,q_{\parallel}} e^{\mp i q_x^{(l,\lambda)} x} )
		(a^{(\lambda)\pm,{\rm in}}_{q_x,\vqparallel} - a^{(\lambda)\pm,{\rm out}\dagger}_{q_x,-\vqparallel}) \nonumber \\ && \mbox{} +
		i \frac{g_0}{c_l} \int d\vqparallel \sqrt{\frac{\Omega_{\vqparallel}^{(R)}}{2 \rho_M}}
		e^{i \vqparallel \cdot \vrparallel} s^{(l,R)\pm}_{q_{\parallel}} e^{\mp i q_x^{(l,R)} x}
		(a^{(R)\pm}_{\vqparallel} - a^{(R)\pm\dagger}_{-\vqparallel}).
		\eea
		Taking the electron wavefunction from Eq.\ \eqref{eq:boundarystte}, we find that the arc-arc electron-phonon matrix elements of Eq.~\eqref{ephmatrixelements} are 
		\bea
		\label{eq:GssB}
		{\cal G}^{(ss\lambda)\pm}_{\vkpparallel,\vkparallel,q_x} &=& 
		i \frac{ g_0}{c_l} 
		\sqrt{\frac{2\Omega^{(\lambda)}_{\vqvector} \kappa_{\pm}(\vkparallel) \kappa_{\pm}(\vkpparallel)}{\rho_M}}
		\left[
		\frac{\delta_{\lambda,l}}{\kappa_{\pm}(\vkpparallel) + \kappa_{\pm}(\vkparallel)+i q_x^{(l,\lambda)}}
		+ 
		\frac{s_{q_x,q_{\parallel}}^{(l,\lambda)\pm}}{\kappa_{\pm}(\vkpparallel) + \kappa_{\pm}(\vkparallel)-i q_x^{(l,\lambda)}}   \right], 
		\\
		{\cal G}^{(ssR)\pm}_{\vkpparallel,\vkparallel} &=& 
		i \frac{ g_0}{c_l} 
		\sqrt{\frac{2\Omega^{(R)}_{\vqparallel} \kappa_{\pm}(\vkparallel) \kappa_{\pm}(\vkpparallel)}{\rho_M}}
		\frac{s_{q_{\parallel}}^{(l,R)\pm}}{\kappa_{\pm}(\vkpparallel) + \kappa_{\pm}(\vkparallel)-i q_x^{(l,R)}},
		\eea
		where $\vqparallel = \vkpparallel - \vkparallel$. Similarly, using Eq.\ \eqref{eq:PhiInterface} for the electronic bulk states, with the reflection amplitudes $r_{k_x,\vkparallel}^{\pm}$ in Eq.\ \eqref{reflmat}, and defining
		\be
		r'^{\pm}_{k_x,\vkparallel} = 
		\frac{\langle \xi_{\mp}(-\alpha)|\xi_{(\pm k_x,\vkparallel)}\rangle}
		{\langle \xi_{\mp}(-\alpha)|\xi_{(\mp k_x,\vkparallel)}\rangle},
		\ee
		we find that the arc-bulk matrix elements are
		\bea
		\label{eq:Gsblambda}
		{\cal G}^{(sb\lambda)\pm}_{\vkpparallel,\vkvector,q_x} &=&
		i \frac{g_0}{c_l} \langle \xi_{\mp}(-\alpha)|\xi_{(\mp k_x,\vkparallel)}\rangle
		\sqrt{\frac{ \Omega^{(\lambda)}_{\vqvector} \kappa_{\pm}(\vkpparallel)}{\rho_M}}
		\nonumber \\ && \mbox{} \times
		\left[ \delta_{\lambda,l} \left(
		\frac{ r'^{\pm}_{k_x,\vkparallel} 
		}{\kappa_{\pm}(\vkpparallel) + i k_x + i q_x^{(l,\lambda)}}
		- \frac{
			r_{k_x,\vkparallel}^{\pm}}
		{\kappa_{\pm}(\vkpparallel) - i k_x + i q_x^{(l,\lambda)}} \right)
		\right. \nonumber \\ && \ \ \ \ \left. \mbox{}
		+ s_{q_x,\vqparallel}^{(l,\lambda)\pm} 
		\left( \frac{ r'^{\pm}_{k_x,\vkparallel}
		}{\kappa_{\pm}(\vkpparallel) + i k_x - i q_x^{(l,\lambda)}}
		- \frac{r_{k_x,\vkparallel}^{\pm}}
		{\kappa_{\pm}(\vkpparallel) - i k_x - i q_x^{(l,\lambda)}} \right)
		\right], \\ \label{eq:GsbR}
		{\cal G}^{(sbR)\pm}_{\vkpparallel,\vkvector} &=& i \frac{g_0}{c_l}
		\langle \xi_{\mp}(-\alpha)|\xi_{(\mp k_x,\vkparallel)}\rangle
		\sqrt{\frac{\Omega^{(R)}_{\vqparallel} \kappa_{\pm}(\vkpparallel)}{\rho_M}}
		\nonumber \\ && \mbox{} \times
		s_{q_{\parallel}}^{(l,R)\pm}
		\left(
		\frac{ r'^{\pm}_{k_x,\vkparallel}
		}{\kappa_{\pm}(\vkpparallel) + i k_x - i q_x^{(l,R)}}
		- \frac{r_{k_x,\vkparallel}^{\pm}}
		{\kappa_{\pm}(\vkpparallel) - i k_x - i q_x^{(l,R)}} \right),
		\eea
	\end{widetext}
	and 
	\bea\nonumber
	{\cal G}^{(bs\lambda)\pm}_{\vkpvector,\vkparallel,q_x} &=& - r^{\pm*}_{k_x',\vkpparallel}{\cal G}^{(sb\lambda)\pm}_{\vkparallel,\vkpvector,q_x}, \\
	{\cal G}^{(bsR)\pm}_{\vkpvector,\vkparallel} &=& - r^{\pm*}_{k_x',\vkpparallel}{\cal G}^{(sbR)\pm}_{\vkparallel,\vkpvector}.
	\label{eq:GbsR}
	\eea  
	The latter two expressions can be verified using the identities $\langle \xi_{\mp}(-\alpha)|\xi_{(k_x,\vkparallel)}\rangle = \langle \xi_{(-k_x,\vkparallel)} | \xi_{\mp}(-\alpha) \rangle$ and $|r^{\pm}_{k_x,\vkparallel}|^2 = 1$. 
	One further has
	\begin{widetext}
		\be
		|\langle \xi_{\mp}(-\alpha)|\xi_{(\mp k_x,\vkparallel)}\rangle|^2 =
		\frac{1}{2}  -   \frac{[(\varepsilon + m(k_z))^2 - v^2 (k_x^2 + k_y^2)] \sin \alpha \mp 2 (\varepsilon + m(k_z)) v k_y \cos \alpha}{2[(\varepsilon + m(k_z))^2 + v^2 (k_x^2 + k_y^2)]}.
		\ee
		
		We next provide matrix elements of the electron-phonon interaction.  In analogy to the arc-bulk matrix element \eqref{ephmatrixelements}, the bulk-arc and arc-arc matrix elements are given by
		\bea
		\langle \Phi_{\vkpparallel}^{(s)\pm} | H_{\rm ep} | \Phi^{(b)\pm}_{k_x,\vkparallel} \rangle
		&=&
		\sum_{\lambda} \int_{0}^{\infty} dq_x
		{\cal G}^{(sb\lambda)\pm}_{\vkpparallel,\vklabel,q_x}
		\left(   
		a^{(\lambda)\pm,{\rm in}}_{q_x,\vqparallel}
		-   
		a^{(\lambda)\pm,{\rm out} \dagger}_{q_x,-\vqparallel} 
		\right)   
		+ {\cal G}^{(sbR)\pm}_{\vkpparallel,\vklabel}
		\left(   
		a^{(R)\pm}_{\vqparallel}
		-
		a^{(R)\pm \dagger}_{-\vqparallel}
		\right)
		, \nonumber \\ 
		\langle \Phi_{\vkpparallel}^{(s)\pm} | H_{\rm ep} | \Phi_{\vkparallel}^{(s)\pm} \rangle
		&=&
		\sum_{\lambda} \int_{0}^{\infty} dq_x
		{\cal G}^{(ss\lambda)\pm}_{\vkpparallel,\vkparallel,q_x}
		\left(
		a^{(\lambda)\pm,{\rm in}}_{q_x,\vqparallel}
		-
		a^{(\lambda)\pm,{\rm out} \dagger}_{q_x,-\vqparallel} 
		\right)  
		+ {\cal G}^{(ssR)\pm}_{\vkpparallel,\vkparallel}
		\left(
		a^{(R)\pm}_{\vqparallel} 
		-
		a^{(R)\pm \dagger}_{-\vqparallel}
		\right),
		\eea
		with $\vqparallel = \vkpparallel-\vkparallel$.
		The corresponding transition rates follow as
		\bea
		\label{eq:Wsb}
		W_{\vkpparallel,\vklabel}^{(sb)\pm} &=& 
		2 \pi \sum_{\lambda} \int_0^{\infty} \frac{dq_x}{2 \pi}
		|{\cal G}^{(sb\lambda)\pm}_{\vkpparallel,\vklabel,q_x}|^2
		\left\{
		n_{\rm B}(\Omega^{(\lambda)}_{\vqvector})
		\delta(\varepsilon^{(s)\pm}_{\vkpparallel}-\varepsilon^{(b)}_{\vklabel} - \Omega^{(\lambda)}_{\vqvector}) 
		+
		[n_{\rm B}(\Omega^{(\lambda)}_{\vqvector})+1]
		\delta(\varepsilon^{(s)\pm}_{\vkpparallel}-\varepsilon^{(b)}_{\vklabel} + \Omega^{(\lambda)}_{\vqvector}) \vphantom{|{\cal G}^{(ss\lambda)\pm}_{\vkpparallel,\vkparallel,q_x}|^2} \right\}
		\nonumber \\ && \mbox{}
		+
		2 \pi |{\cal G}^{(sbR)\pm}_{\vkpparallel,\vklabel}|^2
		\left\{  
		n_{\rm B}(\Omega^{(R)}_{\vqparallel})
		\delta(\varepsilon^{(s)\pm}_{\vkpparallel}-\varepsilon^{(b)}_{\vklabel} - \Omega^{(R)}_{\vqparallel})
		+
		[n_{\rm B}(\Omega^{(R)}_{\vqparallel})+1]
		\delta(\varepsilon^{(s)\pm}_{\vkpparallel}-\varepsilon^{(b)}_{\vklabel} + \Omega^{(R)}_{\vqparallel})
		\right\}  
		,
		\nonumber \\
		W_{\vkpparallel,\vkparallel}^{(ss)\pm} &=&
		2 \pi \sum_{\lambda} \int_0^{\infty} \frac{dq_x}{2 \pi}
		|{\cal G}^{(ss\lambda)\pm}_{\vkpparallel,\vkparallel,q_x}|^2
		\left\{  
		n_{\rm B}(\Omega^{(\lambda)}_{\vqvector})
		\delta(\varepsilon^{(s)\pm}_{\vkpparallel}-\varepsilon^{(s)\pm}_{\vkparallel} - \Omega^{(\lambda)}_{\vqvector}) 
		+
		[n_{\rm B}(\Omega^{(\lambda)}_{\vqvector})+1]
		\delta(\varepsilon^{(s)\pm}_{\vkpparallel}-\varepsilon^{(s)\pm}_{\vkparallel} + \Omega^{(\lambda)}_{\vqvector}) 
		\vphantom{|{\cal G}^{\pm}_{\vkparallel,\vkpparallel,q_x\lambda}|^2} \right\}
		\nonumber \\ && \mbox{}
		+
		2 \pi |{\cal G}^{(ssR)\pm}_{\vkpparallel,\vkparallel}|^2
		\left\{ 
		n_{\rm B}(\Omega^{(R)}_{\vqparallel})
		\delta(\varepsilon^{(s)\pm}_{\vkpparallel}-\varepsilon^{(s)\pm}_{\vkparallel} - \Omega^{(R)}_{\vqparallel})
		+
		[n_{\rm B}(\Omega^{(R)}_{\vqparallel})+1]
		\delta(\varepsilon^{(s)\pm}_{\vkpparallel}-\varepsilon^{(s)\pm}_{\vkparallel} + \Omega^{(R)}_{\vqparallel})
		\right\}  
		, 
		\eea
		where $\vqvector = (q_x,\vqparallel)$.

		\section{Collision integrals} \label{appC}
		
		In analogy to Eq.\ \eqref{collint}, the 
		collision integrals for bulk-arc and arc-arc scattering are defined as
		\bea\nonumber
		{\cal I}^{(sb)\pm}_{\vklabel} &=&
		\int \frac{d\vkpparallel}{(2 \pi)^2}
		\left\{ W^{(bs)\pm}_{\vklabel,\vkpparallel}
		f_{\vkpparallel}^{(s)\pm} (1-f^{(b)}_{\vklabel}) -
		W^{(sb)\pm}_{\vkpparallel,\vklabel}
		f^{(b)}_{\vklabel} (1-f_{\vkpparallel}^{(s)\pm}) \right\}, \\
		{\cal I}_{\vkparallel}^{(ss)\pm} &=&
		\int \frac{d\vkpparallel}{(2 \pi)^2} \left\{
		W_{\vkparallel,\vkpparallel}^{(ss)\pm} f_{\vkpparallel}^{(s)\pm} (1 - f_{\vkparallel}^{(s)\pm})
		- W_{\vkpparallel,\vkparallel}^{(ss)\pm} f_{\vkparallel}^{(s)\pm} (1 - f_{\vkpparallel}^{(s)\pm})
		\right\}.
		\eea
		For the linearized Boltzmann theory, these collision integrals are in analogy to Eq.\ \eqref{eq:calJ} given by
		\bea
		{\cal J}^{(sb)\pm}_{\vklabel} &=&
		\int \frac{d\vkpparallel}{(2 \pi)^2}
		{\cal W}_{\vkpparallel,\vklabel}^{(sb)\pm}
		(\varphi_{\vkpparallel}^{(s)\pm} - \varphi^{(b)}_{\vklabel}), \nonumber \\
		{\cal J}^{(ss)\pm}_{\vkparallel} &=&
		\int \frac{d\vkpparallel}{(2 \pi)^2}
		{\cal W}_{\vkpparallel,\vkparallel}^{(ss)\pm}
		(\varphi_{\vkpparallel}^{(s)\pm} - \varphi_{\vkparallel}^{(s)\pm}).  
		\eea
		The bulk-arc and the arc-arc kernels appearing in these expressions read
		\bea
		{\cal W}^{(sb)\pm}_{\vkpparallel,\vklabel} &=&
		2 \pi
		\sum_{\lambda=l,t} \int_0^{\infty} \frac{dq_x}{2 \pi}
		|{\cal G}^{(sb\lambda)\pm}_{\vkpparallel,\vklabel,q_x}|^2
		\left\{
		[n_{\rm B}(\Omega^{(\lambda)}_{\vqvector}) + n_{\rm F}(\varepsilon^{(b)}_{\vklabel} + \Omega^{(\lambda)}_{\vqvector})] 
		\delta(\varepsilon_{\vkpparallel}^{(s)\pm} - \varepsilon^{(b)}_{\vklabel} - \Omega^{(\lambda)}_{\vqvector})
		\right. \nonumber \\ && \left. \ \ \ \ \mbox{}
		+
		[n_{\rm B}(\Omega^{(\lambda)}_{\vqvector}) + 1 - n_{\rm F}(\varepsilon^{(b)}_{\vklabel} - \Omega^{(\lambda)}_{\vqvector})]
		\delta(\varepsilon_{\vkpparallel}^{(s)\pm} - \varepsilon^{(b)}_{\vklabel} + \Omega^{(\lambda)}_{\vqvector})
		\right\}
		\nonumber \\ && \mbox{} 
		+ 2 \pi |{\cal G}^{(sbR)\pm}_{\vkpparallel,\vklabel}|^2
		\left\{  
		[n_{\rm B}(\Omega^{(R)}_{\vqparallel}) + n_{\rm F}(\varepsilon^{(b)}_{\vklabel} + \Omega^{(R)}_{\vqparallel})] 
		\delta(\varepsilon_{\vkpparallel}^{(s)\pm} - \varepsilon^{(b)}_{\vklabel} - \Omega^{(R)}_{\vqparallel})
		\right. \nonumber \\ && \left. \ \ \ \ \mbox{} 
		+
		[n_{\rm B}(\Omega^{(R)}_{\vqparallel}) + 1 - n_{\rm F}(\varepsilon^{(b)}_{\vklabel} - \Omega^{(R)}_{\vqparallel})]
		\delta(\varepsilon_{\vkpparallel}^{(s)\pm} - \varepsilon^{(b)}_{\vklabel} + \Omega^{(R)}_{\vqparallel})  
		\right\}
		, \nonumber \\
		{\cal W}^{(ss)\pm}_{\vkpparallel,\vkparallel} &=&
		2 \pi
		\sum_{\lambda} \int_0^{\infty} \frac{dq_x}{2 \pi}
		|{\cal G}^{(ss\lambda)\pm}_{\vkpparallel,\vkparallel,q_x}|^2
		\left\{
		[n_{\rm B}(\Omega^{(\lambda)}_{\vqvector}) + n_{\rm F}(\varepsilon_{\vkparallel}^{(s)\pm} + \Omega^{(\lambda)}_{\vqvector})] 
		\delta(\varepsilon_{\vkpparallel}^{(s)\pm} - \varepsilon_{\vkparallel}^{(s)\pm} - \Omega^{(\lambda)}_{\vqvector})
		\right. \nonumber \\ && \left. \ \ \ \ \mbox{}
		+
		[n_{\rm B}(\Omega^{(\lambda)}_{\vqvector}) + 1 - n_{\rm F}(\varepsilon_{\vkparallel}^{(s)\pm} - \Omega^{(\lambda)}_{\vqvector})]
		\delta(\varepsilon_{\vkpparallel}^{(s)\pm} - \varepsilon_{\vkparallel}^{(s)\pm} + \Omega^{(\lambda)}_{\vqvector})  
		\right\}
		\nonumber \\ && \mbox{} 
		+ 2 \pi 
		|{\cal G}^{(ssR)\pm}_{\vkpparallel,\vkparallel}|^2
		\left\{
		[n_{\rm B}(\Omega^{(R)}_{\vqparallel}) + n_{\rm F}(\varepsilon_{\vkparallel}^{(s)\pm} + \Omega^{(R)}_{\vqparallel})] 
		\delta(\varepsilon_{\vkpparallel}^{(s)\pm} - \varepsilon_{\vkparallel}^{(s)\pm} - \Omega^{(R)}_{\vqparallel})
		\right. \nonumber \\ && \left. \ \ \ \ \mbox{} 
		+
		[n_{\rm B}(\Omega^{(R)}_{\vqparallel}) + 1 - n_{\rm F}(\varepsilon_{\vkparallel}^{(s)\pm} - \Omega^{(R)}_{\vqparallel})]
		\delta(\varepsilon_{\vkpparallel}^{(s)\pm} - \varepsilon_{\vkparallel}^{(s)\pm} + \Omega^{(R)}_{\vqparallel})  
		\right\}.
		\label{eq:calWssApp}
		\eea
	\end{widetext}
	The expressions for the bulk-bulk and for the arc-bulk kernel are specified in Eqs.\ \eqref{eq:calWbb} and \eqref{eq:calWbs}, respectively.
	
	We next provide explicit expressions for the collision integrals of the linearized Boltzmann theory.  For the sake of simplicity, 
	we here 
	restrict the arc-arc, bulk-arc, and arc-bulk scattering to the contribution from the Rayleigh surface modes.  However, the other phonon contributions to the collision integrals directly follow along similar lines and have been taken into account in our numerical analysis.
	
	The collision integrals ${\cal J}_{k_z}^{(ss)\pm}$ for arc-arc scattering and ${\cal J}^{(bs)\pm}_{k_z}$ for arc-bulk scattering (omitting the energy argument $\varepsilon$ throughout) read
	\begin{widetext}
		\bea
		\label{eq:JssExpl}
		{\cal J}^{(ss)\pm}_{k_z} &=&  \int \frac{dk_z'}{2 \pi}    \frac{1}{|v^{(s)\pm}_y(k_z')| }
		\frac{4 g_0^2 \xi^{(l)2} c_R q_{\parallel}^2  \kappa(k_z) \kappa(k'_z)}{  \rho_M c_l^2 [\kappa(k'_z) + \kappa (k_z) + q_{\parallel} \sqrt{1 - (c_R/c_l)^2}]^2 }
		{\cal F}(c_R q_{\parallel})
		[\varphi^{(s)\pm}_{k_z'} - \varphi^{(s)\pm}_{k_z}] , \\
		\nonumber
		{\cal J}^{(bs)\pm}_{k_z} &=& \int \frac{d\vkpparallel}{(2 \pi)^2}
		\,  \frac{1}{|v^{(b)}_{x}(\vkpparallel)|}
		\frac{8 g_0^2 \xi^{(l)2} q_{\parallel}^2 c_R \kappa(k_z)}{c_l^2   \rho_M   
			|\langle \xi_{\pm}(\alpha)|\xi_{(\mp k_x',\vkpparallel)}\rangle|^2}
		\left| \mbox{Im}\, \frac{\langle \xi_{\mp}(-\alpha) | \xi_{(\pm k_x',\vkpparallel)} \rangle \langle \xi_{\pm}(\alpha) | \xi_{(\mp k_x',\vkpparallel)} \rangle}{\kappa(k_z) + i k_x' + 
			\gamma_R q_{\parallel} } \right|^2
		\\  && \ \ \ \ \mbox{} \times\label{eq:JbsExpl}
		{\cal F}(c_R q_{\parallel})[\varphi^{(b)}_{\vkpparallel} - \varphi^{(s)\pm}_{k_z}],
		\eea
		where $\xi^{(l)}$ is a dimensionless number, see Eq.\ \eqref{eq:xil1}--\eqref{eq:xil3}, $c_R$ 
		the velocity of the Rayleigh mode,  $\gamma_R=\sqrt{1 - (c_R/c_l)^2}$, $\kappa(k_z)$ the inverse decay length of Fermi arc states, 
		see Eq.\ \eqref{eq:kapp2}. The function ${\cal F}(\Omega)$ has been defined in Eq.\ \eqref{eq:FdefF} and can be written as
		\be
		{\cal F}(\Omega)=  \frac{1}{\sinh(\Omega/ T)} +
		\tanh(\Omega/2 T) \frac{\cosh[(\varepsilon - \mu)/T] - 1}{\cosh[(\varepsilon - \mu)/T] + \cosh(\Omega/T)}. 
		\ee
		In Eq.\ \eqref{eq:JssExpl}, 
		$\vqparallel = (k^\pm_y(k'_z)-k_y^{\pm}(k_z))\hat y + (k'_z-k_z)\hat z$, see Eq.\ \eqref{eq:k2karch}.
		In Eq.\ \eqref{eq:JbsExpl},  $\vqparallel = (k'_y-k_y^{\pm}(k_z))\hat y + (k'_z-k_z)\hat z$ 
		and $k_x'$ is the positive solution of $\varepsilon^{(b)}_{(k'_x,\vkpparallel)} = \varepsilon$. 
		Similarly, the collision integral ${\cal J}^{(sb)\pm}_{\vkparallel}$ for bulk-arc scattering reads
		\bea
		{\cal J}^{(sb)\pm}_{\vkparallel} &=& \int \frac{dk_z'}{2 \pi} \frac{1}{|v^{(s)\pm}_y(k_z')|}
		\frac{8 g_0^2 \xi^{(l)2} q_{\parallel}^2 c_R \kappa(k'_z)}{c_l^2 \rho_M 
			|\langle \xi_{\pm}(\alpha)|\xi_{(\mp k_x,\vkparallel)}\rangle|^2}
		\left| \mbox{Im}\, \frac{\langle \xi_{\mp}(-\alpha) | \xi_{(\pm k_x,\vkparallel)} \rangle \langle \xi_{\pm}(\alpha) | \xi_{(\mp k_x,\vkparallel)} \rangle}{\kappa(k'_z) 
			+ i k_x + \gamma_R q_{\parallel} } \right|^2 \nonumber
		\\ && \ \ \ \ \mbox{} \times
		{\cal F}(c_R q_{\parallel})[\varphi^{(s)\pm}_{k_z'} - \varphi^{(b)}_{\vkparallel}],
		\eea
		where $k_x$ is the positive solution of $\varepsilon^{(b)}_{(k_x,\vkparallel)} = \varepsilon$ 
		and $\vqparallel = (k_y^{\pm}(k'_z)-k_y)\hat y + (k'_z-k_z)\hat z$.
		Finally, ${\cal J}^{(bb)}_{\vkparallel}$ is given by
		\bea
		{\cal J}^{(bb)}_{\vkparallel} &=& 
		\int \frac{d\vkpparallel}{(2 \pi)^2} \frac{1}{|v^{(b)}_x(\vkpparallel)|}
		\frac{g_0^2(  q {\cal F}(c_l q) 
			|\langle \xi_{(k_x',\vkpparallel)}|\xi_{(k_x,\vkparallel)} \rangle|^2 +
			\tilde q {\cal F}(c_l \tilde q) |\langle \xi_{(-k_x',\vkpparallel})|\xi_{(k_x,\vkparallel)} \rangle|^2)}{2 c_l \rho_M}
		[\varphi^{(b)}_{\vkpparallel} - \varphi^{(b)}_{\vkparallel}],
		\eea
		with $k_x$ and $k_x'$ the positive solutions of $\varepsilon^{(b)}_{(k_x,\vkparallel)} = \varepsilon^{(b)}_{(k'_x, \vkpparallel)} = \varepsilon$.
		Moreover, we use $\vqvector = 
		(k'_x-k_x)\hat x +\vkpparallel - \vkparallel$ and $ \tilde \vqvector = 
		(-k'_x-k_x)\hat x +\vkpparallel - \vkparallel.$
	\end{widetext}
	
	We finally summarize symmetry relations of the scattering kernels which follow from the inversion symmetry of the problem.
	For a given energy $\varepsilon$, the integral equations \eqref{eq:BE1} and \eqref{eq:BE2} are invariant under the replacement $\vkvector \to -\vkvector$ combined with interchanging the surfaces at $x = \pm L/2$. This gives the symmetry relations
	\be\label{Wssym}
	{\cal W}^{(ss)\pm}_{k_z',k_z}={\cal W}^{(ss)\mp}_{-k_z',-k_z}
	\ee
	for the arc-arc kernel. Similarly, the bulk-bulk kernel obeys   
	\be \label{Wbbsym}
	{\cal W}^{(bb)}_{\vkparallel',\vkparallel} = {\cal W}^{(bb)}_{-\vkparallel',-\vkparallel}.
	\ee
	The bulk-arc and arc-bulk kernels satisfies the relations 
	\bea
	{\cal W}^{(sb)\pm}_{k_z', \vkparallel } &=& {\cal W}^{(sb)\mp}_{-k_z',-\vkparallel}, \nonumber \\
	{\cal W}^{(bs)\pm}_{\vkparallel',k_z} &=& {\cal W}^{(bs)\mp}_{-\vkparallel',-k_z}. 
	\eea
	Since the velocities appearing on the left-hand side of Eqs.\ \eqref{eq:BE1} and \eqref{eq:BE2} are antisymmetric under inversion $I$, these symmetry relations imply that we may search for a solution of Eqs.\ \eqref{eq:BE4} and \eqref{eq:BE5} that satisfies the antisymmetry properties 
	\bea
	\varphi^{(b)}_{\vkparallel} &=& -\varphi^{(b)}_{-\vkparallel}, \nonumber \\
	\varphi^{(s)\pm}_{k_z} &=& -\varphi^{(s)\mp}_{-k_z}.
	\eea

	\section{On the arc-bulk rate}\label{appD}
	
	Here we outline the derivation of the temperature dependence of the arc-bulk rate 
	for $k_z$ close to the arc termination at $k_z= \bar k_{\rm W}$,
	where $\Omega_{\rm min}\ll T_{\rm BG}^{(b)}$.
	The starting point is Eq.~\eqref{eq:Gammabs}, where we retain only 
	the Rayleigh phonon contribution. 
	We will omit numerical factors of order unity and neglect the differences 
	between the phonon mode velocities, which we will all denote by the symbol $c$.
	For the Rayleigh mode we then have $iq_x^{(l)}\sim -q_\parallel$. Close to the arc 
	edge, the inverse decay length is
	\begin{equation}
	\kappa \sim \kkB (\varepsilon) - k_z \sim \sqrt{\frac{\Omega_{\rm min}\varepsilon}{cv}}.
	\label{kappaestimate}
	\end{equation}
	The integration domain in Eq.~\eqref{eq:Gammabs} is the projection of the bulk states in the $(k_y',k'_z)$-plane, that is
	(in the regime $\varepsilon \ll v \kb$) the disc $k'^2_y+(k_z'-\kb)^2 \leq \varepsilon^2/v^2$.
	We use a rotated coordinate system $(k_1',k_2')$ with the origin on the boundary of the domain at the point closest to 
	the initial arc momentum, so that the domain is parameterized as 
	$0<k_1'< 2 \varepsilon/v$, $|k_2'| \leq \sqrt{\varepsilon^2/v^2-(\varepsilon/v-k_1')^2}$.
	In this frame, the transferred phonon momentum is 
	\begin{equation}
	q_\parallel \sim \sqrt{(k'_1+ \Omega_{\rm min})^2+k_2'^2},
	\end{equation}
	while the transverse momentum $k_x$ and the velocity $v_x$ of the final-state bulk electron  are
	\be
	k_x \sim \sqrt{\frac{k_1'\varepsilon}{v}}, \quad v_x \sim v\sqrt{\frac{vk_1'}{\varepsilon}}. 
	\ee
	We next estimate the matrix element ${\cal G}^{bsR}_{\vkplabel,\vkparallel}$. We first observe that at $k_1'=0$, one has $k_x=0$, hence
	the reflection amplitudes $r_{k_x,\vkpparallel}=r_{k_x,\vkpparallel}'=1$. From Eqs.~\eqref{eq:GsbR} and \eqref{eq:GbsR} 
	we then see that the matrix element vanishes at the boundary of the projection of the bulk states. 
	Upon moving away from the boundary, the difference of the two fractions in Eq.~\eqref{eq:GsbR} 
	becomes nonzero. It becomes nonzero because the denominators are different once $k_x$ is
	nonzero, and because $r_{k_x,\vkpparallel}$ and $r'_{k_x,\vkpparallel}$ are no longer equal. Indeed, to linear order in $k_x$, 
	\be 
	r_{k_x,\vkpparallel} \sim (r_{k_x, \vkpparallel})^* \sim 1 + i\frac{vk_x}{\varepsilon}.
	\ee
	Which of the two mechanisms dominates depends on the relative magnitude of $\kappa+\qqparallel$ and $\varepsilon/v$.
	For $\Omega_{\rm min}\ll T^{(b)}_{\rm BG}$, the estimate \eqref{kappaestimate} implies 
	$\kappa \ll \varepsilon/v$. Since the maximum value of $\qqparallel$ is of order $\varepsilon/v$, 
	we conclude that we may safely restrict ourselves to the $k_x$-dependence arising from the 
	$k_x$ term in the denominators and approximate $r_{k_x,\vkpparallel} \approx r'_{k_x,\vkpparallel} \approx 1$.
	Combining all these estimates, we obtain 
	\begin{widetext}
		\be
		\label{appD:estimate1}
		\Gamma^{(bs)}_{\vkparallel} \sim \frac{g_0^2}{vc\rho_{\rm M}} \int_0^{\varepsilon/v} dk_1' \int^{\varepsilon/v}_{-\varepsilon/v} dk_2' 
		\frac{\kappa q_\parallel^2}{\sinh(cq_\parallel/T)} \sqrt{\frac{\varepsilon}{vk'_1}} \frac{k_x^2}{((\kappa + q_\parallel)^2+k_x^2)^2}.
		\ee
	\end{widetext}
	
	We first consider the case $\Omega_{\rm min}\ll T$. The integrand is sharply peaked for small $k_1'$, 
	but the main contribution to the integral comes from $k_1',|k_2'|\lesssim {\rm min}(T/c,\varepsilon/v)$. The
	integral can be estimated by setting \mbox{$k_1',|k_2'| \sim {\rm min}(T/c,\varepsilon/v)$} in the integrand and multiplying
	by a factor $\sim {\rm min}(T/c,\varepsilon/v)^2$ to account for the integration volume. Then 
	for $\Omega_{\rm min}\ll T_{\rm BG}^{(b)}\ll T$, we have $T/c \gg \varepsilon/v$ and we find
	\begin{align}
	\label{appD:estimate3}
	\Gamma^{(bs)}_{\vkparallel} & \sim \frac{g_0^2}{vc\rho_{\rm M}} (\varepsilon/v)^2 \sqrt{\frac{\Omega_{\rm min} \varepsilon}{cv}} 
	(\varepsilon/v)^2 \frac{Tv}{c\varepsilon} \frac{v^2}{\varepsilon^2} \nonumber \\
	& \sim \Gamma_0 \frac{T \Omega_{\rm min}^{1/2}T^{(b)3/2}_{\rm BG}}{T^3_{\rm BG}}.
	\end{align}
	Similarly, for  $\Omega_{\rm min}\ll T\ll T_{\rm BG}^{(b)}$, we have $T/c \ll \varepsilon/v$ and we get
	\begin{align}
	\label{appD:estimate2}
	\Gamma^{(bs)}_{\vkparallel} & \sim  \frac{g_0^2}{vc\rho_{\rm M}} (T/c)^2 
	\sqrt{\frac{\Omega_{\rm min}\varepsilon}{cv}}(T/c)^2
	\sqrt{\frac{c \varepsilon}{vT}} \frac{cv}{T\varepsilon} \nonumber \\
	& \sim \Gamma_0 \frac{T^{5/2} \Omega_{\rm min}^{1/2}}{T^3_{\rm BG}}.
	\end{align}
	For $T\ll \Omega_{\rm min}$, on the other hand,
	the effective integration range is $k_1' \lesssim T/c$, $|k_2'| \lesssim \sqrt{T \Omega_{\rm min}/c^2}$. 
	Within this range, one may approximate $(\kappa + \qqparallel)^2 + k_x^2 \approx 
	\kappa^2 \sim \Omega_{\rm min} \varepsilon/cv$. To estimate the integral, we set 
	$k_1' \sim T/c$, $k'_2 \sim \sqrt{\Omega_{\rm min}/c^2}$,
	$\qqparallel \sim \Omega_{\rm min}/c$, and multiply with the integration volume, 
	which is $\sim (T/c)^{3/2}(\Omega_{\rm min}/c)^{1/2}$. This gives the estimate
	\begin{widetext}
		\begin{align}
		\label{appD:estimate4}
		\Gamma^{(bs)}_{\vkparallel} & \sim  
		\frac{g_0^2}{vc\rho_{\rm M}} (T/c)^{3/2}(\Omega_{\rm min}/c)^{1/2} \sqrt{\frac{\Omega_{\rm min} \varepsilon}{cv}} 
		(\Omega_{\rm min}/c)^2 e^{-\Omega_{\rm min}/T} \sqrt{\frac{c\varepsilon}{vT}} \frac{T\varepsilon}{cv}
		\left(\frac{cv}{\Omega_{\rm min}\varepsilon}\right)^2
		\nonumber \\
		&\sim \Gamma_0 \frac{ \Omega_{\rm min} T^2}{T_{\rm BG}^3} e^{-\Omega_{\rm min}/T}.
		\end{align}
	\end{widetext}

	\section{Hall response}\label{appE}
	
	We here briefly discuss the Hall response.
	The presence of a net current along the chiral ($\hat y$) direction carried by Fermi-arc surface states comes with a finite Hall voltage $V_{\perp}$. Comparing Eqs.\ \eqref{eq:J_lin} and \eqref{eq:Vperp_lin}, and using $n_{\rm FA}(\varepsilon) v_y^{(s)\pm} \sim \pm \kb$, we estimate 
	\begin{equation}
	e V_{\perp} \sim \frac{J^{(s)+}- J^{(s)-}}{\kb}.    
	\end{equation}
	As a result, the Hall voltage is given by
	\be
	e V_{\perp} \sim \frac{\sigma_{yy}^{(s)}}{\kb} E_y.
	\ee
	There is no transverse response for an electric field applied along the $\hat z$ direction.
	
	\bibliographystyle{aipnum4-1}
	\bibliography{WSM}
	
\end{document}